\documentclass{aa}
\usepackage[varg]{txfonts}
\usepackage[breaklinks=true]{hyperref}
\hypersetup{
    colorlinks,
    linkcolor={blue},
    citecolor={blue},
    urlcolor={blue}
}

\usepackage{natbib,twoopt}
\bibpunct{(}{)}{;}{a}{}{,} 
\makeatletter
\newcommandtwoopt{\citeads}[3][][]{\href{http://adsabs.harvard.edu/abs/#3}%
{\def\hyper@linkstart##1##2{}%
\let\hyper@linkend\@empty\citealp[#1][#2]{#3}}}
\newcommandtwoopt{\citepads}[3][][]{\href{http://adsabs.harvard.edu/abs/#3}%
{\def\hyper@linkstart##1##2{}%
\let\hyper@linkend\@empty\citep[#1][#2]{#3}}}
\newcommandtwoopt{\citetads}[3][][]{\href{http://adsabs.harvard.edu/abs/#3}%
{\def\hyper@linkstart##1##2{}%
\let\hyper@linkend\@empty\citet[#1][#2]{#3}}}
\newcommandtwoopt{\citeyearads}[3][][]%
{\href{http://adsabs.harvard.edu/abs/#3}
{\def\hyper@linkstart##1##2{}%
\let\hyper@linkend\@empty\citeyear[#1][#2]{#3}}}
\makeatother

\newcommand\hst{{HST}\xspace}
\newcommand\chandra{\textit{Chandra}\xspace}

\graphicspath{{./figures/}}

\begin{document}

\title{A MUSE view of the massive merging galaxy cluster ACT-CL~J0102$-$4915 (El Gordo) at $z$ = 0.87}
\subtitle{Robust strong lensing model and data release}

\author{G.~B.~Caminha       \inst{\ref{tum},\ref{mpa}}                        
                            \thanks{e-mail address: \href{mailto:gb.caminha@tum.de}{gb.caminha@tum.de}. Table \ref{tab:full_z_cat} containing the full redshift catalogue and lens model files are only available in electronic form at the CDS via anonymous ftp to cdsarc.u-strasbg.fr (130.79.128.5) or via \url{http://cdsweb.u-strasbg.fr/cgi-bin/qcat?J/A+A/}.} \and
        C.~Grillo \inst{\ref{unimi},\ref{inafmilano}} \and
        P.~Rosati \inst{\ref{unife},\ref{inafbo}} \and
        A.~Liu \inst{\ref{mpe}} \and
        A.~Acebron \inst{\ref{unimi},\ref{inafmilano}} \and
        P.~Bergamini \inst{\ref{unimi}, \ref{inafbo}} \and
        K.~I.~Caputi        \inst{\ref{Kapteyn},\,\ref{dawn}} \and
        A.~Mercurio \inst{\ref{inafna}} \and
        P.~Tozzi \inst{\ref{arcetri}} \and
        E.~Vanzella \inst{\ref{inafbo}} \and
        R.~Demarco \inst{\ref{concepcion}} \and
        B.~Frye \inst{\ref{arizona}} \and
        G.~Rosani \inst{\ref{digital_groningen},\,\ref{Kapteyn}} \and
        K.~Sharon \inst{\ref{annarbor}}
        }
\institute{
Technical University of Munich, TUM School of Natural Sciences, Department of Physics, James-Franck-Str 1, 85748 Garching, Germany \label{tum} \and
Max-Planck-Institut f\"ur Astrophysik, Karl-Schwarzschild-Str. 1, D-85748 Garching, Germany \label{mpa} \and
Dipartimento di Fisica, Universit\`a  degli Studi di Milano, via Celoria 16, I-20133 Milano, Italy \label{unimi} \and
INAF - IASF Milano, via A. Corti 12, I-20133 Milano, Italy \label{inafmilano} \and
Dipartimento di Fisica e Scienze della Terra, Universit\`a degli Studi di Ferrara, via Saragat 1, I-44122 Ferrara, Italy \label{unife} \and
INAF -- OAS, Osservatorio di Astrofisica e Scienza dello Spazio di Bologna, via Gobetti 93/3, I-40129 Bologna, Italy \label{inafbo} \and
Max Planck Institute for Extraterrestrial Physics, Giessenbach-strasse 1, 85748 Garching, Germany \label{mpe} \and
Kapteyn Astronomical Institute, University of Groningen, Postbus 800, 9700 AV Groningen, The Netherlands \label{Kapteyn}\and
The Cosmic Dawn Center, Niels Bohr Institute, University of Copenhagen, Juliane Maries Vej 30, DK-2100 Copenhagen {\O}, Denmark\label{dawn} \and
INAF -- Osservatorio Astronomico di Capodimonte, Via Moiariello 16, I-80131 Napoli, Italy \label{inafna} \and
INAF - Osservatorio Astrofisico di Arcetri, Largo E. Fermi, I-50125 Firenze, Italy \label{arcetri} \and
Departamento de Astronom\'ia, Facultad de Ciencias F\'isicas y Matem\'aticas, Universidad de Concepci\'on, Concepci\'on, Chile \label{concepcion} \and
Department of Astronomy/Steward Observatory, University of Arizona, 933 N. Cherry Avenue, Tucson, AZ 85721, USA \label{arizona} \and
Digital Competence Centre, University of Groningen, Centre for Information Technology (CIT), Nettelbosje 1, 9747AJ Groningen, The Netherlands \label{digital_groningen} \and
Department of Astronomy, University of Michigan, 1085 S. University Ave, Ann Arbor, MI 48109, USA \label{annarbor}
}

\abstract{
We present a detailed strong lensing analysis of the massive and distant ($z=0.870$) galaxy cluster ACT-CL~J0102$-$4915 (ACT0102, also known as El Gordo), taking advantage of new spectroscopic data from the Multi Unit Spectroscopic Explorer (MUSE) on the Very Large Telescope and archival imaging from the \textit{Hubble} Space Telescope.
Thanks to the MUSE data, we were able to measure secure redshifts for 374 single objects, including 23 multiply lensed galaxies, and 167 cluster members of ACT0102.
We used the observed positions of 56 multiple images, along with their new spectroscopic redshift measurements, as constraints for our strong lensing model.
Remarkably, some multiple images are detected out to a large projected distance of $\approx 1$~Mpc from the brightest cluster galaxy, allowing us to estimate a projected total mass value of $1.84_{-0.04}^{+0.03} \times 10^{15}\, \rm M_{\odot}$ within that radius.
We find that we need two extended cluster mass components, the mass contributions from the cluster members and the additional lensing effect of a foreground ($z=0.633$) group of galaxies, to predict the positions of all multiple images with a root mean square offset of $0\farcs75$.
The main cluster-scale mass component is centred very close to the brightest cluster galaxy, and the other extended mass component is located in the north-west region of the cluster. These two mass components have very similar values of mass projected within 300~kpc of their centres, namely $2.29_{-0.10}^{+0.09}\times10^{14}\,\rm M_{\odot}$ and $2.10_{-0.09}^{+0.08}\times10^{14}\,\rm M_{\odot}$, in agreement with the major merging scenario of ACT0102.
We make publicly available the lens model, including the magnification maps and posterior distributions of the model parameter values, as well as the full spectroscopic catalogue containing all redshift measurements obtained with MUSE.}


\maketitle

\section{Introduction}
\label{sec:introduction}

\begin{figure*}
  \centering
  
  \includegraphics[width = 1.0\textwidth]{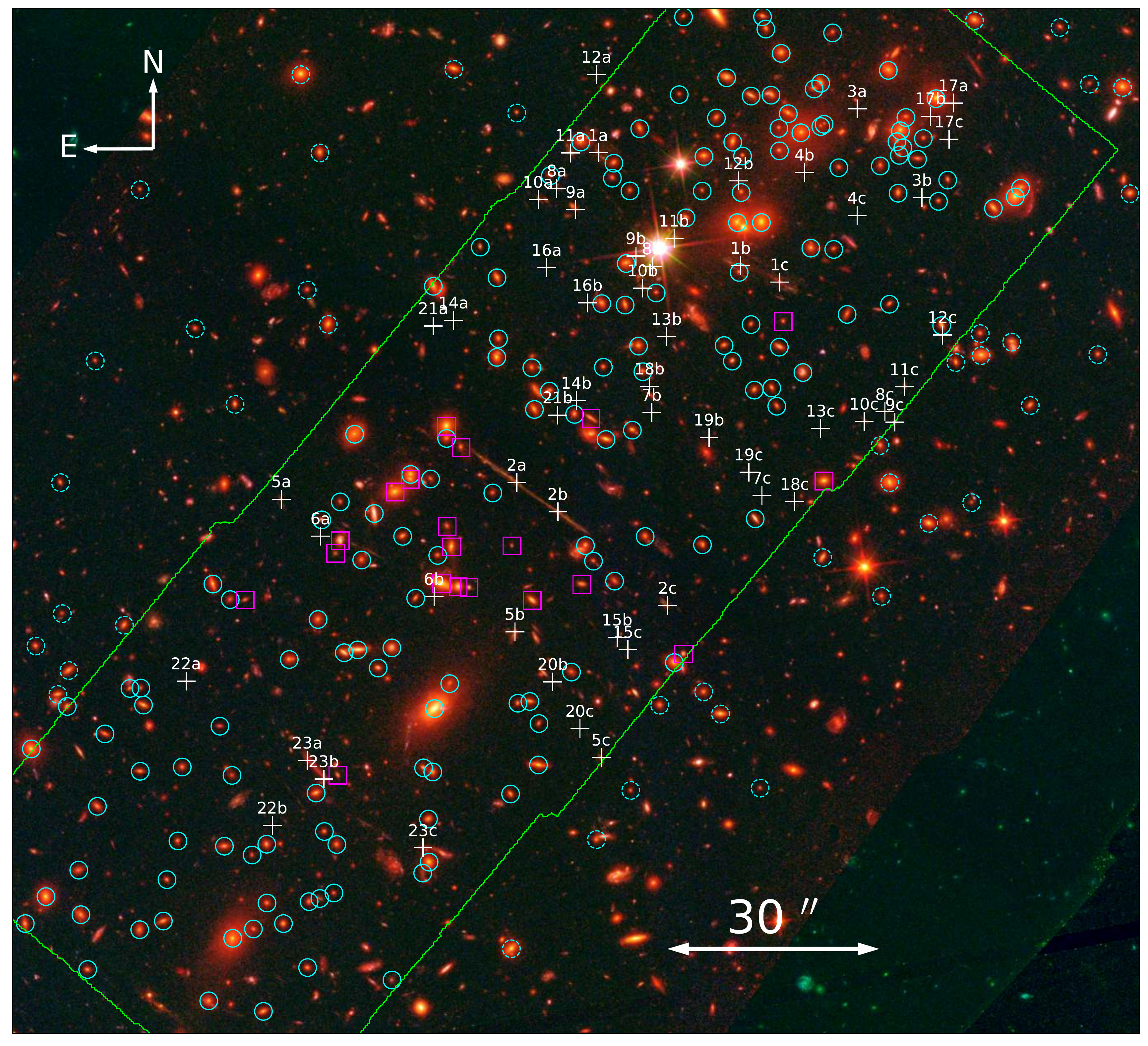}

  \caption{ACT0102 MUSE field of view overlaid on a \hst colour image (where the filter F435W is blue; F606W+F625W+F775W+F850LP is green; and F105W+F125W+F140W+F160W is red). Green lines show the MUSE mosaic footprint, which is composed of three pointings with an exposure time of $\approx 2.3$~hours. Cyan (dashed) circles indicate spectroscopically (photometrically) selected cluster members (see Fig. \ref{fig:redsequence}). Galaxies marked with magenta boxes belong to the group at $z=0.63$ (see Fig. \ref{fig:cluster_histogram}). The multiple images used in our strong lensing model are shown by white crosses, and all the multiple image families have secure spectroscopic redshift measurements.}
  \label{fig:members_all}
\end{figure*}

The well-established scenario of a hierarchical structure formation of the Universe predicts that small overdensities undergo merging events, growing in mass across cosmic time and forming massive clusters of galaxies \citep[see e.g.][]{1991ApJ...379...52W, 1996ApJ...462..563N}.
In such a scenario, clusters act as crossroads between cosmology and astrophysics, and thus carry precious cosmological and astrophysical information.
Their abundance is mainly driven by the amplitude of mass density fluctuations (parameterised by the quantity $\sigma_8$) and the total mass density of the Universe, $\Omega_m$ \citep{1998ApJ...508..483W,Rosati_02,2020PhRvD.102b3509A}.
Moreover, the merging history is also a factor that shapes the properties of galaxy clusters.
For instance, galaxy clusters with no recent merging event are expected to have a more regular distribution of satellite galaxies and a very massive and bright central galaxy compared to other cluster members \citep[see e.g.][]{2005ApJ...630L.109D, 2021A&A...655A.103Z}.
On the other hand, numerous recent merging events tend to disturb the dynamical state of galaxies in clusters and produce a spatial offset between the dark-matter distribution and the intracluster hot gas, traced by the X-ray emission \citep{2004ApJ...604..596C, 2006ApJ...652..937B}.
Thus, an accurate description of the total mass distribution in clusters is crucial to better understanding how these structures evolve across cosmic time.

Gravitational lensing is one of the most direct ways to measure the total mass in galaxies \citep{2012ApJ...747L..15G,2014MNRAS.439.2494O} and galaxy clusters \citep{2016ApJ...821..116U,2019A&A...632A..36C} because it does not depend on baryonic or dynamical processes \citep{2011A&ARv..19...47K, 2021LNP...956.....M}.
In the very inner cores of galaxy clusters, that is, the region characterised by the strong lensing regime (a few hundred kiloparsecs), a detailed total mass map can be obtained using the model constraints provided by a large set of spectroscopically confirmed multiple images from background galaxies \citep[see e.g.][]{2017MNRAS.469.3946L,2017A&A...607A..93C,2022arXiv220709416B}.

All these studies, supported by high-precision lens models, have mostly focused on clusters at relatively low redshifts, mostly in the range 0.3--0.5.
Their inferred internal total mass distribution, particularly that from their sub-halo mass component, has recently been compared to state-of-the-art cosmological simulations, revealing intriguing tensions with expectations in the $\Lambda$ cold dark matter (CDM) scenario \citep[see][and references therein]{Meneghetti_2020,2022arXiv220409065M,2022arXiv220409067R}. By extending these studies to similarly massive systems at higher redshifts, to $z\sim\! 1$, one can further test $\Lambda$CDM predictions on structure formation.
In addition, a detailed characterisation of the internal mass structure of clusters at $z\sim 1$, and possibly beyond, can help complete our knowledge of the evolution of structures at earlier ages, from protoclusters at $z>2$ to local massive galaxy clusters.

However, this has been difficult to date due to the rarity of high-z ($z\gtrsim 0.8$) cluster lenses and the general decrease in the number of strong lensing features in clusters at progressively higher redshifts \citep[see e.g.][]{2018ApJ...863..154P, 2019ApJ...874..132A,2020ApJ...894..150M}. In particular, adequate supporting spectroscopic datasets needed for high-precision strong lensing modelling are lacking for the few high-$z$ systems known. 

The galaxy cluster ACT-CL~J0102-4915 (hereafter ACT0102) is one of the most massive and gravitationally bound structures at $z\approx1.0$, when the Universe was approximately half of its current age ($\approx6.3$~Gyr after the Big Bang).
It was first identified via the Sunyaev-Zeldovich effect by the Atacama Cosmology Telescope \citep{2010ApJ...723.1523M}. 
Subsequent photometric, spectroscopic, and X-ray follow-up observations found that ACT0102 is a very massive galaxy cluster ($M_{200} \approx 2 \times 10^{15}~M_{\odot}$)  that underwent a major merging event at $z\approx 0.87$ \citep{2012ApJ...748....7M,2014ApJ...785...20J}.
The exceptional conditions of such a merger at a high redshift have been investigated in the context of $\Lambda$CDM expectations \citep{2015ApJ...813..129Z, 2018ApJ...855...36Z, 2021MNRAS.500.5249A} and with dedicated X-ray and radio studies \citep{2014ApJ...786...49L, 2016MNRAS.463.1534B, 2016ApJ...829L..23B}.

The spectroscopic campaign of ACT0102 with the Multi Unit Spectroscopic Explorer (MUSE) presented in this work, combined with archival \textit{Hubble} Space Telescope (\hst) high-resolution imaging, allowed us to develop, for the first time, a high-precision and accurate lens model of a distant, massive cluster. This works enables a detailed characterisation of the total mass distribution of the cluster lens, as similarly performed for low-redshift systems.
Given its high mass and redshift, this system is a unique laboratory for studying cosmology and galaxy evolution.

In this work we take advantage of the latest spectroscopic and photometric MUSE and \hst data to measure the mass distribution of ACT0102 by constructing the finest strong lens model of this cluster so far. This paper is organised as follows. In Sect. \ref{sec:data_sets} we present the photometric and spectroscopic data used in our analyses. In Sect. \ref{sec:sl_model} we describe our lens model in detail, and in Sect. \ref{sec:discussions_and_comparison} we discuss the results and compare our total mass reconstruction with those from previous works.
In Sect. \ref{sec:conclusions} we summarise our conclusions and future perspectives.
Finally, in Appendix \ref{ap:multiple_image_spectra} we present the MUSE spectra of multiple images.
Figures are oriented with north to the top and east to the left.
Throughout this work, we adopt a flat $\Lambda$CDM cosmology with $\Omega_m=0.3$.
For this cosmology, $1\arcsec$ corresponds to 7.714~kpc at the cluster redshift, $z_{cluster} = 0.8704$.

\section{Datasets}
\label{sec:data_sets}

\begin{table*}[!]
\centering
\small
\caption{Complete MUSE redshift catalogue  (extract).}
\begin{tabular}{c c c c c c c l} \hline \hline
ID &  RA & Dec & $z_{\rm spec}$ & QF & mult.\\
(1)&  (2)& (3) & (4)            & (5)& (6)\\
\hline
ACT0102-J$010301.08$$-491559.58$ & 15.7544997 & $-$49.2665507 & 0.0000 & 4 & 1 \\
ACT0102-J$010301.90$$-491659.86$ & 15.7579086 & $-$49.2832932 & 0.0000 & 4 & 1 \\
ACT0102-J$010252.00$$-491429.73$ & 15.7166775 & $-$49.2415930 & 0.0000 & 4 & 1 \\
ACT0102-J$010254.20$$-491502.07$ & 15.7258268 & $-$49.2505744 & 0.0000 & 4 & 1 \\
ACT0102-J$010254.50$$-491514.05$ & 15.7270819 & $-$49.2539034 & 0.0000 & 4 & 1 \\
ACT0102-J$010253.29$$-491511.04$ & 15.7220268 & $-$49.2530679 & 0.0000 & 4 & 1 \\
ACT0102-J$010256.24$$-491530.92$ & 15.7343249 & $-$49.2585899 & 0.1330 & 9 & 1 \\
ACT0102-J$010256.83$$-491528.75$ & 15.7368105 & $-$49.2579870 & 0.1932 & 2 & 1 \\
ACT0102-J$010304.58$$-491636.96$ & 15.7690796 & $-$49.2769329 & 0.2082 & 3 & 1 \\
ACT0102-J$010259.94$$-491714.34$ & 15.7497641 & $-$49.2873157 & 0.2219 & 3 & 1 \\
\vdots & \vdots & \vdots & \vdots & \vdots & \vdots \\
\hline
\end{tabular}
\label{tab:full_z_cat}
\tablefoot{The complete MUSE redshift catalogue is available at the CDS. The columns correspond to: (1) the ID built from the cluster name and object RA and Dec; (2) and (3) are the observed right ascension and declination in degrees (J2000) using as a reference the RELICS public images \citep{2019ApJ...884...85C}. The astrometry of these photometric data is calibrated with the Wide-field Infrared Survey Explorer point source
catalogue \citep{2010AJ....140.1868W}; (4) and (5) are the spectroscopic redshift value and its QF; (6) is the number of entries of the same object in this catalogue used to indicate multiply lensed sources.
}
\end{table*}

In this section we describe the data used in this work. It consists of multi-wavelength, high-resolution imaging from the \hst and deep spectroscopy from MUSE at the Very Large Telescope (VLT).

\subsection{\hst photometry}
\label{sec:hst_photometry}

\begin{figure}
  \centering
      \includegraphics[width = 1.0\columnwidth]{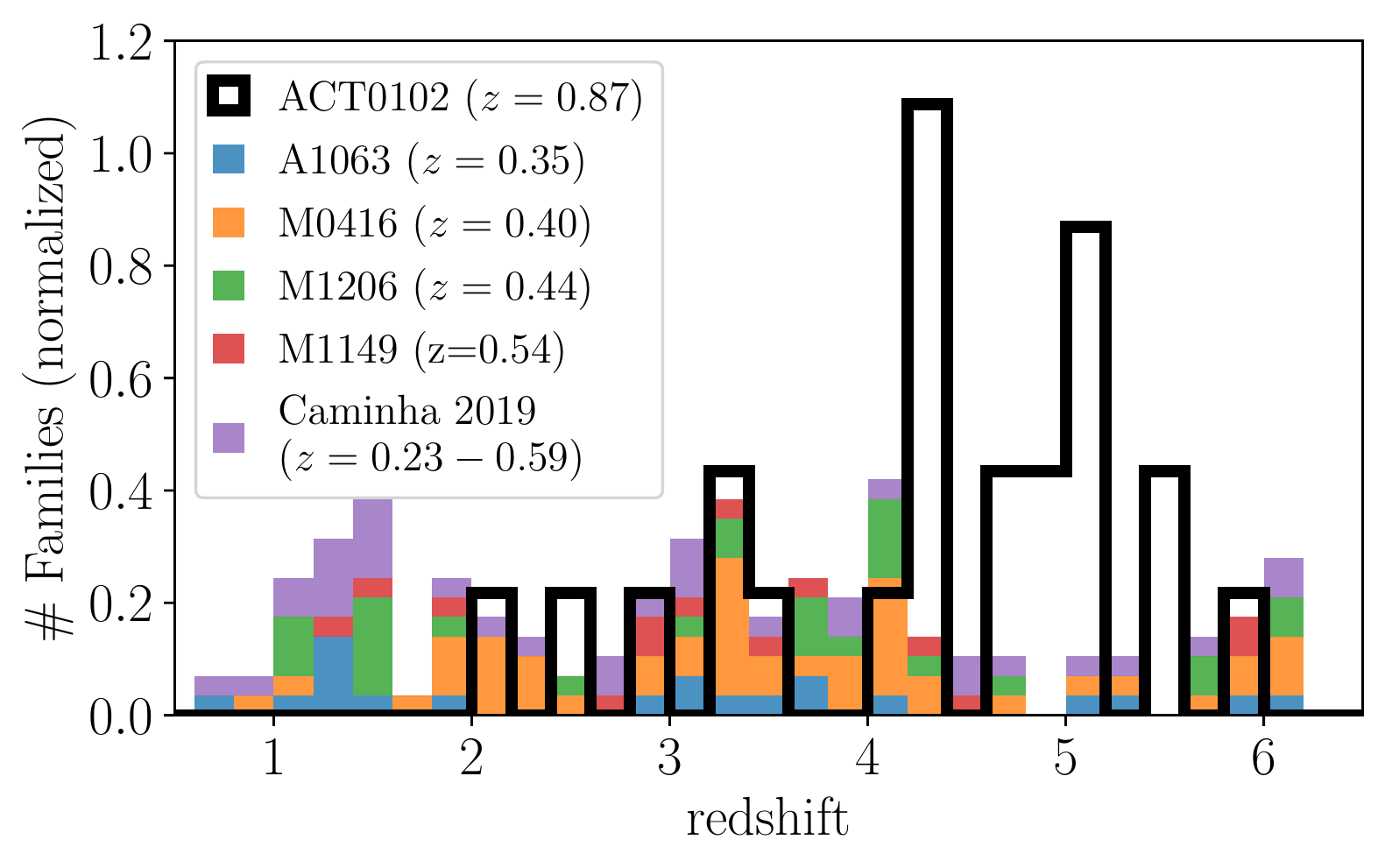}
  \caption{Normalised redshift distributions of the multiply lensed sources of ACT0102 (back histogram, accounting for multiplicity) and of other clusters with extensive spectroscopic data (coloured stacked histogram). The multiple images of the previous sample are presented in \citet{2016ApJ...822...78G}, \citet{2017A&A...607A..93C}, \citet{2019A&A...632A..36C}, and \citet{2021A&A...645A.140B}.}
  \label{fig:families_z_hist}
\end{figure}

The central region of ACT0102 was observed by the \hst under programmes 12755 (P.I.: J. Hughes) and 12477 (P.I. High, F.) in the optical bands F606W, F625W, F775W, F814W, and F850LP.
Additional data in the F435W filter, and the infra-red bands F105W, F125W, F140W, and F160W were obtained by the \hst Treasury programme Reionization Lensing Cluster Survey \citep[RELICS; ID 14096,][]{2019ApJ...884...85C}.
We used the data products (i.e. the reduced images and photometric catalogues) made publicly available by the RELICS team via the Mikulski Archive for Space Telescope (MAST\footnote{\url{https://archive.stsci.edu/prepds/relics/}}).
The imaging depths, considering a $5\sigma$ detection of point sources, vary within the range $\approx27.2$ and $\approx 26.5$ mag, from the blue to the red filters \citep{2019ApJ...884...85C}.
In Fig. \ref{fig:members_all} we show a \hst colour composite image generated using the software {\tt Trilogy} \citep{2012ApJ...757...22C}.

The depth of the \hst imaging is especially important to detect faint objects and multiple image candidates.
Moreover, its spatial resolution allows us to identify and measure the precise positions of the peaks of the surface brightness distribution of extended multiple images, used as input in our strong lensing models (see Sect. \ref{sec:sl_model}).
We note that some Lyman-$\alpha$ emitters are not clearly detected in the \hst photometry; however, MUSE provides a secure confirmation of such \hst-`dark' objects at $z>2.9$, as we briefly discuss in the following section.

\subsection{MUSE spectroscopy}
\label{sec:muse_spectroscopy}

\begin{table*}[!htbp]
\centering           
\caption{\label{tab:multiple_images} Spectroscopic redshift catalogue of the multiple images in ACT0102.}
\begin{tabular}{lccccccccccc}
\hline
ID & RA & Dec. & $z_{\rm MUSE}$ & $\rm ID^{\rm Zitrin}$ & $z_{\rm model}^{\rm Zitrin}$ & $\rm ID^{\rm Cerny}$ & $z_{\rm model}^{\rm Cerny}$ & $\rm ID^{\rm Diego}$ & $z_{\rm model}^{\rm Diego}$\\

\hline
1a    &  15.7307953  &  $-$49.2500932  &  2.5636  & 1.3 & $2.69_{-1.54}^{+0.69}$ & 1.3/10.3 & $2.99_{-0.15}^{+0.29}$ &  1.3 & [3]    \\
1b    &  15.7222014  &  $-$49.2545452  &  2.5636  & 1.1 & $''$                              & 1.1/10.1 & $''$                   &  1.1 & $''$ \\
1c    &  15.7198393  &  $-$49.2551912  &  2.5636  & 1.2 & $''$                                  & 1.2/10.2 & $''$                   &  1.2 & $''$ \\
2a    &  15.7357164  &  $-$49.2630878  &  2.8254  & 2.2 & $2.11_{-0.26}^{+0.96}$   & 2.2/20.2 & [3.3] & 2.2 & [3.3]\\
2b    &  15.7332368  &  $-$49.2642377  &  2.8254  & 2.1 & ---   & 2.1/20.1 & $''$  & 2.1 & $''$ \\
2c    &  15.7266000  &  $-$49.2679302  &  2.8254  & 2.3 & ---   & 2.3/20.3 & $''$  & 2.3 & $''$ \\
3a$*$ &  15.7151505  &  $-$49.2483621  &  3.3300  & --- & ---   & --- & --- & --- & --- \\
3b$*$ &  15.7112463  &  $-$49.2518560  &  3.3300  & --- & ---   & --- & --- & --- & --- \\
4b        &  15.7183394  &  $-$49.2508679  &  3.3339  & --- & ---   & --- & --- & --- & --- \\
4c    &  15.7151687  &  $-$49.2525681  &  3.3339  & --- & ---   & --- & --- & --- & --- \\
5a    &  15.7499300  &  $-$49.2637435  &  3.5376  & 4.1 & $2.11_{-0.29}^{+0.76}$  & 4.2 & $4.61^{+2.29}_{-0.74}$ & 4.1 & [3.2] \\
5b    &  15.7358293  &  $-$49.2689701  &  3.5376  & 4.5 & $''$                    & 4.3 & $''$                   & 4.3 & $''$ \\
5c    &  15.7306204  &  $-$49.2739172  &  3.5376  & 4.4 & $''$                    & 4.1 & $''$                   & 4.2 & $''$ \\
6a    &  15.7475700  &  $-$49.2652013  &  4.1879  & --- & ---   & --- & --- & --- & --- \\
6b    &  15.7407187  &  $-$49.2675811  &  4.1879  & --- & ---   & --- & --- & --- & --- \\
7b$*$ &  15.7275678  &  $-$49.2603268  &  4.2306  & --- & ---   & --- & --- & --- & --- \\
7c$*$ &  15.7209037  &  $-$49.2635983  &  4.2306  & --- & ---   & --- & --- & --- & --- \\
8a    &  15.7332986  &  $-$49.2515002  &  4.3175  & --- & ---   & --- & --- & --- & --- \\
8b    &  15.7275328  &  $-$49.2545678  &  4.3175  & --- & ---   & --- & --- & --- & --- \\
8c    &  15.7134793  &  $-$49.2602988  &  4.3175  & --- & ---   & --- & --- & 6.3$\dagger$ & [4.3] \\
9a    &  15.7321595  &  $-$49.2523270  &  4.3196  & --- & ---   & --- & --- & --- & --- \\
9b    &  15.7285210  &  $-$49.2541661  &  4.3196  & --- & ---   & --- & --- & --- & --- \\
9c    &  15.7128838  &  $-$49.2607093  &  4.3196  & --- & ---   & --- & --- & --- & --- \\
10a   &  15.7344194  &  $-$49.2519422  &  4.3275  & 3.1 & [4.16]& 3.1 & $7.42^{+0.58}_{-1.72}$ & 3.1 & [4.4] \\
10b   &  15.7281160  &  $-$49.2554323  &  4.3275  & 3.2 & $''$  & 3.2 & $''$                   & 3.2 & $''$ \\
10c   &  15.7147336  &  $-$49.2606804  &  4.3275  & 3.3 & $''$  & 3.3 & $''$                   & 3.3 & $''$ \\
11a   &  15.7324855  &  $-$49.2501018  &  4.3278  & --- & ---   & --- & --- & 17.1 & --- \\
11b   &  15.7262164  &  $-$49.2534759  &  4.3278  & --- & ---   & --- & --- & 17.2 & --- \\
11c   &  15.7123036  &  $-$49.2593173  &  4.3278  & --- & ---   & --- & --- & 17.3& --- \\
12a   &  15.7309031  &  $-$49.2470175  &  4.7042  & --- & ---   & --- & --- & --- & --- \\
12b   &  15.7223336  &  $-$49.2512054  &  4.7042  & --- & ---   & --- & --- & --- & --- \\
12c   &  15.7100095  &  $-$49.2572749  &  4.7042  & --- & ---   & --- & --- & --- & --- \\
13b$*$&  15.7266712  &  $-$49.2573340  &  4.7528  & --- & ---   & --- & --- & --- & --- \\
13c$*$&  15.7173614  &  $-$49.2609511  &  4.7528  & --- & ---   & --- & --- & --- & --- \\
14a$*$&  15.7395296  &  $-$49.2566966  &  4.9486  & --- & ---   & --- & --- & --- & --- \\
14b$*$&  15.7321155  &  $-$49.2598586  &  4.9486  & --- & ---   & --- & --- & --- & --- \\
15b   &  15.7296464  &  $-$49.2691880  &  4.9770  & --- & ---   & --- & --- & --- & --- \\
15c   &  15.7290102  &  $-$49.2696687  &  4.9770  & --- & ---   & --- & --- & --- & --- \\
16a$*$&  15.7339095  &  $-$49.2546160  &  5.0880  & --- & ---   & --- & --- & --- & --- \\
16b$*$&  15.7314569  &  $-$49.2560088  &  5.0880  & --- & ---   & --- & --- & --- & --- \\
17a$*$&  15.7093400  &  $-$49.2481341  &  5.0929  & --- & ---   & --- & --- & --- & --- \\
17b$*$&  15.7107380  &  $-$49.2486519  &  5.0929  & --- & ---   & --- & --- & --- & --- \\
17c$*$&  15.7096130  &  $-$49.2495637  &  5.0929  & --- & ---   & --- & --- & --- & --- \\
18b$*$&  15.7276976  &  $-$49.2593015  &  5.1173  & --- & ---   & --- & --- & --- & --- \\
18c$*$&  15.7189278  &  $-$49.2638429  &  5.1173  & --- & ---   & --- & --- & --- & --- \\
19b   &  15.7241041  &  $-$49.2613159  &  5.1198  & --- & ---   & --- & --- & --- & --- \\
19c   &  15.7217032  &  $-$49.2626827  &  5.1198  & --- & ---   & --- & --- & --- & --- \\
20b$*$&  15.7335442  &  $-$49.2709446  &  5.4845  & --- & ---   & --- & --- & --- & --- \\
20c$*$&  15.7319066  &  $-$49.2727819  &  5.4845  & --- & ---   & --- & --- & --- & --- \\
21a$*$&  15.7407597  &  $-$49.2569208  &  5.5811  & --- & ---   & --- & --- & --- & --- \\
21b$*$&  15.7332464  &  $-$49.2604359  &  5.5811  & --- & ---   & --- & --- & --- & --- \\
22a   &  15.7556999  &  $-$49.2709142  &  5.9520  & --- & ---   & --- & --- & --- & --- \\
22b   &  15.7504956  &  $-$49.2765971  &  5.9520  & --- & ---   & --- & --- & --- & --- \\
23a   &  15.7483751  &  $-$49.2740433  &  2.1887  & c5.1& $2.21_{-0.30}^{+1.83}$ & --- & --- & 11.1 & [3.1] \\
23b   &  15.7473758  &  $-$49.2747730  &  2.1887  & c5.2& $''$                   & --- & --- & 11.2 & $''$ \\
23c   &  15.7413883  &  $-$49.2774839  &  2.1887  & --- & ---                    & --- & --- & 11.3 & $''$ \\
\hline

\end{tabular}
\tablefoot{Multiple images marked with asterisks are Lyman-$\alpha$ emitters with no clear \hst photometric counterpart (see Fig. \ref{fig:specs}). In these cases, the observed positions are measured from the MUSE datacube. Redshifts in square brackets were fixed in the corresponding lens models.}
\end{table*}

\begin{figure}
  \centering
      \includegraphics[width = 1.0\columnwidth]{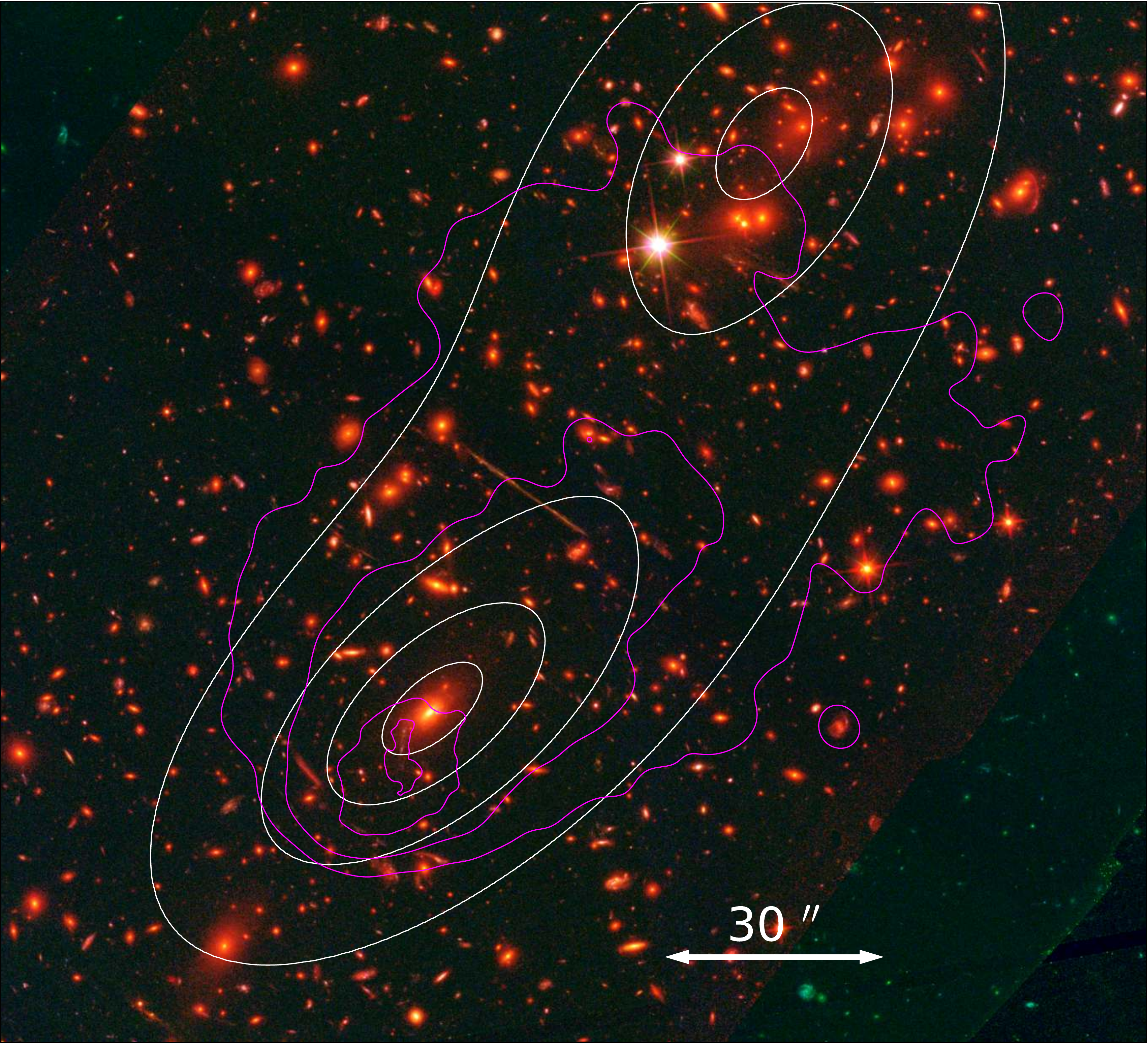}
  \caption{ACT0102 X-ray emission and mass distribution overlaid on the \hst colour image (same as in Fig. \ref{fig:members_all}). The magenta lines show the \chandra X-ray surface brightness isophotes, and the white lines represent the projected total mass isocontours of the smooth component from our best-fit strong lensing model.}
  \label{fig:xrays}
\end{figure}
\begin{figure}
  \centering
      \includegraphics[width = 1.0\columnwidth]{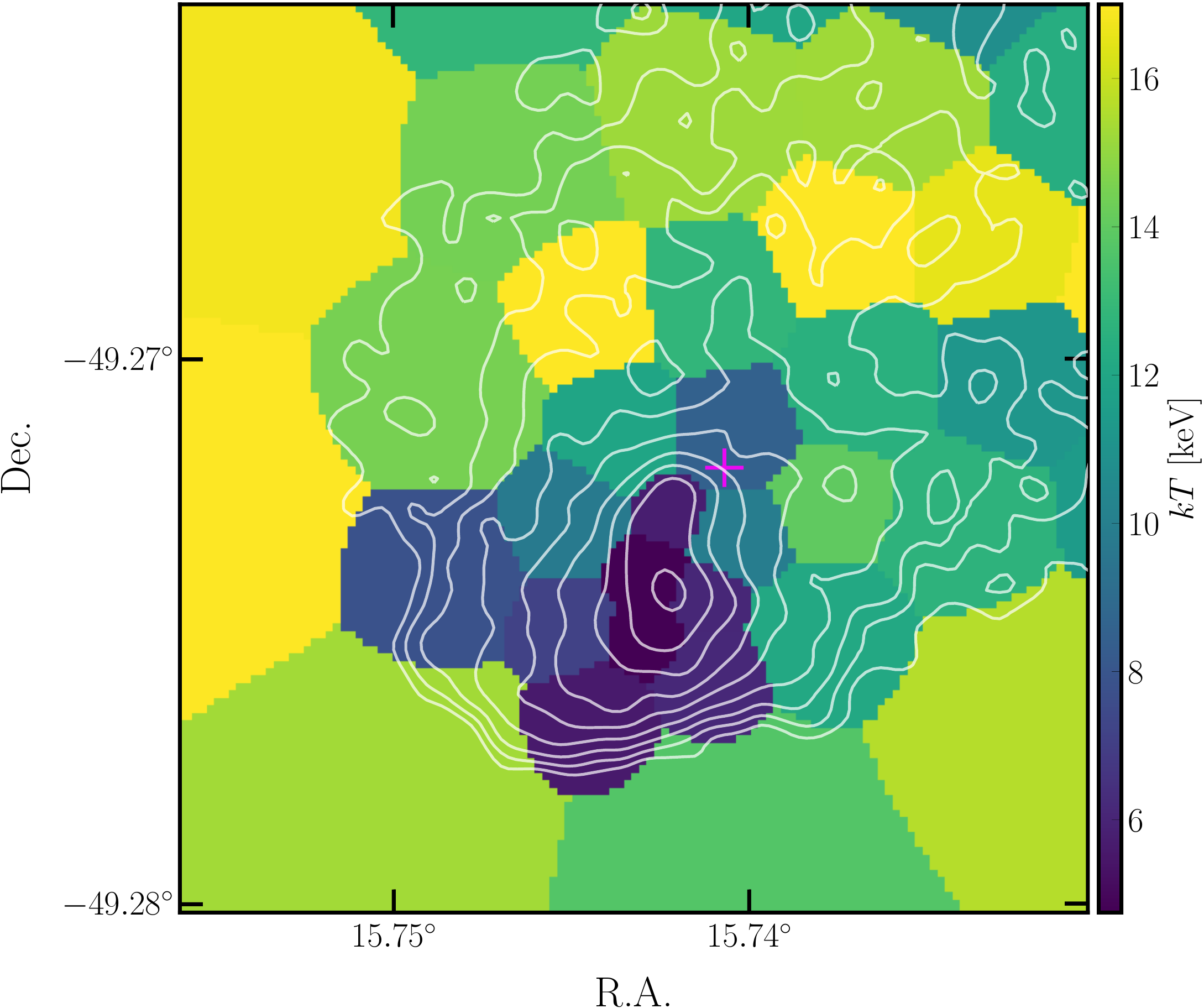}
  \caption{Gas temperature distribution from the \chandra X-ray analysis. The relative error on the temperature value is $\sim10$\%. The magenta cross indicates the position of the BCG, and white isocontours represent the \chandra X-ray emission.}
  \label{fig:temperature_map}
\end{figure}

In addition to the multi-band \hst imaging, we made use of high-quality spectroscopic data from MUSE.
This spectroscopic dataset is especially important when measuring the redshifts of (1) several multiple images, some of which have no \hst detection, (2) cluster members, and (3) intervening deflectors that we used to build an accurate strong lensing model.
The observations were carried out between December 2018 and September 2019 under the ESO programme ID 0102.A-0266 (P.I.: G. B. Caminha), and consist of three pointings of $\approx2.3$~hours each.
All exposures were performed using the ground layer adaptive optics (GLAO) in order to correct for first-order atmospheric dispersion and improve the final image quality.

We employed the standard MUSE reduction pipeline version 2.6 \citep{2020A&A...641A..28W} to apply all corrections and calibrations and to create the final datacube.
Moreover, we used the self-calibration method, implemented in the reduction pipeline, to mitigate the instrumental variations across each integral field unit slice and to improve the background subtraction.
We also made use of the {\tt Zurich Atmosphere Purge} \citep[ZAP;][]{2016MNRAS.458.3210S} to remove instrumental and sky residuals not fully corrected for by the standard reduction recipes.
The final datacube covers the wavelength range $4700\,\AA - 9350\,\AA$, with a gap in the narrow region of $5805\,\AA$ to $5965\,\AA$ that is masked because of a strong sodium emission from the GLAO laser guiding system.
The field of view of the three pointings in the cluster core is shown in Fig. \ref{fig:members_all}. It covers an area of $\approx 3$~arcmin$^2$, with a final point-spread function of $\approx 0\farcs55 - 0\farcs60$ full width at half maximum, measured from stars in the pseudo MUSE white image.
Finally, we used compact sources detected in both the \hst filter F606W and MUSE white images to match the astrometry of both datasets, obtaining a positional r.m.s. of $0\farcs04$, much smaller than the MUSE pixel scale of $0\farcs2$.

Akin to our previous works \citep[see e.g.][]{2017A&A...607A..93C,2019A&A...632A..36C, 2022ApJ...926...86A}, we extracted the spectra of all \hst detections in order to measure their redshifts.
In this step, we adopted circular apertures of $0\farcs8$ in radius to obtain one-dimensional spectra of all detections.
We carefully inspected all spectra, and in the cases with continuum detection we cross-correlated the data with templates in order to obtain precise redshift measurements.
To have a spectral coverage from rest frame ultraviolet to optical wavelengths and maximise the success rate of our measurements, we used empirical and stacked spectral templates from different surveys, for instance zCOSMOS \citep{2007ApJS..172...70L}, the Galaxy Mass Assembly ultra-deep Spectroscopic Survey \citep[GMASS;][]{2013A&A...549A..63K}, the Visible Multi-object Spectrograph (VIMOS) VLT Deep Survey \citep[VVDS;][]{2013A&A...559A..14L}, and our previous MUSE observations \citep{2016A&A...587A..80C, 2016A&A...595A.100C, 2017A&A...600A..90C, 2017A&A...607A..93C, 2019A&A...632A..36C}.

For spectra with no continuum or with very low signal-to-noise, we searched for emission lines to assign redshifts.
Moreover, we performed a blind search to identify emission lines of objects with faint continuum emission that are not clearly detected in the photometric data.
The blind search was performed in two stages.
The first one was done automatically, by applying the difference of Gaussians with an algorithm implemented in the Pyhton \texttt{scikit-image} package \citep{scikit-image} on pseudo-narrow-images of the continuum-subtracted MUSE datacube.
Detections that persist in two or more wavelengths were inspected visually to confirm whether they are real emissions or not.
The second stage consisted of a visual inspection of the entire continuum-subtracted datacube to capture possible emissions missed by the automatic step.
In total, we detect $\sim30$ Lyman-$\alpha$ emitters, of which some are multiply lensed (some examples are shown in Fig. \ref{fig:specs}).
Such a population of sources, with a very faint UV-continuum emission, will be studied in more detail in future works.

We assigned a `quality flag' (QF) to each redshift measurement.
Similarly to our previous works, we used the following convention: QF=3 is a secure confirmation with the identification of several spectral features, or where the nature of one single emission line can be clearly characterised, for instance the \ion{O}{II} and \ion{C}{III} doublets or the Lyman-$\alpha$ shape; QF=2 is a measurement obtained using only one or noisy spectral features, usually absorption lines; QF=9 indicates that the redshift was measured from one narrow or noisy emission line with no secure identification of its nature; finally, stars have QF=4.
In Table \ref{tab:full_z_cat}, we present the first entries of the full redshift catalogue.
This contains 402 secure redshift measurements (i.e. with QFs higher than one) from 374 single objects, after accounting for multiple images of the same lensed sources.
The full version of the catalogue is publicly available in the online version of this manuscript.
Given the MUSE line spread function and and calibration, the typical uncertainty of our redshift measurements is of the order of $\delta z \approx 5 \times 10^{-4}$, in line with what was previously found in other works \citep[see e.g.][]{2017A&A...599A..28K, 2017A&A...608A...2I}.

After a careful inspection of our spectroscopic catalogue, we were able to identify a total of 56 multiple images with secure redshift measurements, from 23 single background sources.
In Table \ref{tab:multiple_images}, we provide the coordinates and MUSE redshift values of all multiple images used as input for the lens model presented in this work (see Sect. \ref{sec:sl_model}).
The multiple images span a redshift range from $z=2.19$ up to $z=5.95$ and their distribution is illustrated in Fig. \ref{fig:families_z_hist}.
There, we also show the distribution of spectroscopically confirmed multiple image families from works with similar datasets \citep{2016ApJ...822...78G, 2017A&A...607A..93C, 2019A&A...632A..36C, 2021A&A...645A.140B}.
Remarkably, the very elongated total mass distribution and the high total mass value of ACT0102 make it very efficient at producing multiple images of high redshift sources. The overdensity of image confirmations in the redshift range $z\approx 4 - 5$ can be at least partially associated with the high redshift value of our cluster, $z=0.87$. The particular peak at $z\approx4.2$ contains the galaxy group studied in \citet{2021ApJ...908..146C}, and the additional overdensities at $z\approx5.0$ and their physical properties will be explored in future works.
The sample of lensing clusters taken from the literature are on average at lower redshifts ($z\approx 0.2 - 0.6$), thus producing a more uniform redshift distributions of multiple images, as confirmed with MUSE.

\subsection{\chandra X-ray data}
\label{sec:chandra_data}

\begin{figure}
  \centering
      \includegraphics[width = 1.0\columnwidth]{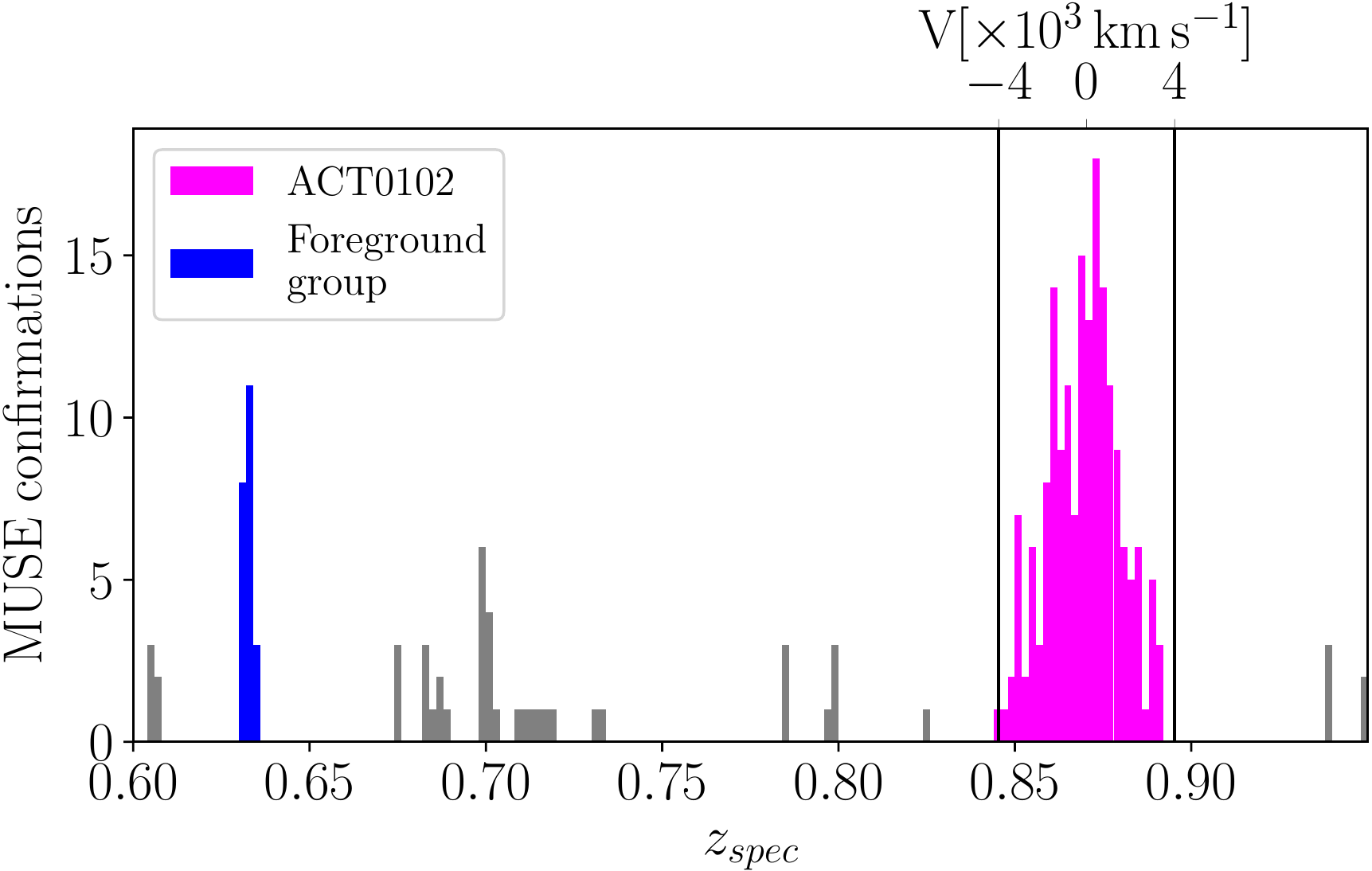}
      \includegraphics[width = 1.0\columnwidth]{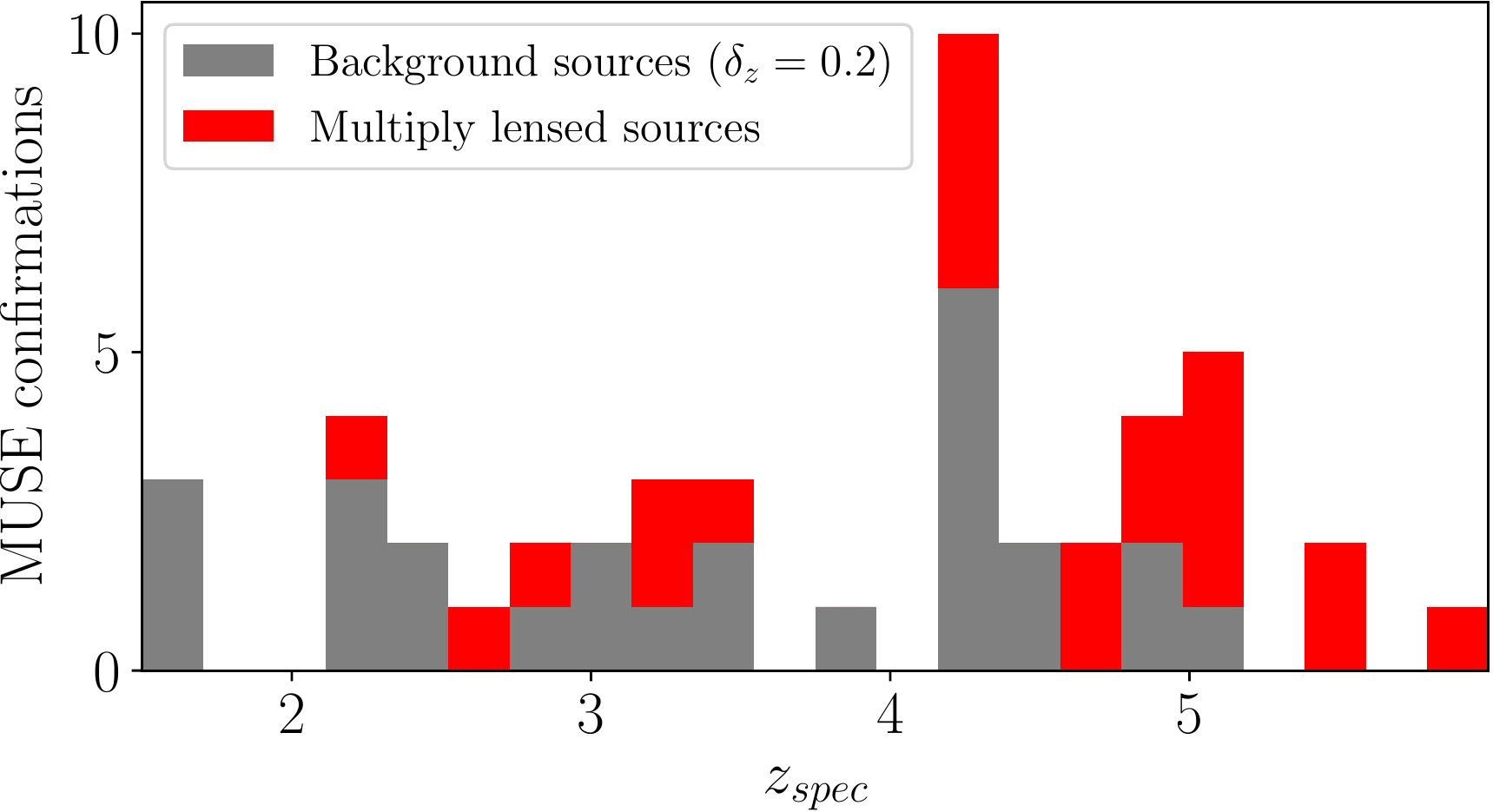}
  \caption{Spectroscopic confirmations with MUSE. Top panel: Redshift distribution of the objects in the MUSE catalogue in bins of $\delta z = 0.002$. The magenta and blue histograms indicate, respectively, the ACT0102 cluster members (167 galaxies) and the foreground group (20 galaxies) used in our strong lensing modelling. The upper $x$-axis shows the line-of-sight rest-frame galaxy velocity with respect to the cluster member median redshift value of $z=0.8704$. Bottom panel: Redshift distribution of background single sources (i.e. corrected by image multiplicity) with $\delta z = 0.2$, where the overdensity at $z\approx4.2$ is clearly noted.}
  \label{fig:cluster_histogram}
\end{figure}

We investigated the properties of the intracluster medium (ICM) in the core of ACT0102, with the aim of finding possible hints of a correlation between the bright cluster galaxy (BCG) and its surrounding ICM. We used the \chandra observation IDs 12258, 14022, and 14023 (P.I.: Hughes) for the X-ray analysis of the ICM.
The reduction of the \chandra data was performed using the software {\tt CIAO v4.13}, with the latest release of the \chandra Calibration Database at the time of writing ({\tt CALDB v4.9}).
Time intervals with a high background level were filtered out by performing a 2$\sigma$ clipping of the light curve in the 2.3--7.3~keV band. The total cleaned exposure time is 273.4~ks. The ancillary response file (ARF) and redistribution matrix file (RMF) for each observation were computed with the commands {\tt mkarf} and {\tt mkacisrmf}. The background spectra were extracted from source-free regions on the same CCD chip as for the cluster.

The regions for spatially resolved spectral analysis are selected using the Voronoi tessellations method \citep{2003Cappellari}. Each region contains $\sim$500 net counts in the energy range 0.5--7~keV. The spectral fitting for each region is performed with {\tt Xspec 12.12.0} \citep{1996Arnaud} using C-statistics \citep{1979Cash} and the solar abundance table from \citet{2009Asplund}. Galactic hydrogen absorption is described by the model {\tt tbabs}, where the column density of hydrogen $n_{\rm H}$ is fixed at $1.265\times10^{20}~{\rm cm}^{-2}$ \citep{2016HI4PI}. The ICM spectrum in the 0.5--7~keV band is fitted with the {\tt apec} thermal plasma emission model \citep{2001Smith}, where the redshift is fixed at 0.87, and the temperature, abundance, and normalisation are set as free parameters.
The X-ray emission and two-dimensional temperature maps are shown, respectively, in Figs. \ref{fig:xrays} and \ref{fig:temperature_map}, and discussed in the following sections.

\section{Strong lens modelling}
\label{sec:sl_model}

\begin{figure}
    \includegraphics[width=1.0\columnwidth]{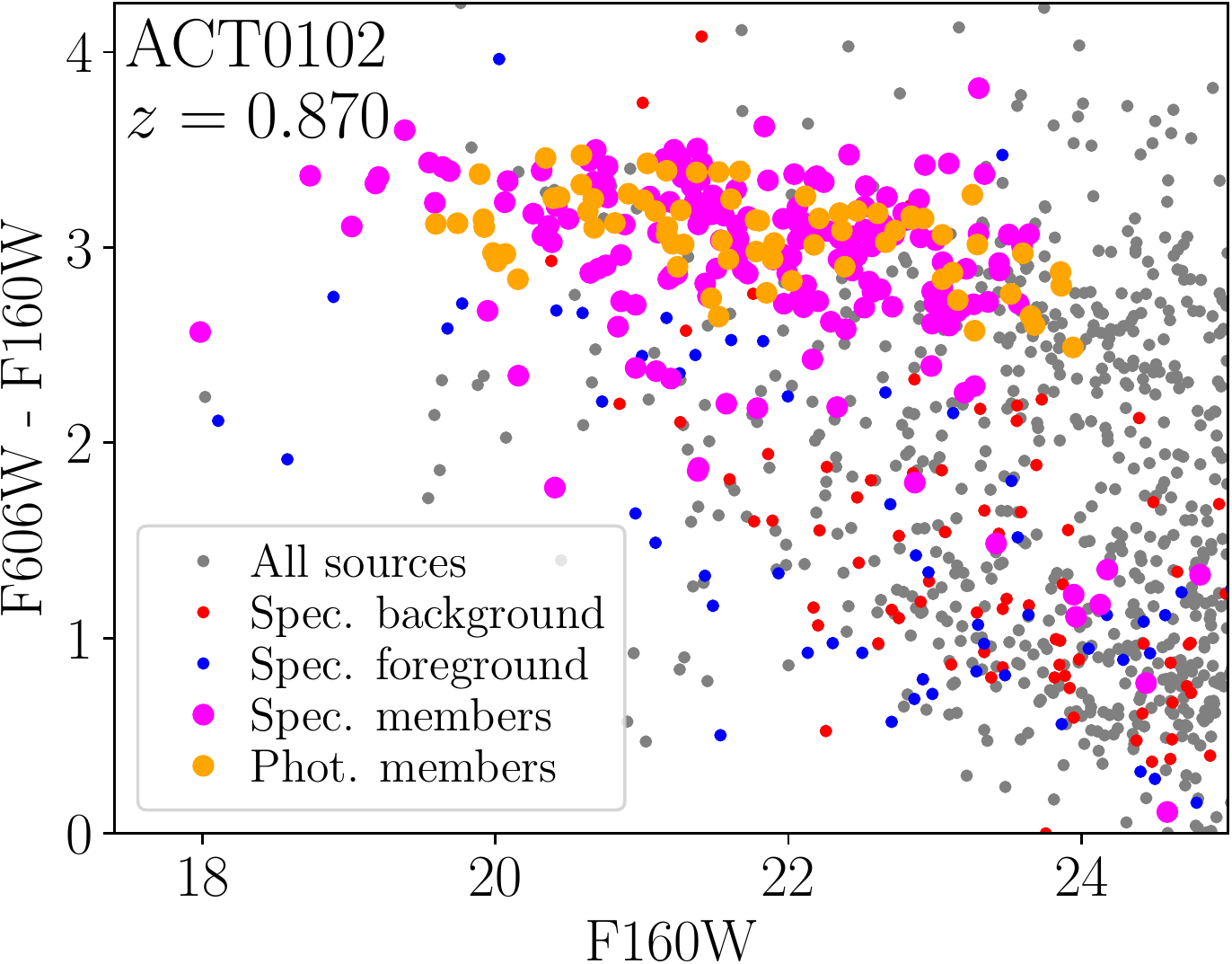}
  \caption{Colour magnitude diagram of ACT0102. Spectroscopic members are shown in magenta and photometrically selected ones in orange (see Sect. \ref{sec:sl_model} for more details on our selection criteria). Spectroscopically confirmed background and foreground objects are indicated by red and blue dots, respectively.}
  \label{fig:redsequence}
\end{figure}

We used the publicly available software \texttt{lenstool} \citep{1996ApJ...471..643K, 2007NJPh....9..447J, 2009MNRAS.395.1319J} to model the cluster total mass distribution.
This is characterised by two main mass components represented with parametric mass density profiles.
The first one, defined over the extended cluster scale, is dominated by dark matter and has a small contribution from the hot-gas \citep[see e.g.][]{2017ApJ...842..132B, 2018ApJ...864...98B} and the intracluster light.
The second component accounts for the total mass distribution of the galaxies, mainly cluster members and a few foreground perturbers, that are shown to be relevant to accurately reproduce the observed multiple images in the field of ACT0102.
Moreover, we tested some perturbations to the cluster total mass distribution, making it deviate from the perfect elliptical symmetry commonly adopted in parametric models, to improve the reconstruction of the positions of the multiple images. To do this, we followed the approach presented in \citet{2021MNRAS.506.2002B} and its implementations in \texttt{lenstool}.

\subsection{ACT0102 members and line-of-sight structure}
\label{sec:members_and_los}

The selection of the cluster members is based on the spectroscopic confirmation from our MUSE catalogue (see Sect. \ref{sec:muse_spectroscopy}).
Figure~\ref{fig:cluster_histogram} clearly shows the galaxies associated with ACT0102 as an overdensity around $z=0.87$.
In detail, we selected as cluster members those galaxies with a rest-frame velocity value within 4000~$\rm km\,s^{-1}$ of the peak of this distribution (i.e. in the range $z=[0.835, 0.907]$).
This velocity limit ensures that we did not miss any galaxy relevant to our strong lensing study in the core of ACT0102, and it is in line with the spectroscopic values observed in other clusters \citep{2015A&A...579A...4G, 2016ApJS..224...33B,2021A&A...656A.147M}, especially in the innermost regions (<500~kpc).
According to this criterion, 167 objects are defined as spectroscopically confirmed members.
Since the MUSE field of view covers only the central regions of the cluster, we used the \hst colour and photometric redshift information of these spectroscopic members to select members with no spectroscopic information.
From the colour (see Fig. \ref{fig:redsequence}) and photometric-redshift (from the publicly available RELICS catalogue) distributions of spectroscopically confirmed members, we computed the 68\% confidence levels and used these limits to select the photometric members.
We limited the magnitude of photometric members to galaxies brighter than $mag_{\rm F160W} = 24$ in order to minimise possible contaminations.
In this way, we included 76 additional cluster members in the modelling, all of them located outside the MUSE field of view, as shown in Fig. \ref{fig:members_all}.
In Fig. \ref{fig:redsequence} we present the colour-magnitude diagram of all the objects detected in the \hst data.
We show here the magnitudes in the filters F606W and F160W because these bands best sample the Balmer break at $\approx 4000\AA$ rest-frame.
The spectroscopically confirmed members define a clear red sequence \citep{2000AJ....120.2148G} around $\rm F606W - F160W \approx (3.0 \pm 0.6)$~mag, which is followed by the photometrically selected members. Objects within the red sequence not selected as cluster members are excluded by our photometric-redshift selection described above or because they have spectroscopic redshifts whose values are outside the range we defined for the members.

From the spectroscopic data (see Fig. \ref{fig:cluster_histogram}), it is possible to identify a secondary peak in the redshift distribution, at $z\approx 0.63$, composed of 20 elliptical galaxies.
Their relatively concentrated spatial distribution is shown in Fig. \ref{fig:members_all}.
We illustrate in Sect.~\ref{sec:total_mass} that this foreground group has a non negligible effect on the model predicted positions of the multiple images, and it thus must be included in the modelling.
In total, we have 263 galaxies, of which 243 belong to ACT0102 and 20 to the foreground group, that we take into account in our lens model.

\subsection{Mass components and parameterisation}
\label{sec:mass_param}

\begin{table*}[!]
\small
\caption{Model summary.}
\centering                                      
\begin{tabular}{l c c c c c c l}
\hline \hline
Model ID & N. par. & DOF & $\Delta_{\rm rms} ['']$ & $\rm \chi^2_{min}$  & BIC & AICc & Description \\
\hline
No-foreground     & 14 & 52 & 0.87 & 103& 281 &  257 & Foreground galaxies are not included\\
\bf Reference     & 14 & 52 & 0.75 & 80 & 242 &  220 & Foreground galaxies included in member scalings\\
Two-scalings      & 16 & 50 & 0.75 & 79 & 249 &  226 & Foreground galaxies with different scaling relations\\
Free-scaling      & 16 & 50 & 0.74 & 80 & 250 &  226 & Member scaling relation slopes, $\alpha$ and $\gamma$, are free\\
External-shear    & 16 & 50 & 0.75 & 80 & 250 &  226 & Same as reference plus external shear\\
\it 3-PIEMD-circ.& 18 & 48 & 0.72 & 52 & 231 &  206 & Third smooth component with zero ellipticity\\
\it 3-PIEMD      & 20 & 46 & 0.69 & 48 & 235 &  210 & Three smooth components\\
\hline
Models with B-spline perturbation \\
\hline
PertBS-3-1DM & 21 & 45 & 1.26 &208 & 369 & 349 & --- \\ 
PertBS-4-1DM & 28 & 38 & 1.28 &241 & 462 & 444 & --- \\
PertBS-5-1DM & 37 & 29 & 1.09 &197 & 455 & 475 & --- \\
PertBS-2-2DM & 22 & 44 & 0.69 &63  & 259 & 234 & ---  \\
PertBS-3-2DM & 27 & 39 & 0.53 &35  & 252 & 232 & ---  \\
PertBS-4-2DM & 34 & 32 & 0.50 &27  & 273 & 275 & ---  \\
PertBS-5-2DM & 43 & 23 & 0.41 &22  & 306 & 384 & ---  \\
\hline
\end{tabular}
\label{tab:summary_models_all}
\tablefoot{For each model, we present the number of free parameters (N. par.), the number of degrees of freedom (DOF), the root-mean-square difference between the model predicted and observed multiple image positions ($\Delta_{\rm rms}$), the minimum $\chi^2$, the BIC and AICc values, and a short description of the model parameterisation.
}
\end{table*}

The small-scale mass components (i.e. each cluster member and foreground perturber) are modelled with axially symmetric dual pseudo isothermal mass density profiles \citep{2007arXiv0710.5636E, 2010A&A...524A..94S}.
This mass density distribution is characterised in projection by the values of two free parameters, a central velocity dispersion, $\sigma_{v,gal}$, and a cut radius, $r_{cut,gal}$, and is given by the following expression:
\begin{equation}
\Sigma(R) = \frac{\sigma_{v,gal}^2}{2G}\left( \frac{1}{R} - \frac{1}{\sqrt{R^2-r_{cut,gal}^2}}  \right),
\end{equation}
where $R$ is the radial coordinate and $G$ the Newtonian constant of gravitation.

Since it is computationally unfeasible to optimise the values of two free parameters for each cluster and foreground group member (i.e. to have a total $2\times 263$ free parameters), we assumed that the mass parameter values are related to those of the galaxy luminosities according to the relations
\begin{equation}
\sigma_{v,gal}^i = \sigma_{v,gal}^{ref}\left( \frac{L_i}{L_\star} \right)^{1/\alpha},\; r_{cut,gal}^i = r_{cut,gal}^{ref} \left( \frac{L_i}{L_\star} \right)^{1/\gamma},
\label{eq:scaling_relation}
\end{equation}
where $\sigma_{v,gal}^{ref}$ and $r_{cut,gal}^{ref}$ are the only two free parameters to be optimised in the scaling relations.
Unless otherwise specified, in our models we adopted a constant total mass-to-light ratio, which can be obtained with $\alpha=4$ and $\gamma=2$.
The luminosity $L_{\star}$ was chosen to be that of the BCG and corresponds to the magnitude $mag_{\rm F160W} =17.99$.
In our lens models (see Sect. \ref{sec:total_mass}), we considered two cases: either we adopted two different sets of scaling relations for the cluster and foreground group members, or all galaxies follow the same relation.

For each cluster-scale mass component (i.e. mainly the dark matter component), we assumed that its projected mass density distribution follows that of a pseudo-isothermal elliptical mass model \citep[PIEMD;][]{1993ApJ...417..450K}:
\begin{equation}
\Sigma(R) = \frac{\sigma_{v}^2}{2G} \left( \frac{1}{\sqrt{{R(\varepsilon)}^2 + r_{core}^2}} \right),
\end{equation}
where $r_{core}$ is the so-called core radius and $R(\varepsilon)$ is constant over ellipses with ellipticity $\varepsilon$ defined as $(a^2 - b^2)/(a^2+b^2)$ ($a$ and $b$ are the semi-major and semi-minor axes, respectively).
In addition to these three parameters ($\sigma_v$, $r_{core}$ and $\varepsilon$), the orientation angle, $\theta$ and the central position ($x_0$ and $y_0$) fully describe this mass distribution and they are all optimised in our lens modelling.
As we discuss in the following sections, in order to be able to predict the multiplicity of all multiple images, we had to adopt two PIEMD halos.
One located near the BCG and the other in the north-west region.
Such a complex mass distribution is expected because ACT0102 is a prominent merging cluster  \citep{2021arXiv210600031K}, as it can be seen in Fig. \ref{fig:members_all}.
Moreover, the X-ray emission presented in Fig. \ref{fig:xrays} shows a clear peak nearby the BCG in the south-east region, with a large tail extending towards the north-west.

\subsection{Best-fit models and the total mass distribution}
\label{sec:total_mass}

\begin{figure}
      \includegraphics[width = 0.498\columnwidth]{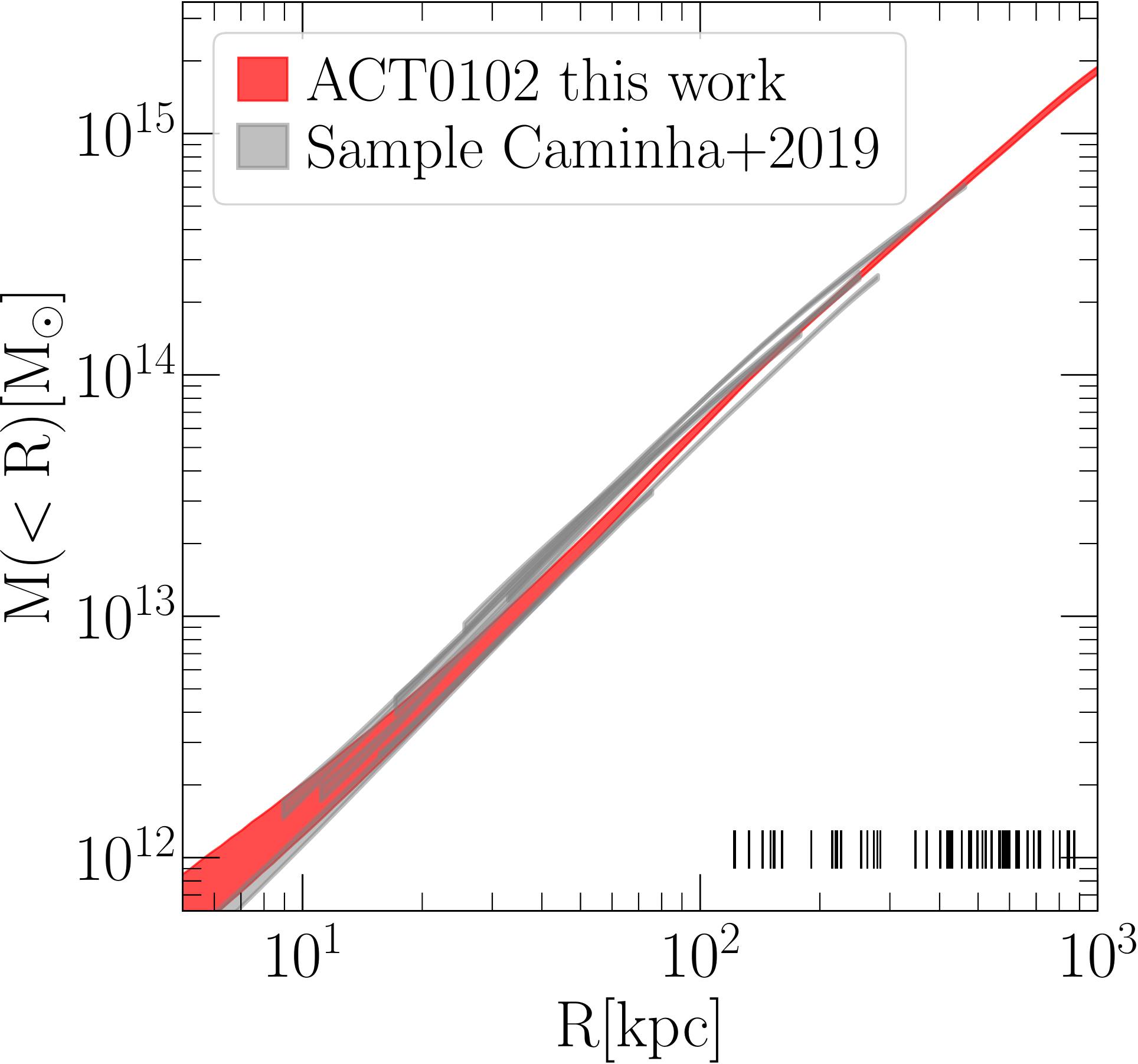}\includegraphics[width = 0.493\columnwidth]{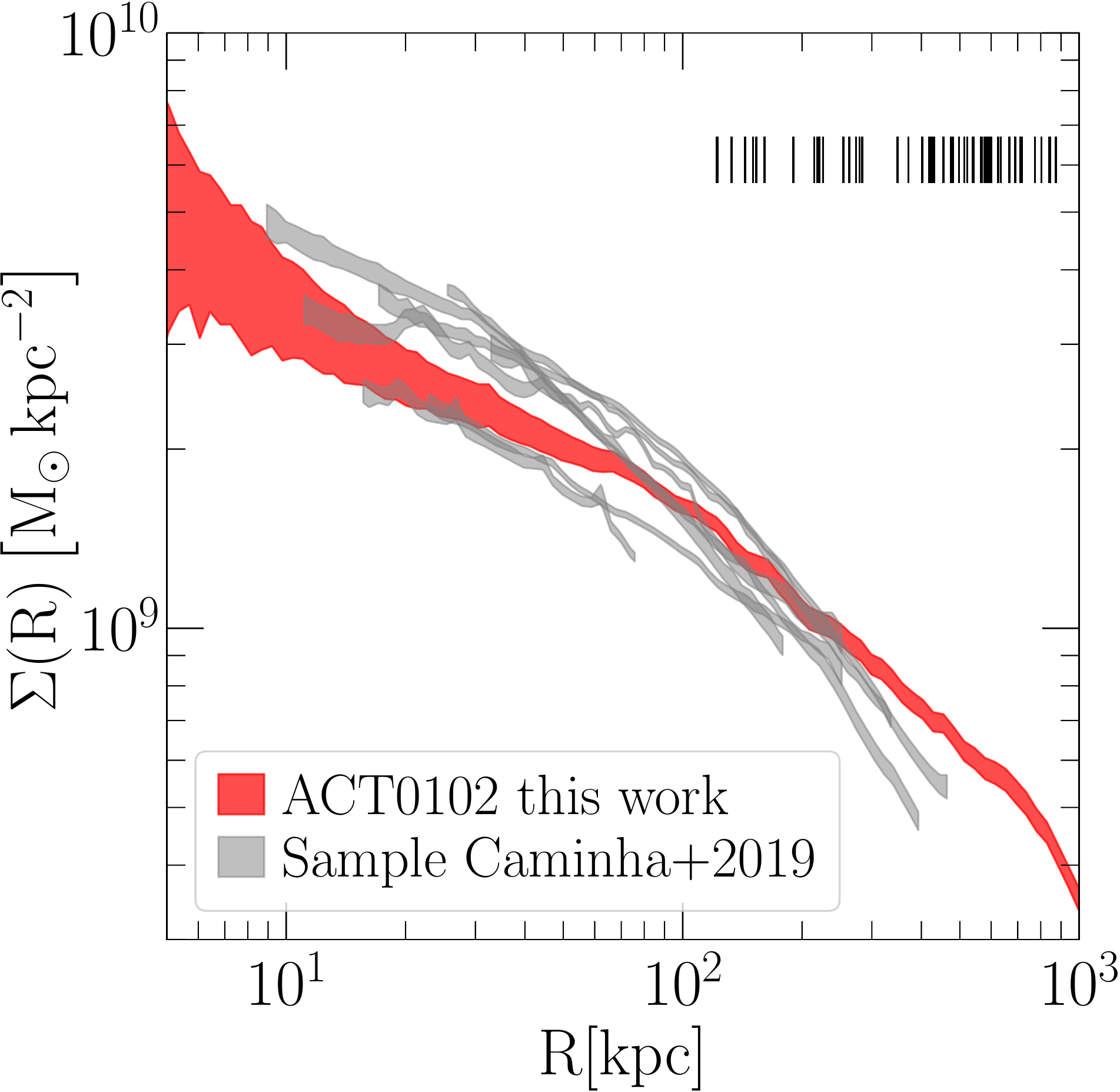}
      \includegraphics[width = 0.49\columnwidth]{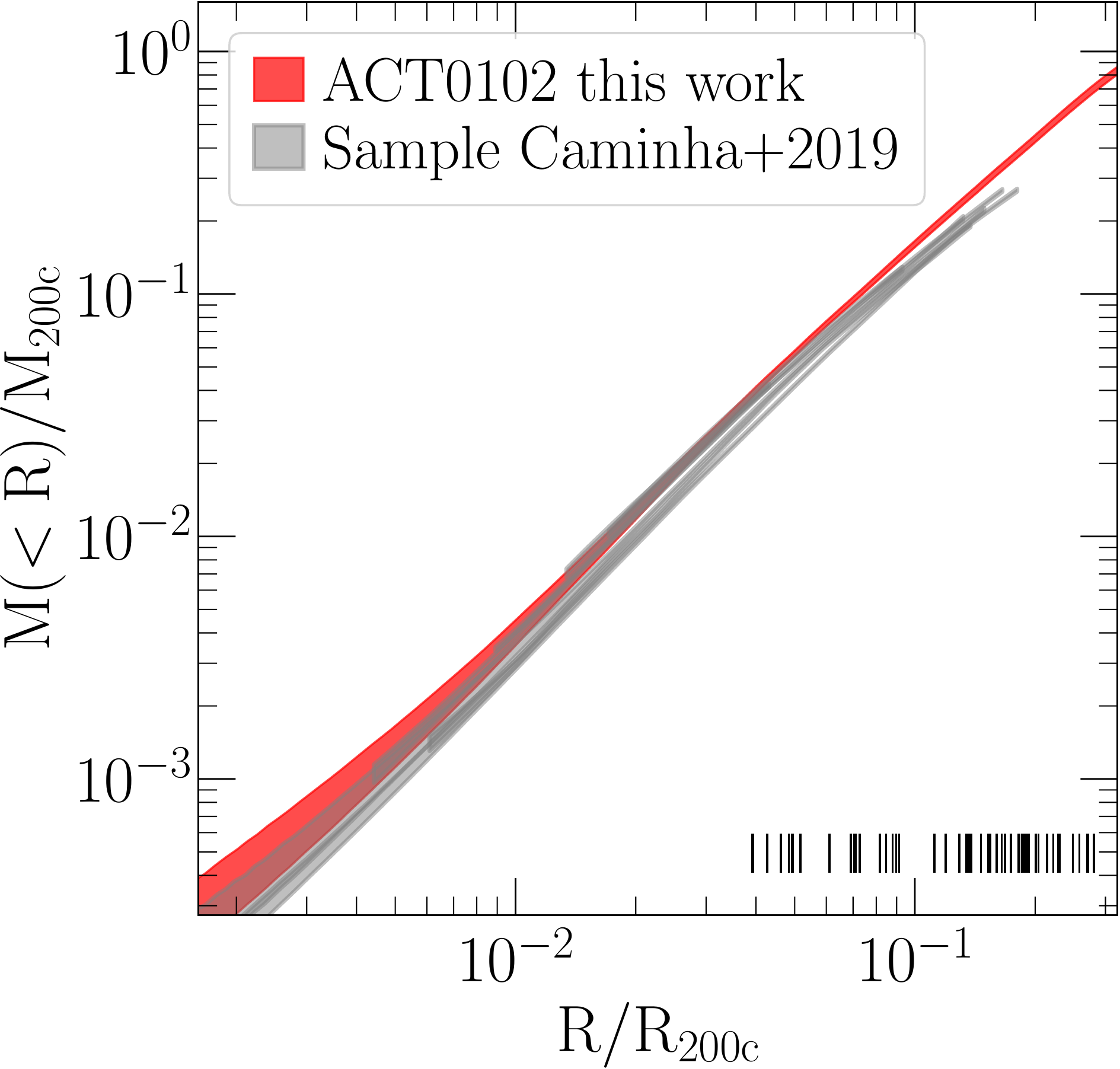}\hspace{0.2cm}\includegraphics[width = 0.465\columnwidth]{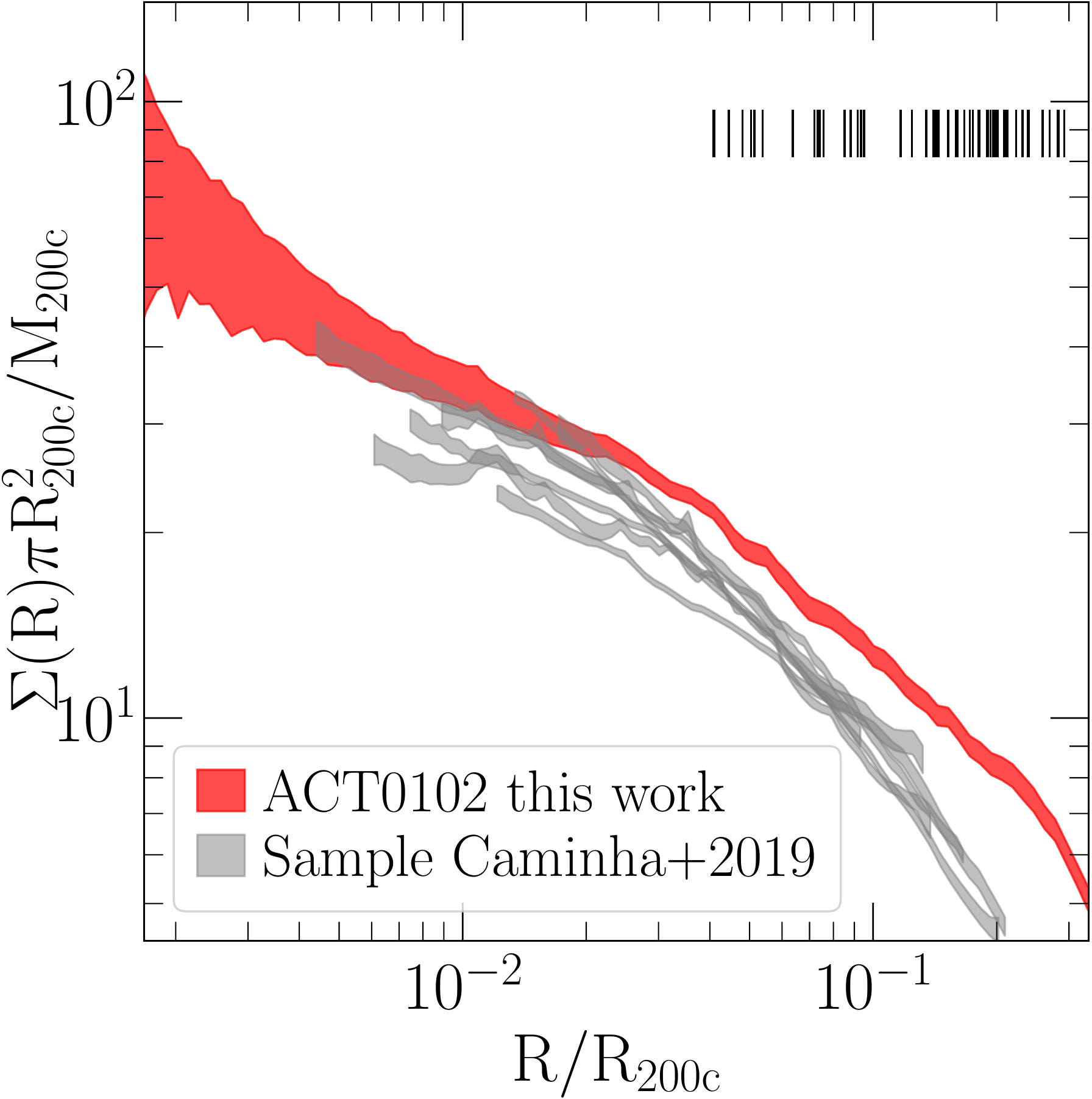}
      
  \caption{Cumulative projected total mass profile (left) and total surface mass density profile (right) computed from the centre of the main cluster component, close to the BCG. The areas in red correspond to the 95\% confidence level intervals of our reference lens model for ACT0102. Vertical lines show the radial distances of the multiple images used to reconstruct the cluster total mass distribution. The profiles of other clusters are plotted in grey and are limited to the radial distances over which the multiple images are visible. The top panels show the absolute values, and the bottom panels show the values rescaled to those of $M_{200c}$ and $R_{200c}$ (see the axis labels).}
  \label{fig:mass_profiles}
\end{figure}

\begin{figure}
      \includegraphics[width = 0.498\columnwidth]{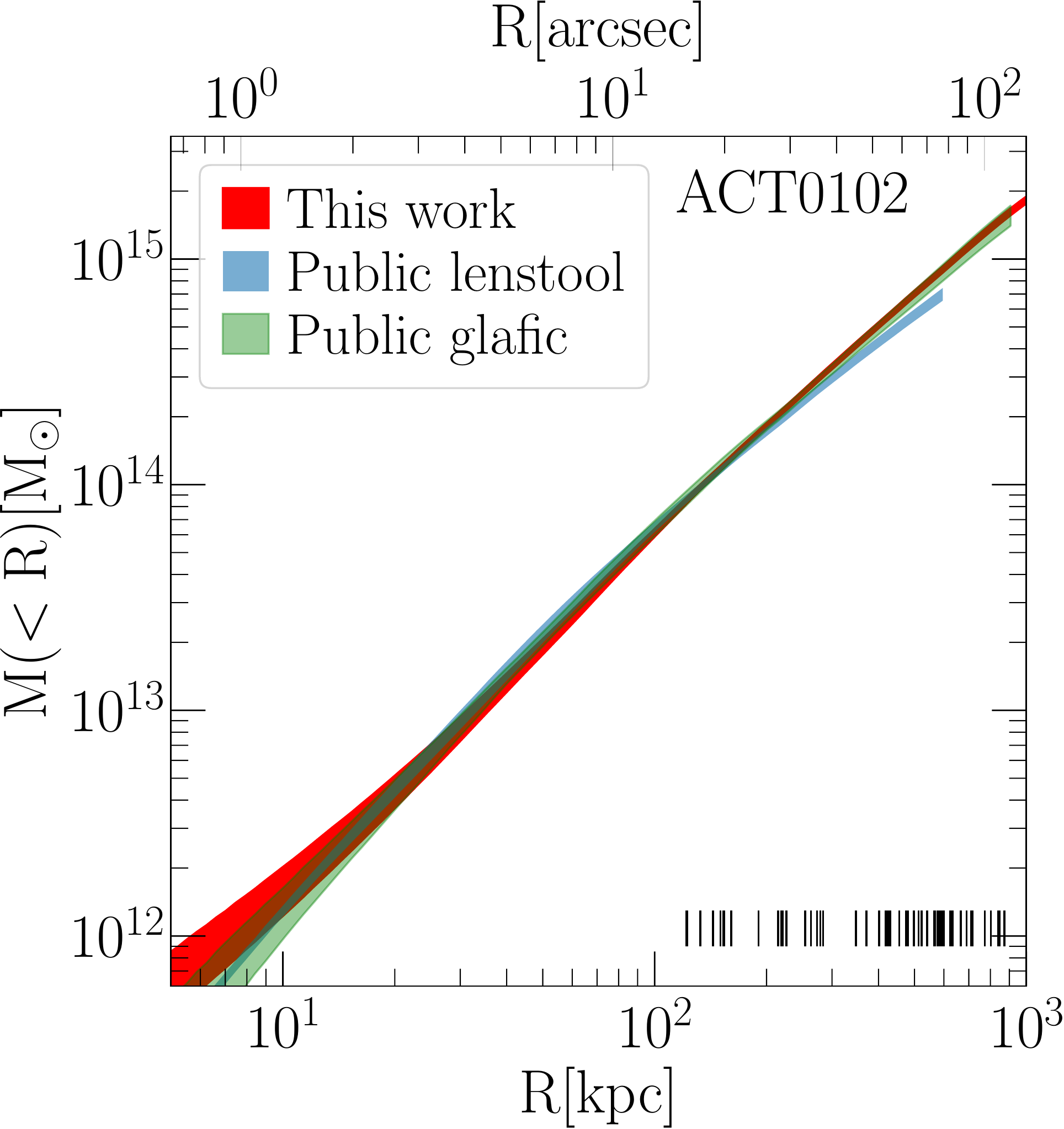} \includegraphics[width = 0.493\columnwidth]{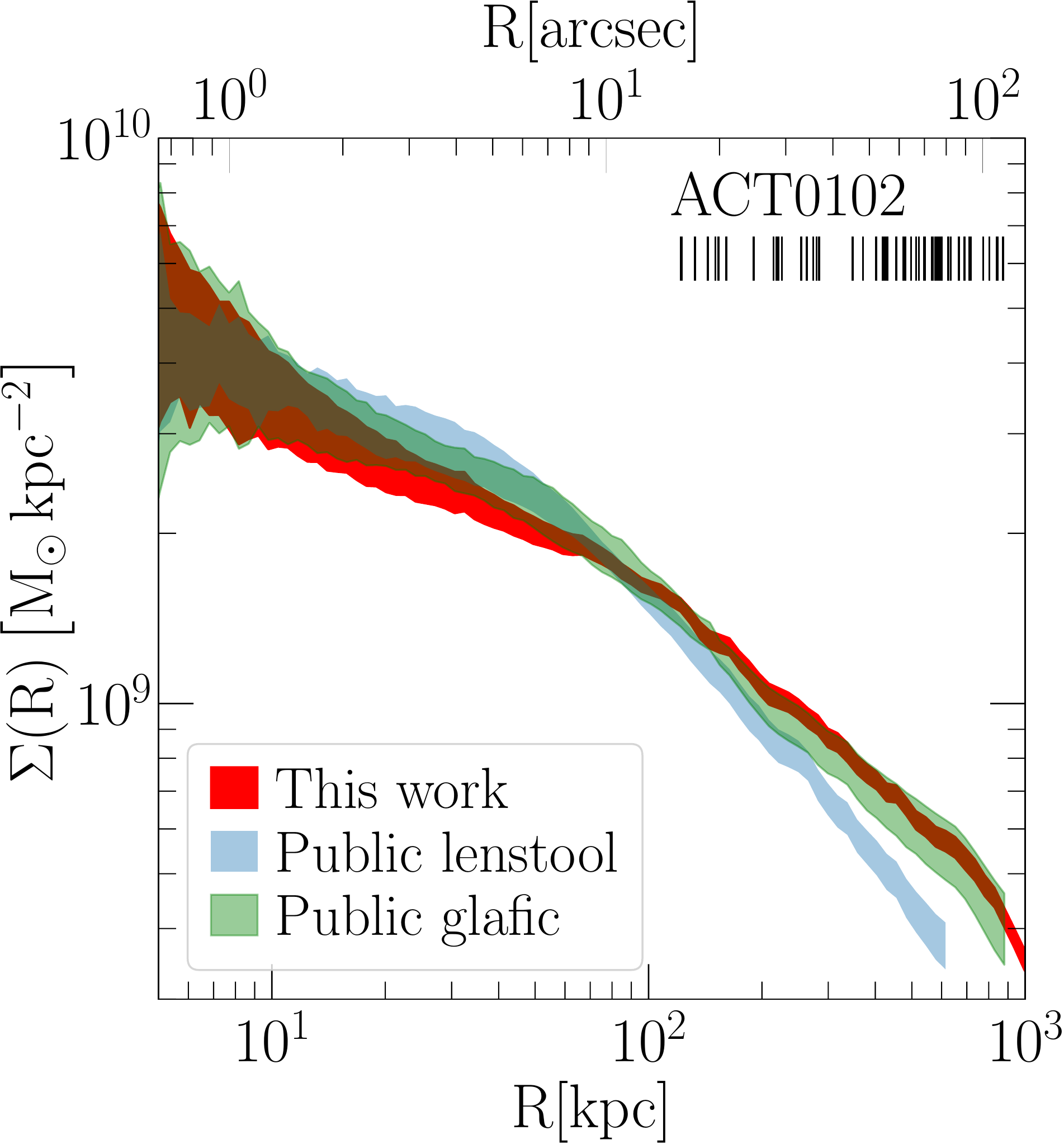}
      
  \caption{Cumulative projected total mass profile (left) and total surface mass density profile (right)  compared with the results of public RELICS models. The coloured regions correspond to the 95\% confidence level intervals from our reference lens model (in red) and from the two public models made available by the RELICS team (in cyan and green). The radial distances of the multiple images used in our model  to reconstruct the cluster total mass distribution are indicated by vertical black lines.}
  \label{fig:mass_comparison}
\end{figure}

In our strong lensing model, we used the positions of multiply lensed sources to constrain the total mass distribution of the cluster.
The distance between the observed and model predicted multiple image positions is quantified with a $\chi^2$ function given by
\begin{equation}
\label{eq:chi2}
\chi^2(\vec{\Pi}) \equiv \sum_{j}^{N_{images}} \frac{\left| \vec{\theta}^{observed}_{j} - \vec{\theta}^{model}_{j}(\vec{\Pi}) \right|^2} {\sigma_{j}^{2}},
\end{equation}
where $\vec{\theta}$ indicates the positions of the multiple images (observed and predicted by the model), $\sigma$ is the positional uncertainty, $\vec{\Pi}$ represents the free parameters of the model and $N_{images}$ is the number of multiple images used as model constraints.
For multiple images with \hst detections, we adopted a positional error of $\sigma_j = 0\farcs5$ to account for small and large-scale perturbations along the line of sight that are not incorporated in parametric models \citep[see e.g.][]{2012MNRAS.420L..18H,2018A&A...614A...8C}.
The best-fit model is given by the values of the model parameters that minimise the $\chi^2$ function.
We also defined the root-mean-square difference ($\Delta_{rms}$) between the observed and best-fit, model predicted positions of the multiple images ($\Delta_{rms}^2\equiv \sum{\left|\vec{\theta}^{obs} - \vec{\theta}^{bf}\right|^2} / N_{images}$) to quantify the goodness of a model.

In this work, we consider only multiple images with secure confirmations and spectroscopic redshift measurements.
The only exception is multiple image 12a, which lies outside the MUSE field of view; however, it has clear \hst detection with similar colours and shape of images 12b and 12b, which are spectroscopically confirmed.
This ensures that we do not have any multiple image misidentification or systematic effects introduced by unknown or uncertain photometric redshifts.
Thanks to the high-efficiency of MUSE in detecting emission lines and the possibility of integrating the spectra of extended objects over large areas, we collected a final sample of 56 multiple images from 23 spectroscopically confirmed sources.
Interestingly, 19 multiple images from 9 different sources are Lyman-$\alpha$ emitters with no clear \hst photometric counterparts (see Table \ref{tab:multiple_images} and Fig. \ref{fig:specs}). Model predicted multiple images located outside the MUSE field of view with no clear HST counter part are not considered in this work.
For multiple images with \hst detections, we used the F160W filter to determine their precise positions and employ them as input to the lens model. For MUSE only detections instead, we make use of a pseudo narrow-band image created by stacking $\approx 12$ spectral pixels (i.e. $\approx 15\AA$) around the Lyman-$\alpha$ emission to estimate the image positions.
Because of the lower MUSE spatial resolution compared to that of \hst, we chose a positional uncertainty $\sigma_j$ of MUSE only detections two times larger than that of F160W measurements.

We considered different parameterisations to describe the total mass distribution of ACT0102.
In order to compare these different models, we ran \texttt{lenstool} in the optimisation mode to find the best-fit values of the model parameters and compute the quantities $\chi^2$, $\Delta_{rms}$, and the Bayesian \citep[BIC;][]{schwarz1978} and corrected Akaike \citep[AICc;][]{1974ITAC...19..716A} information criteria.
Here, we adopted positional uncertainties of $0\farcs5$ and $1\farcs0$ for \hst and MUSE only detections, respectively.
These statistical estimators are used to quantify the goodness of each model, taking into account the number of free parameters, thus indicating some possible overfitting due to increased flexibility of the models.
In Table \ref{tab:summary_models_all}, we list the different models tested in this work and the corresponding information.

The simplest model capable of reproducing all multiple images consists of two extended PIEMD profiles plus the cluster members (see Sect. \ref{sec:mass_param}).
In this model, we do not consider the foreground group of galaxies, only the 243 cluster members discussed in Sect. \ref{sec:members_and_los} are included.
This parameterisation has 14 free parameters and the number of degrees of freedom (${\rm DOF} \equiv$ number of model constraints $-$ number of free parameters) is 52.
We name this model with the ID `No-foreground' in Table \ref{tab:summary_models_all}.
The best-fit $\Delta_{rms}$ value is $0\farcs87$, with $\chi^2/{\rm DOF}=2.0$.
These values are slightly higher than those obtained from the strong lensing models of other merging clusters, such as MACS~J0416 with $\Delta_{rms} = 0\farcs59$ \citep{2017A&A...600A..90C}, Abell~370 with $\Delta_{rms} \approx 0\farcs7$ \citep{2019MNRAS.485.3738L} and Abell~2744 with $\Delta_{rms} = 0\farcs37$ \citep{2022arXiv220709416B}, to mention a few examples.
However, our model is capable of reproducing the positions of the multiple images created by ACT0102 with a much higher precision than previous models.
For instance, \citet{2013ApJ...770L..15Z} and \citet{2018ApJ...859..159C} obtained $\Delta_{rms}$ values of $3\farcs2$ and $\approx 1\farcs1$, respectively.
We attribute this improvement to a careful analysis of the spectroscopic data that has provided clean samples of multiple images and cluster members, and to the fact that we did not make use of photometric redshifts as priors in our model.
A more detailed comparison with other publicly available lens models of ACT0102 is presented in Sect. \ref{sec:comparison_with_previous_models}.

Next, we included the foreground group of 20 galaxies in the modelling. This group, at $z=0.63$, is located in projection between the two cluster BCGs and angularly close to some multiple images, as shown in Fig. \ref{fig:members_all}.
First, we assigned a mass value to these perturbers following the cluster member scaling relations, as if they were located at the cluster redshift, thus not increasing the number of free parameters of the model.
Within this approximation, the best-fit $\Delta_{rms}$ value is reduced to $=0\farcs75$ compared to the model without the foreground group, and we find $\chi^2/{\rm DOF} = 1.54$.
The improved value of $\Delta_{rms}$, BIC and AICc clearly indicates that the presence of this foreground group of galaxies must be included in the lens model.

Then, for the foreground galaxies we introduced two additional normalisation factors, $\sigma^{ref}_{v, fore}$ and $r^{ref}_{cut, fore}$, free to vary independently of those of the cluster members.
Despite having two additional free parameters, this model has the same $\Delta_{rms}$ value as the previous one, and is disfavoured by the values of the BIC and AICc information criteria (see model IDs `Reference' and `Two-scalings' in Table \ref{tab:summary_models_all}).
We also investigated a model in which the values of the exponent of the scaling relations (i.e. $\alpha$ and $\gamma$ in Eq. \ref{eq:scaling_relation}) are optimised.
In this model, ID `Free-scaling', we obtain $\Delta_{rms}=0\farcs74$, with best-fit values of $\alpha=3.9$ and $\gamma=1.7$.
From these tests, we conclude that more freedom in the models to describe the cluster members and foreground galaxies does not improve significantly the overall goodness of the fit.

We also checked whether a more complex parameterisation of the cluster-scale mass component can refine the image position reconstruction.
To do this, we included an additional PIEMD mass component in the model and allowed its position to vary in a square region with $200\arcsec$ per side ($\approx 1.5$~Mpc) over the lens plane centred around the BCG. We considered two cases: first with an axially symmetric distribution (i.e. for $\varepsilon=0$), then with the values of ellipticity and position angle of this new component free to vary. These two parameterisations add, respectively, four and six additional free parameters to the reference model.
The best-fit models have $\Delta_{rms}$ values of $0\farcs72$ and $0\farcs69$ for the circular and elliptical distributions, respectively.
The two models with a third PIEMD component are shown in Table \ref{tab:summary_models_all}.
Even though the position of the additional profile can vary across a large area, its best-fit centre is found very close to those of the main cluster mass components, located near the BCG.
The distance between the extra mass component and the main cluster halo is $7\farcs4$ and $12\farcs0$ for the circular and elliptical models, respectively, with similar values offset by $\approx 3\arcsec$.
This might point to the fact that ACT0102 has a complex total mass distribution, with some deviations from simple elliptical symmetry, that cannot be easily captured by parametric models.
Moreover, from the Markov chain Monte Carlo (MCMC)\ sampling, we see that the position is not well constrained and the posterior distributions are multi-modal.
For instance, the centre coordinates of the additional elliptical PIEMD component is not well constrained and has very large uncertainties where $x_0=-7\,_{-20}^{+15}$\arcsec \, and $y_0=6\,_{-5}^{+49}$\arcsec, at the 68\% confidence level, making it difficult to obtain a converged MCMC chain.
We note that reducing the allowed region for the third component does not improve the convergence and might bias the posterior distribution obtained from the sampling.
This indicates that the additional PIEMD mass profile might not represent a real third mass component, but rather the model trying to compensate for additional asymmetries of the mass distribution.
It is worth mentioning that the position of  this third component does not show correlation with the foreground structure, thus suggesting that this line-of-sight perturber does not have a significant dark matter halo.
Therefore, we did not use the models with additional mass components in our analyses because additional halos might not represent real mass components and are challenging to constrain with our current strong lens model inputs.

In the attempt to further improve the cluster mass model, we tried introducing perturbations to the elliptical PIEMD profiles to account for higher-order asymmetries in the extended mass profile (i.e. mainly for dark matter). To do this, we used the perturbative approach presented in \citet{2021MNRAS.506.2002B} and implemented in the \texttt{lenstool} software.
This method starts from two-dimensional B-spline basis functions, placed on a squared grid on the lens plane, to perturb the PIEMD profile.
The priors on the basis function parameters are set in order to ensure that the perturbations are small and preserve the total mass of the cluster.
For more details, we refer the reader to the works by \citet{2021MNRAS.506.2002B}, where the method is described in detail and tested with a simulated lens cluster, and by \citet{2022arXiv220202992L}, in which this method is applied to real data for the first time.

The additional free parameters describing the perturbations are the central position and the orientation angle of the grid, the distance between each node of the grid, and the amplitude of each perturbation.
For example, with a $3\times 3$ grid, the number of associated free parameters is $2 + 1 + 1 + 9 = 13$.
Given the number of additional free parameters increasing rapidly with the grid size, this method must be limited to grids with a relatively small number of nodes.
In Table \ref{tab:summary_models_all}, the models with the described perturbative approach have their IDs starting with `PertBS' followed by a number that indicates the number of nodes on each side of the grid (e.g. PertBS-3 denotes a $3\times3$ grid). Models with one or two PIEMD mass components are also identified with `1DM' and `2DM' in the model IDs.

Interestingly, models with one perturbed PIEMD component are capable of reproducing the multiplicity of all strongly lensed sources, if the grid has a size of $3 \times 3$ or larger (see the best-fit values in Table \ref{tab:summary_models_all}).
However, the values of the $\Delta_{\rm rms}$ are always high ($> 1 \arcsec$) and the increased number of free parameters is not justified by the values of the BIC and AICc criteria. For models with two PIEMD components plus perturbations, the lowest $\Delta_{\rm rms}$ value obtained is $0\farcs41$, for a grid with a size of $5 \times 5$.
We note that the model with a $2 \times 2$ grid size has similar $\Delta_{\rm rms}$ value as the model {\it 3-PIEMD}, indicating that a third PIEMD component might not have a physical origin and could just account for asymmetries in the cluster total mass distribution. We remark that the large number of free parameters and the BIC and AICc values suggest the presence of overfitting of the observed data.

\section{Discussion and comparison with previous strong lensing models}
\label{sec:discussions_and_comparison}

\begin{figure*}
      \centering
      \includegraphics[width = 0.33\textwidth]{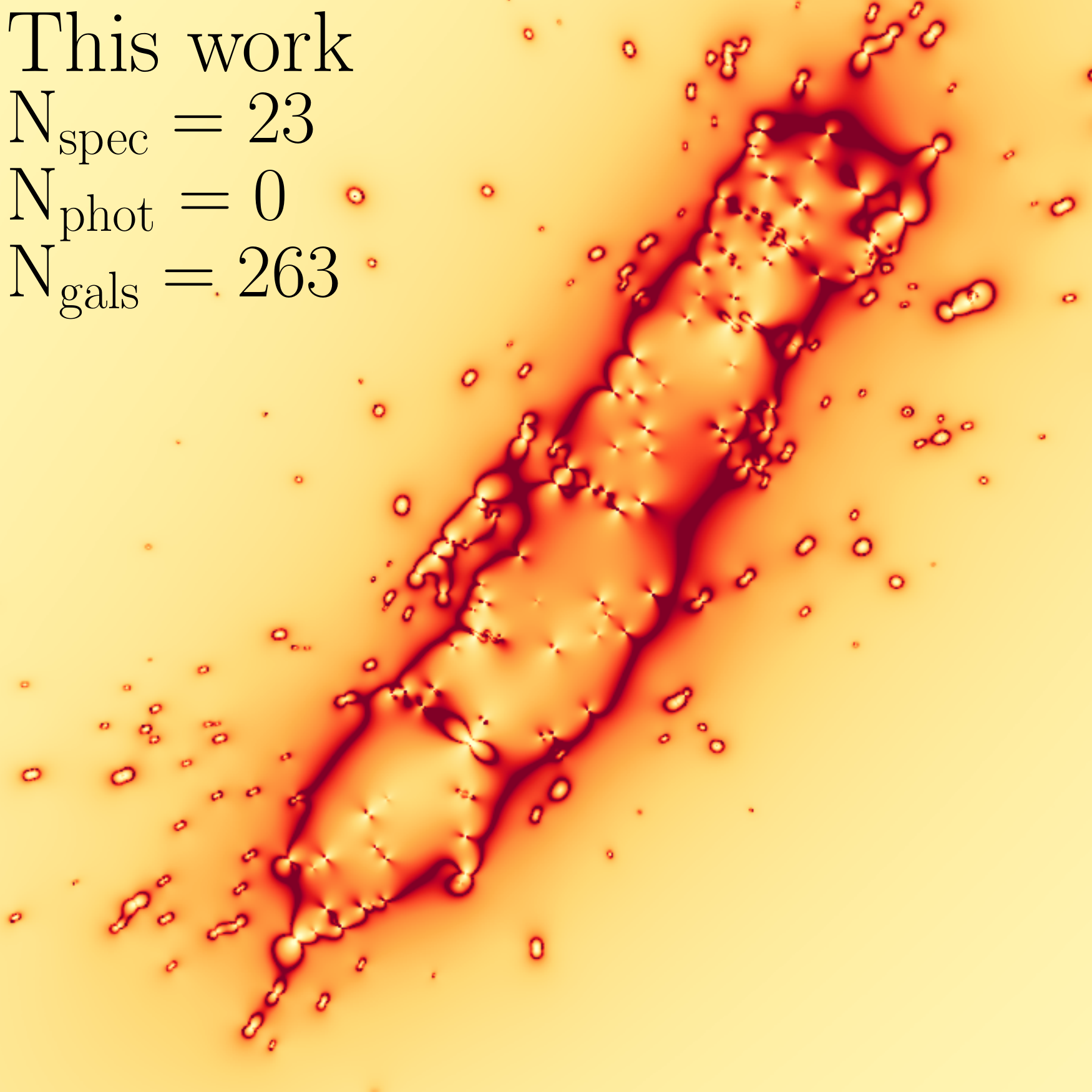}
      \includegraphics[width = 0.33\textwidth]{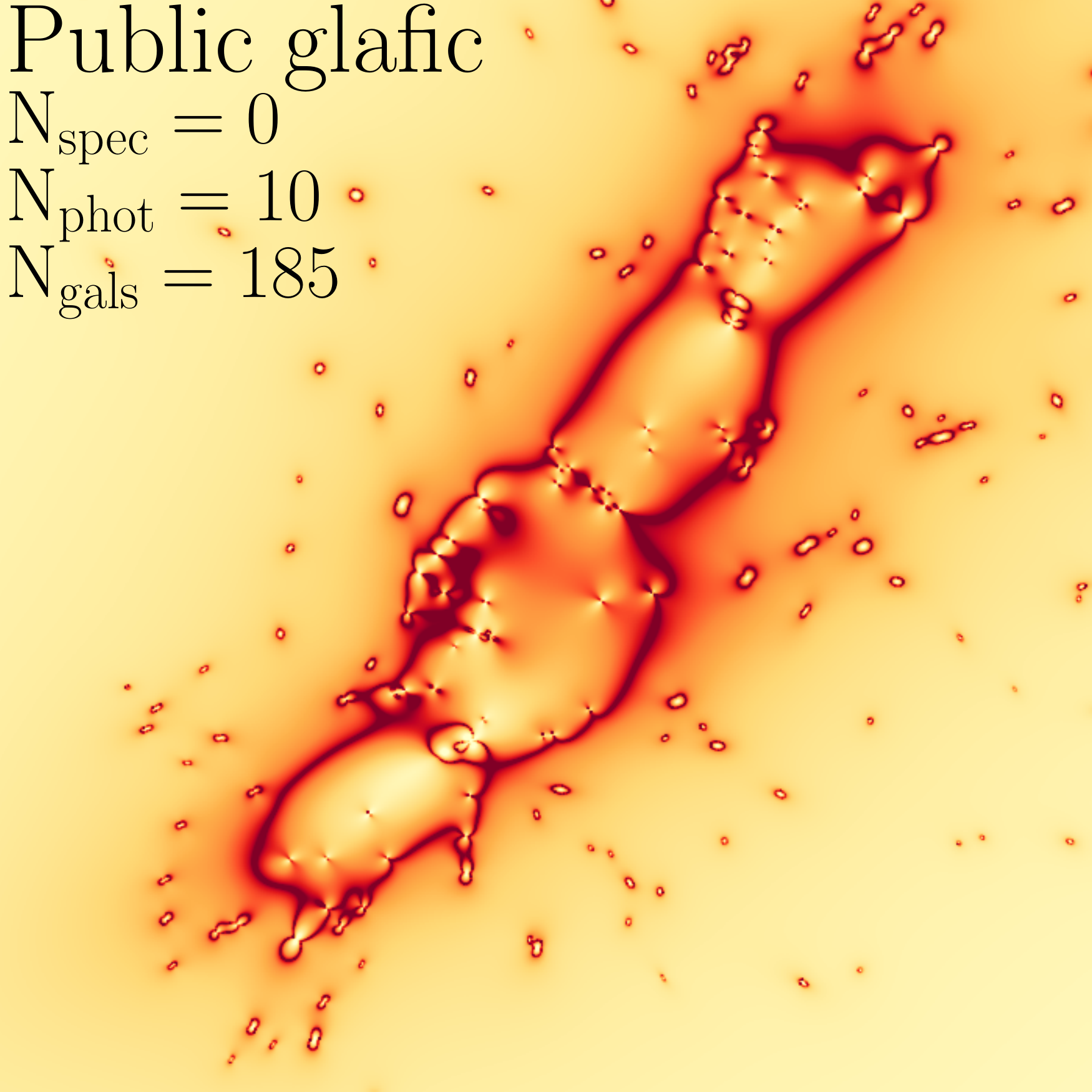}
      \includegraphics[width = 0.33\textwidth]{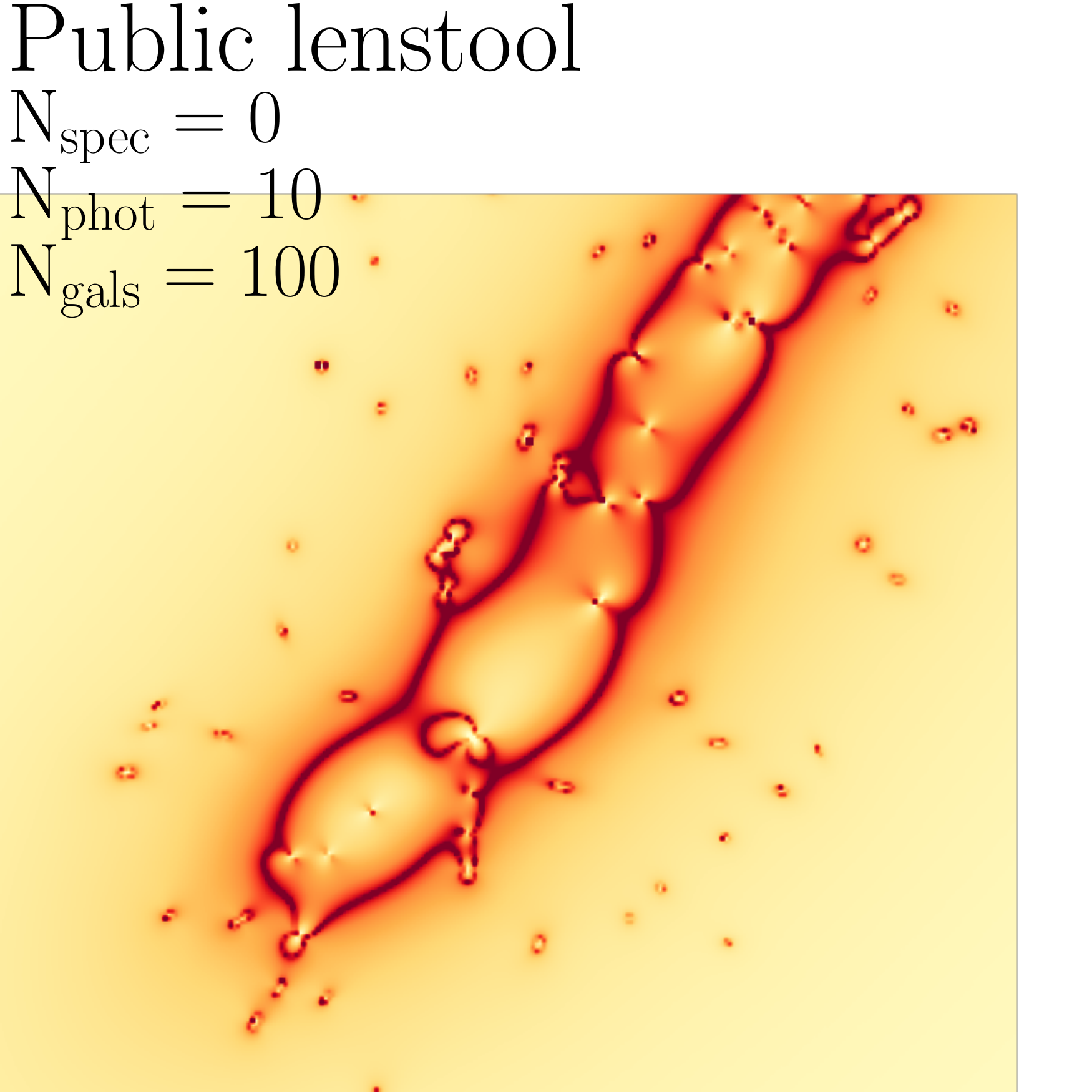}

      \vspace{0.05cm}
      \includegraphics[width = 0.33\textwidth]{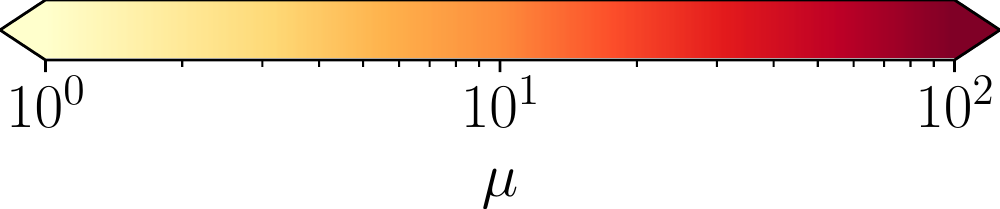}
  \caption{Magnification map comparison for the public models of ACT0102 for a source at $z_s=4$. The numbers at the top of each panel indicate the number of families of multiple images with spectroscopic redshifts ($\rm N_{spec}$) or selected from photometry only ($\rm N_{phot}$), and the number of cluster members and line-of-sight perturbers included in the lens total mass models ($\rm N_{gals}$).}
  \label{fig:magnification_map_comparison}
\end{figure*}

In this section we study the properties of the total mass distribution of ACT0102 and compare it with the results of previous works. We refer only to our reference model (ID Reference) for the sake of simplicity and with no impact on the validity of the general discussion.

\begin{table}[h]
\centering
    \caption{Mass parameters of ACT0102.}
    \begin{tabular}{c c c c c} \hline \hline
    ~ & Median & 68\% CL & 95\% CL & 99.7\% CL \\
\hline
$x_1~[\arcsec]$           & $-0.23$  & $_{-0.42}^{+0.38}$ & $_{-0.92}^{+0.74}$ &$_{-1.62}^{+1.08}$ \\
$y_1~[\arcsec]$           & $-0.08$  & $_{-0.50}^{+0.44}$ & $_{-1.06}^{+0.85}$ &$_{-1.68}^{+1.28}$ \\
$\varepsilon_1$           & $0.62$   & $_{-0.02}^{+0.02}$ & $_{-0.04}^{+0.06}$ &$_{-0.07}^{+0.11}$ \\
$\theta_1$~[deg]          & $41.0$  & $_{-1.0}^{+1.0}$ & $_{-2.0}^{+2.0}$ &$_{-3.2}^{+3.0}$ \\
$r_{core,1}~[\arcsec]$    & $14.0$  & $_{-1.5}^{+1.6}$ & $_{-3.1}^{+3.2}$ &$_{-5.4}^{+4.8}$ \\
$\sigma_{v,1}$~[km/s]     & $1041$   & $_{-37}^{+37}$ & $_{-76}^{+75}$ &$_{-129}^{+113}$ \\
$x_2~[\arcsec]$           & $47.0$  & $_{-0.9}^{+1.0}$ & $_{-1.7}^{+2.2}$ &$_{-2.4}^{+3.6}$ \\
$y_2~[\arcsec]$           & $80.6$  & $_{-1.1}^{+1.3}$ & $_{-2.2}^{+2.8}$ &$_{-3.1}^{+4.7}$ \\
$\varepsilon_2$           & $0.43$   & $_{-0.07}^{+0.07}$ & $_{-0.14}^{+0.15}$ &$_{-0.20}^{+0.21}$ \\
$\theta_2$~[deg]          & $53.2$  & $_{-1.3}^{+1.2}$ & $_{-2.6}^{+2.6}$ &$_{-4.1}^{+4.4}$ \\
$r_{core,2}~[\arcsec]$    & $17.2$  & $_{-1.6}^{+1.7}$ & $_{-3.0}^{+3.6}$ &$_{-4.2}^{+5.7}$ \\
$\sigma_{v,2}$~[km/s]     & $1010$   & $_{-37}^{+39}$ & $_{-72}^{+79}$ &$_{-105}^{+122}$ \\
$r_{cut,gals}~[\arcsec]$  & $13.5$  & $_{-4.2}^{+5.7}$ & $_{-7.5}^{+12.6}$ &$_{-9.6}^{+18.8}$ \\
$\sigma_{v,gal}$~[km/s]   & $290$ & $_{-30}^{+39}$ & $_{-52}^{+94}$ &$_{-69}^{+172}$ \\
\hline
\hline\hline
    \end{tabular}
    \tablefoot{Positions are in arcseconds relative to the BCG luminosity centre (RA=$15.7406934$ and Dec=$-49.2719924$). Angles are referred to the $x$-axis and increase going anti-clockwise.}
\label{tab:model_params}
\end{table}

\subsection{The total mass distribution of ACT0102 from strong lensing}
\label{sec:the_mass_distribution}

From the previous sections, we find that the parameterisation that best represents the cluster total mass distribution and does not overfit the data is composed of two PIEMD halos plus cluster members and the group of foreground galaxies following the same total mass-to-light scaling relation.
For this model, we ran \texttt{lenstool} in the sampling mode to compute the posterior distribution of all free parameters.
In this step, we rescaled the positional errors $\sigma_j$ (see Eq. \ref{eq:chi2}) in order to have $\chi^2/\rm DOF=1$ to obtain realistic statistical uncertainties.
The recovered values of all 14 free parameters in this model, along with their confidence level intervals, are listed in Table \ref{tab:model_params}.
The position of the main PIEMD halo is in very good agreement with that of the BCG, and the second halo is located along the extended X-ray emission in the north-west direction.
In Fig. \ref{fig:xrays}, we compare the \chandra X-ray isophotes with the projected mass isocontours of the cluster smooth mass component (i.e. removing the contribution of the cluster members) from our strong lens model.
The peak of the X-ray emission has an offset of $\sim 5 \arcsec$ from the BCG, in agreement with what is commonly found in merging clusters \citep{2016MNRAS.457.4515R}.
Moreover, in Fig. \ref{fig:temperature_map} we show that the gas temperature map has low temperatures (in projection) in the region around the BCG, indicating the presence of a cool-core.
It is found that such spacial offsets between the X-ray and BCG, and the presence of a cool-core is associated with major merger systems \citep{2010A&A...513A..37H}.
Not surprisingly, from our lens model we obtain very similar values for both PIEMD mass components $\rm M_{main}(<300~kpc) = 2.29_{-0.10}^{+0.09} \times 10^{14}M_{\odot}$ for the main mass component, and $\rm M_{north-west}(<300~kpc) = 2.10_{-0.09}^{+0.08} \times 10^{14}M_{\odot}$ for the north-west component.

In Fig. \ref{fig:mass_profiles} we show the  cumulative projected total mass and total surface mass density profiles.
The same profiles for another seven clusters from \citet{2016A&A...587A..80C, 2017A&A...600A..90C, 2017A&A...607A..93C, 2019A&A...632A..36C} with similar datasets are also included.
Remarkably, ACT0102 reveals multiple images out to distances of $\approx 1$~Mpc from its BCG, becoming the lens cluster with the most extended region over which strong lensing observations are available to map the cluster total mass distribution. In Fig. \ref{fig:mass_profiles} we also plot the same profiles rescaled to the values of $R_{200c}$ and $M_{200c}$, obtained from independent weak lensing analyses. These quantities are defined as, respectively, the radius inside which the cluster mean density value is equal to 200 times that of the critical density of the Universe at the cluster redshift and the corresponding mass enclosed within a sphere with that radius.
For ACT0102, we refer to the recent weak lensing study by \citet{2021arXiv210600031K}, performed by using \hst imaging. For the sample in \citet{2019A&A...632A..36C}, the weak lensing mass reconstructions were presented in \citet{2018ApJ...860..104U}, by analysing deep ground base imaging.

The rescaled profiles of ACT0102 deviate slightly from the general trends presented in \citet{2019A&A...632A..36C}, with differences of the order of $\approx 10\%$ to $30\%$ between $0.01\times R_{200c}$ and $0.1\times R_{200c}$ for the total mass.
Regarding the slope (i.e. $\frac{{\rm d}M}{{\rm d}R_{200c}}$), the difference in this region varies from $5\%$ to $30\%$, with ACT0102 being steeper compared to the sample average, especially at large radii.
We partially attribute this to the different weak lensing methodologies and datasets used for the different clusters.
Moreover, the complex merging state of ACT0102 might also be responsible for some deviations from the overall homologous profiles. For instance, in the sample of \citet{2019A&A...632A..36C}, the merging cluster MACS~J0416 deviates the most from the general trend. Interestingly, the X-ray emission shows a front near the BCG and a long tail towards the north-west region, suggesting a recent merging event.

\subsection{Comparison with previous models}
\label{sec:comparison_with_previous_models}

The first strong lensing analysis of ACT0102 was presented in \citet{2013ApJ...770L..15Z}, where the authors made use of relatively shallow ($\approx 40$ minutes) \hst imaging in the F625W, F775W and F850LP filters.
In that work, the authors identified multiple images of nine strongly lensed background sources, which they used to constrain the cluster total mass distribution, obtaining a $\Delta_{\rm rms}$ value of $3\farcs2$.
Such a high value of $\Delta_{\rm rms}$ might be mostly explained by the lack of spectroscopic redshifts for both the multiple images and the cluster members, and the large uncertainties associated with the photometric redshifts used as priors in the model.

Following the acquisition of additional \hst data under the RELICS programme, updated models were presented in \citet{2018ApJ...859..159C} and \citet{2020ApJ...904..106D}.
All those works lacked spectroscopic information, especially for the multiple image systems.
In Table \ref{tab:multiple_images}, we list the previous identifications that match with our spectroscopically confirmed multiple images.
A total of five multiple image families from previous works have now been secured with our MUSE data.
Remarkably, 41 multiple images from 18 different sources are new identifications presented for the first time here and all have reliable redshift measurements.
Therefore, the cluster total mass reconstruction obtained in this work is much less subject to possible systematic effects related to multiple image (mis)identifications, based only on photometry. In fact, our spectroscopic measurements show that the multiple families 6.1 and 6.2 used in \citet{2020ApJ...904..106D} have turned out to be wrongly identified (see also Table \ref{tab:multiple_images} and our family ID 8).

In Fig. \ref{fig:mass_comparison} we compare our cumulative projected total mass and total surface mass density profiles with those of the two publicly available models, through the MAST portal.
The first one is that already mentioned by \citet{2018ApJ...859..159C} and the second one was obtained by using the software \texttt{glafic} \citep{2010PASJ...62.1017O, 2020MNRAS.496.2591O}.
These models have $\Delta_{rms}$ values of $\approx 0\farcs82$ \citep{2018ApJ...859..159C} and $0\farcs52$ \citep{2020MNRAS.496.2591O} and they both considered 10 families of multiple images, with no spectroscopic measurements, as model constraints.
From Fig. \ref{fig:mass_comparison} we conclude that \citet{2018ApJ...859..159C} underestimated the cluster total mass and, more clearly, the projected mass density in the radial range between 100~kpc and 1~Mpc. Such a discrepancy could be due to a general overestimate (see Table \ref{tab:multiple_images}) of the source redshifts, which were optimised with all the other model parameters in the previous work, and the intrinsic degeneracy (i.e. an anti-correlation) between the redshift of a source and the total mass of a lens.
Moreover, the north-west cluster-scale mass component was unconstrained because of the low number of multiple images in that region in previous works.
We note that the total mass and mass density profiles obtained with the \texttt{glafic} code have shapes very similar to ours, but larger statistical uncertainties.

While the total projected mass density enclosed within the very core of the cluster, where strong lensing constraints exist, is robust to within a few percent even with only a handful of constraints \citep[e.g.][]{2021ApJ...920...98R,2021ApJ...910..146R}, this may not be the case for other lensing outputs.
For instance, \citet{2016ApJ...832...82J} show that accuracy of recovering the mass distribution and lensing magnification increases significantly with the number of spectroscopic redshifts of lensed sources used to constrain the lens model.
We also contrast our magnification maps with those of the public models. In Fig. \ref{fig:magnification_map_comparison} we show the magnification maps for a source at redshift $z_s=4.0$. We note that the available map by \citet{2018ApJ...859..159C} does not cover the entire tangential critical line at that redshift. The overall shape of the critical lines is comparable for all models, but our reconstruction is more detailed in the regions with high magnification values, thanks to the additional spectroscopic  information exploited for the selection of the cluster members.
Accurate magnification maps are crucial for studies of highly magnified sources \citep[see e.g.][]{2021ApJ...908..146C} and will be of great value for future works using upcoming\textit{ James Webb }Space Telescope (JWST) data.
Our lens model, along with the magnification maps and full posterior distribution of mass parameters are publicly available in the electronic version of this work.

\section{Summary and conclusions}
\label{sec:conclusions}

In this work we have presented a strong gravitational lens model of ACT0102 based on 56 new spectroscopic confirmations of multiple images.
The positions of the multiple images, which are our model constraints, expand across a region of 1~Mpc in the cluster core, an area remarkably larger than that of any other strong lens cluster.
In addition to the multiple images, we have also measured precise spectroscopic redshifts for 167 cluster members, and we have identified a foreground group of galaxies with a significant impact on the gravitational lensing deflection of background galaxies.
We summarise the main results of this work as follows:

\begin{itemize}
 \item We used a sample of 56 multiple images, from 23 background galaxies with secure spectroscopic redshifts, to constrain the total mass distribution of ACT0102. The observed positions of the multiple images are reproduced with a root mean square value of $\Delta_{\rm rms} = 0.75$. In this reference model, the mass distribution is parameterised with two cluster-scale components (mainly dark matter) plus the cluster members and a foreground group of galaxies.
 Introducing perturbations to the total mass distribution following the methodology in \citet{2021MNRAS.506.2002B} can also improve the model predictions. However, the very large number of free parameters in this approach disfavours these models according to the BIC and AICc.

 \item Thanks to the capabilities of MUSE, in addition to the 56 images of multiply lensed sources, we have spectroscopically confirmed 167 cluster members and a foreground group of 20 galaxies at $z=0.63$. Such a large number of confirmations, and the identification of the intervening mass component in the foreground not considered in previous works, is crucial to reducing the value of $\Delta_{\rm rms}$, as indicated in Table \ref{tab:summary_models_all}.
 
 \item We included constraints out to $\approx 1$~Mpc from the BCG in our lens model, and we have estimated a total mass value within this radius of $\rm M_{\rm total}(<1~Mpc) = 1.84_{-0.04}^{+0.03} \times 10^{15} M_{\odot}$. The main cluster-scale component is located close to the BCG, and the second is located in the north-west region (see Fig. \ref{fig:xrays}); they have comparable mass values of $\rm M_{main}(<300~kpc) = 2.29_{-0.10}^{+0.09} \times 10^{14}M_{\odot}$ and $\rm M_{north-west}(<300~kpc) = 2.10_{-0.09}^{+0.08} \times 10^{14}M_{\odot}$ within 300~kpc of their centres. This is in very good agreement with the major merging scenario of ACT0102 \citep[see e.g.][]{2014ApJ...785...20J,2021arXiv210600031K}.

 \item We have found a very small offset of $0\farcs5_{-0.3}^{+0.9}$ between the main cluster mass component and the BCG, in contrast to the offset of $\approx 5\arcsec$ between these two components and the X-ray emission. This offset between the components with small interaction cross-sections (i.e. stars and dark matter) and cluster hot gas (traced by the X-ray emission) is commonly found in merging systems and is a signature of cool-core clusters.
 
 \item We compared our total mass and density profiles with those from previous strong lens models of ACT0102 (see Fig. \ref{fig:mass_comparison}), finding that the work of \citet[][which uses the software \texttt{lenstool}]{2018ApJ...859..159C} underestimates the cluster total mass and density in the outer regions ($\rm R \gtrsim 200~kpc$). In addition, the model uncertainties of \citet[][who used \texttt{glafic}]{2020MNRAS.496.2591O} are much larger than those from our total mass reconstruction. These results further stress the importance of including secure multiple images and cluster members with spectroscopic confirmations in the lens models  \citep[see also][]{2015ApJ...800...38G, 2016ApJ...832...82J}.
 
\end{itemize}

The strong lens model together with the MUSE redshift catalogue presented in this work will be extremely valuable for future works, especially in view of the new near-infrared imaging data obtained under the GTO/JWST PEARLS programme ID  1176 (P.I.: R. A. Windhorst).
JWST photometry allows the identification of several new multiple image systems \citep[see e.g.][]{2022arXiv220707567C}, but new spectroscopic confirmations of faint galaxies will be challenging. Therefore, our spectroscopic confirmations will be crucial in `anchoring' any successor lens model based on these new data.

Moreover, the lens model presented in this work provides accurate magnification maps that can be used to characterise the faint and magnified population of galaxies.
We note that ACT0102 is especially efficient at strongly lensing galaxies at larger cosmological distances because of its high redshift ($z=0.87$) when compared to the current sample of strong lens clusters (see, for instance, Fig. \ref{fig:families_z_hist}).
We have made  the lens model presented in this work publicly available, including the magnification maps and \texttt{lenstool} configuration file, along with the full redshift catalogue built using the MUSE data.

\begin{acknowledgements}
The authors thank the anonymous referee for the useful comments that helped to improve the manuscript.
GBC acknowledge the Max Planck Society for financial support through the Max Planck Research Group for S. H. Suyu and the academic support from the German Centre for Cosmological Lensing.
CG, PR, AA, PB, AM, PT and EV acknowledge financial support through grants PRIN-MIUR 2017WSCC32 and 2020SKSTHZ.
KIC acknowledges the Dutch Research Council (NWO) for the award of the Vici Grant VI.C.212.036.
RD gratefully acknowledges support by the ANID BASAL projects ACE210002 and FB210003.
Based on observations made with the NASA/ESA \hst, obtained at the Space Telescope Science Institute, which is operated by the Association of Universities for Research in Astronomy, Inc., under NASA contract NAS 5-26555. These observations are associated with programs GO-12755, GO-14096.
High level science products from the RELICS program were obtained from Mikulski Archive for Space Telescopes (MAST).
Based on VLT/MUSE observations collected at the European Southern Observatory under ESO programme ID 0102.A-0266(A), P.I.: G. B. Caminha.
This research made use of {\tt Astropy},\footnote{\url{http://www.astropy.org}} a community-developed core Python package for Astronomy \citep{astropy:2013, astropy:2018}, {\tt NumPy} \citep{harris2020array} and {\tt Matplotlib} \citep{Hunter:2007}.
\end{acknowledgements}

\bibliographystyle{aa}
\bibliography{references}

\begin{thebibliography}{94}
\expandafter\ifx\csname natexlab\endcsname\relax\def\natexlab#1{#1}\fi

\bibitem[{{Abbott} {et~al.}(2020){Abbott}, {Aguena}, {Alarcon}, {Allam},
  {Allen}, {Annis}, {Avila}, {Bacon}, {Bechtol}, {Bermeo}, {Bernstein},
  {Bertin}, {Bhargava}, {Bocquet}, {Brooks}, {Brout}, {Buckley-Geer}, {Burke},
  {Carnero Rosell}, {Carrasco Kind}, {Carretero}, {Castander}, {Cawthon},
  {Chang}, {Chen}, {Choi}, {Costanzi}, {Crocce}, {da Costa}, {Davis}, {De
  Vicente}, {DeRose}, {Desai}, {Diehl}, {Dietrich}, {Dodelson}, {Doel},
  {Drlica-Wagner}, {Eckert}, {Eifler}, {Elvin-Poole}, {Estrada}, {Everett},
  {Evrard}, {Farahi}, {Ferrero}, {Flaugher}, {Fosalba}, {Frieman},
  {Garc{\'\i}a-Bellido}, {Gatti}, {Gaztanaga}, {Gerdes}, {Giannantonio},
  {Giles}, {Grandis}, {Gruen}, {Gruendl}, {Gschwend}, {Gutierrez}, {Hartley},
  {Hinton}, {Hollowood}, {Honscheid}, {Hoyle}, {Huterer}, {James}, {Jarvis},
  {Jeltema}, {Johnson}, {Johnson}, {Kent}, {Krause}, {Kron}, {Kuehn},
  {Kuropatkin}, {Lahav}, {Li}, {Lidman}, {Lima}, {Lin}, {MacCrann}, {Maia},
  {Mantz}, {Marshall}, {Martini}, {Mayers}, {Melchior}, {Mena-Fern{\'a}ndez},
  {Menanteau}, {Miquel}, {Mohr}, {Nichol}, {Nord}, {Ogando}, {Palmese},
  {Paz-Chinch{\'o}n}, {Plazas}, {Prat}, {Rau}, {Romer}, {Roodman}, {Rooney},
  {Rozo}, {Rykoff}, {Sako}, {Samuroff}, {S{\'a}nchez}, {Sanchez}, {Saro},
  {Scarpine}, {Schubnell}, {Scolnic}, {Serrano}, {Sevilla-Noarbe}, {Sheldon},
  {Smith}, {Smith}, {Suchyta}, {Swanson}, {Tarle}, {Thomas}, {To}, {Troxel},
  {Tucker}, {Varga}, {von der Linden}, {Walker}, {Wechsler}, {Weller},
  {Wilkinson}, {Wu}, {Yanny}, {Zhang}, {Zhang}, {Zuntz}, \& {DES
  Collaboration}}]{2020PhRvD.102b3509A}
{Abbott}, T.~M.~C., {Aguena}, M., {Alarcon}, A., {et~al.} 2020, \prd, 102,
  023509

\bibitem[{{Acebron} {et~al.}(2019){Acebron}, {Alon}, {Zitrin}, {Mahler}, {Coe},
  {Sharon}, {Cibirka}, {Brada{\v{c}}}, {Trenti}, {Umetsu}, {Andrade-Santos},
  {Avila}, {Bradley}, {Carrasco}, {Cerny}, {Czakon}, {Dawson}, {Frye}, {Hoag},
  {Huang}, {Johnson}, {Jones}, {Kikuchihara}, {Lam}, {Livermore}, {Lovisari},
  {Mainali}, {Oesch}, {Ogaz}, {Ouchi}, {Past}, {Paterno-Mahler}, {Peterson},
  {Ryan}, {Salmon}, {Sendra-Server}, {Stark}, {Strait}, {Toft}, \&
  {Vulcani}}]{2019ApJ...874..132A}
{Acebron}, A., {Alon}, M., {Zitrin}, A., {et~al.} 2019, \apj, 874, 132

\bibitem[{{Acebron} {et~al.}(2022){Acebron}, {Grillo}, {Bergamini}, {Mercurio},
  {Rosati}, {Caminha}, {Tozzi}, {Brammer}, {Meneghetti}, {Morelli}, {Nonino},
  \& {Vanzella}}]{2022ApJ...926...86A}
{Acebron}, A., {Grillo}, C., {Bergamini}, P., {et~al.} 2022, \apj, 926, 86

\bibitem[{{Akaike}(1974)}]{1974ITAC...19..716A}
{Akaike}, H. 1974, IEEE Transactions on Automatic Control, 19, 716

\bibitem[{{Arnaud}(1996)}]{1996Arnaud}
{Arnaud}, K.~A. 1996, in Astronomical Society of the Pacific Conference Series,
  Vol. 101, Astronomical Data Analysis Software and Systems V, ed.
  {G.~H.~Jacoby \& J.~Barnes}, 17--+

\bibitem[{{Asencio} {et~al.}(2021){Asencio}, {Banik}, \&
  {Kroupa}}]{2021MNRAS.500.5249A}
{Asencio}, E., {Banik}, I., \& {Kroupa}, P. 2021, \mnras, 500, 5249

\bibitem[{{Asplund} {et~al.}(2009){Asplund}, {Grevesse}, {Sauval}, \&
  {Scott}}]{2009Asplund}
{Asplund}, M., {Grevesse}, N., {Sauval}, A.~J., \& {Scott}, P. 2009, \araa, 47,
  481

\bibitem[{{Astropy Collaboration} {et~al.}(2018){Astropy Collaboration},
  {Price-Whelan}, {Sip{\H{o}}cz}, {G{\"u}nther}, {Lim}, {Crawford}, {Conseil},
  {Shupe}, {Craig}, {Dencheva}, {Ginsburg}, {Vand erPlas}, {Bradley},
  {P{\'e}rez-Su{\'a}rez}, {de Val-Borro}, {Aldcroft}, {Cruz}, {Robitaille},
  {Tollerud}, {Ardelean}, {Babej}, {Bach}, {Bachetti}, {Bakanov}, {Bamford},
  {Barentsen}, {Barmby}, {Baumbach}, {Berry}, {Biscani}, {Boquien}, {Bostroem},
  {Bouma}, {Brammer}, {Bray}, {Breytenbach}, {Buddelmeijer}, {Burke},
  {Calderone}, {Cano Rodr{\'\i}guez}, {Cara}, {Cardoso}, {Cheedella}, {Copin},
  {Corrales}, {Crichton}, {D'Avella}, {Deil}, {Depagne}, {Dietrich}, {Donath},
  {Droettboom}, {Earl}, {Erben}, {Fabbro}, {Ferreira}, {Finethy}, {Fox},
  {Garrison}, {Gibbons}, {Goldstein}, {Gommers}, {Greco}, {Greenfield},
  {Groener}, {Grollier}, {Hagen}, {Hirst}, {Homeier}, {Horton}, {Hosseinzadeh},
  {Hu}, {Hunkeler}, {Ivezi{\'c}}, {Jain}, {Jenness}, {Kanarek}, {Kendrew},
  {Kern}, {Kerzendorf}, {Khvalko}, {King}, {Kirkby}, {Kulkarni}, {Kumar},
  {Lee}, {Lenz}, {Littlefair}, {Ma}, {Macleod}, {Mastropietro}, {McCully},
  {Montagnac}, {Morris}, {Mueller}, {Mumford}, {Muna}, {Murphy}, {Nelson},
  {Nguyen}, {Ninan}, {N{\"o}the}, {Ogaz}, {Oh}, {Parejko}, {Parley}, {Pascual},
  {Patil}, {Patil}, {Plunkett}, {Prochaska}, {Rastogi}, {Reddy Janga},
  {Sabater}, {Sakurikar}, {Seifert}, {Sherbert}, {Sherwood-Taylor}, {Shih},
  {Sick}, {Silbiger}, {Singanamalla}, {Singer}, {Sladen}, {Sooley},
  {Sornarajah}, {Streicher}, {Teuben}, {Thomas}, {Tremblay}, {Turner},
  {Terr{\'o}n}, {van Kerkwijk}, {de la Vega}, {Watkins}, {Weaver}, {Whitmore},
  {Woillez}, {Zabalza}, \& {Astropy Contributors}}]{astropy:2018}
{Astropy Collaboration}, {Price-Whelan}, A.~M., {Sip{\H{o}}cz}, B.~M., {et~al.}
  2018, \aj, 156, 123

\bibitem[{{Astropy Collaboration} {et~al.}(2013){Astropy Collaboration},
  {Robitaille}, {Tollerud}, {Greenfield}, {Droettboom}, {Bray}, {Aldcroft},
  {Davis}, {Ginsburg}, {Price-Whelan}, {Kerzendorf}, {Conley}, {Crighton},
  {Barbary}, {Muna}, {Ferguson}, {Grollier}, {Parikh}, {Nair}, {Unther},
  {Deil}, {Woillez}, {Conseil}, {Kramer}, {Turner}, {Singer}, {Fox}, {Weaver},
  {Zabalza}, {Edwards}, {Azalee Bostroem}, {Burke}, {Casey}, {Crawford},
  {Dencheva}, {Ely}, {Jenness}, {Labrie}, {Lim}, {Pierfederici}, {Pontzen},
  {Ptak}, {Refsdal}, {Servillat}, \& {Streicher}}]{astropy:2013}
{Astropy Collaboration}, {Robitaille}, T.~P., {Tollerud}, E.~J., {et~al.} 2013,
  \aap, 558, A33

\bibitem[{{Balestra} {et~al.}(2016){Balestra}, {Mercurio}, {Sartoris},
  {Girardi}, {Grillo}, {Nonino}, {Rosati}, {Biviano}, {Ettori}, {Forman},
  {Jones}, {Koekemoer}, {Medezinski}, {Merten}, {Ogrean}, {Tozzi}, {Umetsu},
  {Vanzella}, {van Weeren}, {Zitrin}, {Annunziatella}, {Caminha}, {Broadhurst},
  {Coe}, {Donahue}, {Fritz}, {Frye}, {Kelson}, {Lombardi}, {Maier},
  {Meneghetti}, {Monna}, {Postman}, {Scodeggio}, {Seitz}, \&
  {Ziegler}}]{2016ApJS..224...33B}
{Balestra}, I., {Mercurio}, A., {Sartoris}, B., {et~al.} 2016, \apjs, 224, 33

\bibitem[{{Basu} {et~al.}(2016){Basu}, {Sommer}, {Erler}, {Eckert}, {Vazza},
  {Magnelli}, {Bertoldi}, \& {Tozzi}}]{2016ApJ...829L..23B}
{Basu}, K., {Sommer}, M., {Erler}, J., {et~al.} 2016, \apjl, 829, L23

\bibitem[{{Beauchesne} {et~al.}(2021){Beauchesne}, {Cl{\'e}ment}, {Richard}, \&
  {Kneib}}]{2021MNRAS.506.2002B}
{Beauchesne}, B., {Cl{\'e}ment}, B., {Richard}, J., \& {Kneib}, J.-P. 2021,
  \mnras, 506, 2002

\bibitem[{{Bergamini} {et~al.}(2023){Bergamini}, {Acebron}, {Grillo}, {Rosati},
  {Caminha}, {Mercurio}, {Vanzella}, {Angora}, {Brammer}, {Meneghetti}, \&
  {Nonino}}]{2022arXiv220709416B}
{Bergamini}, P., {Acebron}, A., {Grillo}, C., {et~al.} 2023, \aap, 670, A60

\bibitem[{{Bergamini} {et~al.}(2021){Bergamini}, {Rosati}, {Vanzella},
  {Caminha}, {Grillo}, {Mercurio}, {Meneghetti}, {Angora}, {Calura}, {Nonino},
  \& {Tozzi}}]{2021A&A...645A.140B}
{Bergamini}, P., {Rosati}, P., {Vanzella}, E., {et~al.} 2021, \aap, 645, A140

\bibitem[{{Bonamigo} {et~al.}(2017){Bonamigo}, {Grillo}, {Ettori}, {Caminha},
  {Rosati}, {Mercurio}, {Annunziatella}, {Balestra}, \&
  {Lombardi}}]{2017ApJ...842..132B}
{Bonamigo}, M., {Grillo}, C., {Ettori}, S., {et~al.} 2017, \apj, 842, 132

\bibitem[{{Bonamigo} {et~al.}(2018){Bonamigo}, {Grillo}, {Ettori}, {Caminha},
  {Rosati}, {Mercurio}, {Munari}, {Annunziatella}, {Balestra}, \&
  {Lombardi}}]{2018ApJ...864...98B}
{Bonamigo}, M., {Grillo}, C., {Ettori}, S., {et~al.} 2018, \apj, 864, 98

\bibitem[{{Botteon} {et~al.}(2016){Botteon}, {Gastaldello}, {Brunetti}, \&
  {Kale}}]{2016MNRAS.463.1534B}
{Botteon}, A., {Gastaldello}, F., {Brunetti}, G., \& {Kale}, R. 2016, \mnras,
  463, 1534

\bibitem[{{Brada{\v{c}}} {et~al.}(2006){Brada{\v{c}}}, {Clowe}, {Gonzalez},
  {Marshall}, {Forman}, {Jones}, {Markevitch}, {Randall}, {Schrabback}, \&
  {Zaritsky}}]{2006ApJ...652..937B}
{Brada{\v{c}}}, M., {Clowe}, D., {Gonzalez}, A.~H., {et~al.} 2006, \apj, 652,
  937

\bibitem[{{Caminha} {et~al.}(2016{\natexlab{a}}){Caminha}, {Grillo}, {Rosati},
  {Balestra}, {Karman}, {Lombardi}, {Mercurio}, {Nonino}, {Tozzi}, {Zitrin},
  {Biviano}, {Girardi}, {Koekemoer}, {Melchior}, {Meneghetti}, {Munari},
  {Suyu}, {Umetsu}, {Annunziatella}, {Borgani}, {Broadhurst}, {Caputi}, {Coe},
  {Delgado-Correal}, {Ettori}, {Fritz}, {Frye}, {Gobat}, {Maier}, {Monna},
  {Postman}, {Sartoris}, {Seitz}, {Vanzella}, \&
  {Ziegler}}]{2016A&A...587A..80C}
{Caminha}, G.~B., {Grillo}, C., {Rosati}, P., {et~al.} 2016{\natexlab{a}},
  \aap, 587, A80

\bibitem[{{Caminha} {et~al.}(2017{\natexlab{a}}){Caminha}, {Grillo}, {Rosati},
  {Balestra}, {Mercurio}, {Vanzella}, {Biviano}, {Caputi}, {Delgado-Correal},
  {Karman}, {Lombardi}, {Meneghetti}, {Sartoris}, \&
  {Tozzi}}]{2017A&A...600A..90C}
{Caminha}, G.~B., {Grillo}, C., {Rosati}, P., {et~al.} 2017{\natexlab{a}},
  \aap, 600, A90

\bibitem[{{Caminha} {et~al.}(2017{\natexlab{b}}){Caminha}, {Grillo}, {Rosati},
  {Meneghetti}, {Mercurio}, {Ettori}, {Balestra}, {Biviano}, {Umetsu},
  {Vanzella}, {Annunziatella}, {Bonamigo}, {Delgado-Correal}, {Girardi},
  {Lombardi}, {Nonino}, {Sartoris}, {Tozzi}, {Bartelmann}, {Bradley}, {Caputi},
  {Coe}, {Ford}, {Fritz}, {Gobat}, {Postman}, {Seitz}, \&
  {Zitrin}}]{2017A&A...607A..93C}
{Caminha}, G.~B., {Grillo}, C., {Rosati}, P., {et~al.} 2017{\natexlab{b}},
  \aap, 607, A93

\bibitem[{{Caminha} {et~al.}(2016{\natexlab{b}}){Caminha}, {Karman}, {Rosati},
  {Caputi}, {Arrigoni Battaia}, {Balestra}, {Grillo}, {Mercurio}, {Nonino}, \&
  {Vanzella}}]{2016A&A...595A.100C}
{Caminha}, G.~B., {Karman}, W., {Rosati}, P., {et~al.} 2016{\natexlab{b}},
  \aap, 595, A100

\bibitem[{{Caminha} {et~al.}(2019){Caminha}, {Rosati}, {Grillo}, {Rosani},
  {Caputi}, {Meneghetti}, {Mercurio}, {Balestra}, {Bergamini}, {Biviano},
  {Nonino}, {Umetsu}, {Vanzella}, {Annunziatella}, {Broadhurst},
  {Delgado-Correal}, {Demarco}, {Koekemoer}, {Lombardi}, {Maier}, {Verdugo}, \&
  {Zitrin}}]{2019A&A...632A..36C}
{Caminha}, G.~B., {Rosati}, P., {Grillo}, C., {et~al.} 2019, \aap, 632, A36

\bibitem[{{Caminha} {et~al.}(2022){Caminha}, {Suyu}, {Mercurio}, {Brammer},
  {Bergamini}, {Acebron}, \& {Vanzella}}]{2022arXiv220707567C}
{Caminha}, G.~B., {Suyu}, S.~H., {Mercurio}, A., {et~al.} 2022, \aap, 666, L9

\bibitem[{{Cappellari} \& {Copin}(2003)}]{2003Cappellari}
{Cappellari}, M. \& {Copin}, Y. 2003, \mnras, 342, 345

\bibitem[{{Caputi} {et~al.}(2021){Caputi}, {Caminha}, {Fujimoto}, {Kohno},
  {Sun}, {Egami}, {Deshmukh}, {Tang}, {Ao}, {Bradley}, {Coe}, {Espada},
  {Grillo}, {Hatsukade}, {Knudsen}, {Lee}, {Magdis}, {Morokuma-Matsui},
  {Oesch}, {Ouchi}, {Rosati}, {Umehata}, {Valentino}, {Vanzella}, {Wang}, {Wu},
  \& {Zitrin}}]{2021ApJ...908..146C}
{Caputi}, K.~I., {Caminha}, G.~B., {Fujimoto}, S., {et~al.} 2021, \apj, 908,
  146

\bibitem[{{Cash}(1979)}]{1979Cash}
{Cash}, W. 1979, \apj, 228, 939

\bibitem[{{Cerny} {et~al.}(2018){Cerny}, {Sharon}, {Andrade-Santos}, {Avila},
  {Brada{\v{c}}}, {Bradley}, {Carrasco}, {Coe}, {Czakon}, {Dawson}, {Frye},
  {Hoag}, {Huang}, {Johnson}, {Jones}, {Lam}, {Lovisari}, {Mainali}, {Oesch},
  {Ogaz}, {Past}, {Paterno-Mahler}, {Peterson}, {Riess}, {Rodney}, {Ryan},
  {Salmon}, {Sendra-Server}, {Stark}, {Strolger}, {Trenti}, {Umetsu},
  {Vulcani}, \& {Zitrin}}]{2018ApJ...859..159C}
{Cerny}, C., {Sharon}, K., {Andrade-Santos}, F., {et~al.} 2018, \apj, 859, 159

\bibitem[{{Chiriv{\`\i}} {et~al.}(2018){Chiriv{\`\i}}, {Suyu}, {Grillo},
  {Halkola}, {Balestra}, {Caminha}, {Mercurio}, \&
  {Rosati}}]{2018A&A...614A...8C}
{Chiriv{\`\i}}, G., {Suyu}, S.~H., {Grillo}, C., {et~al.} 2018, \aap, 614, A8

\bibitem[{{Clowe} {et~al.}(2004){Clowe}, {Gonzalez}, \&
  {Markevitch}}]{2004ApJ...604..596C}
{Clowe}, D., {Gonzalez}, A., \& {Markevitch}, M. 2004, \apj, 604, 596

\bibitem[{{Coe} {et~al.}(2019){Coe}, {Salmon}, {Brada{\v{c}}}, {Bradley},
  {Sharon}, {Zitrin}, {Acebron}, {Cerny}, {Cibirka}, {Strait},
  {Paterno-Mahler}, {Mahler}, {Avila}, {Ogaz}, {Huang}, {Pelliccia}, {Stark},
  {Mainali}, {Oesch}, {Trenti}, {Carrasco}, {Dawson}, {Rodney}, {Strolger},
  {Riess}, {Jones}, {Frye}, {Czakon}, {Umetsu}, {Vulcani}, {Graur}, {Jha},
  {Graham}, {Molino}, {Nonino}, {Hjorth}, {Selsing}, {Christensen},
  {Kikuchihara}, {Ouchi}, {Oguri}, {Welch}, {Lemaux}, {Andrade-Santos}, {Hoag},
  {Johnson}, {Peterson}, {Past}, {Fox}, {Agulli}, {Livermore}, {Ryan}, {Lam},
  {Sendra-Server}, {Toft}, {Lovisari}, \& {Su}}]{2019ApJ...884...85C}
{Coe}, D., {Salmon}, B., {Brada{\v{c}}}, M., {et~al.} 2019, \apj, 884, 85

\bibitem[{{Coe} {et~al.}(2012){Coe}, {Umetsu}, {Zitrin}, {Donahue},
  {Medezinski}, {Postman}, {Carrasco}, {Anguita}, {Geller}, {Rines},
  {Diaferio}, {Kurtz}, {Bradley}, {Koekemoer}, {Zheng}, {Nonino}, {Molino},
  {Mahdavi}, {Lemze}, {Infante}, {Ogaz}, {Melchior}, {Host}, {Ford}, {Grillo},
  {Rosati}, {Jim{\'e}nez-Teja}, {Moustakas}, {Broadhurst}, {Ascaso}, {Lahav},
  {Bartelmann}, {Ben{\'\i}tez}, {Bouwens}, {Graur}, {Graves}, {Jha}, {Jouvel},
  {Kelson}, {Moustakas}, {Maoz}, {Meneghetti}, {Merten}, {Riess}, {Rodney}, \&
  {Seitz}}]{2012ApJ...757...22C}
{Coe}, D., {Umetsu}, K., {Zitrin}, A., {et~al.} 2012, \apj, 757, 22

\bibitem[{{Diego} {et~al.}(2020){Diego}, {Molnar}, {Cerny}, {Broadhurst},
  {Windhorst}, {Zitrin}, {Bouwens}, {Coe}, {Conselice}, \&
  {Sharon}}]{2020ApJ...904..106D}
{Diego}, J.~M., {Molnar}, S.~M., {Cerny}, C., {et~al.} 2020, \apj, 904, 106

\bibitem[{{D'Onghia} {et~al.}(2005){D'Onghia}, {Sommer-Larsen}, {Romeo},
  {Burkert}, {Pedersen}, {Portinari}, \& {Rasmussen}}]{2005ApJ...630L.109D}
{D'Onghia}, E., {Sommer-Larsen}, J., {Romeo}, A.~D., {et~al.} 2005, \apjl, 630,
  L109

\bibitem[{{El{\'\i}asd{\'o}ttir} {et~al.}(2007){El{\'\i}asd{\'o}ttir},
  {Limousin}, {Richard}, {Hjorth}, {Kneib}, {Natarajan}, {Pedersen}, {Jullo},
  \& {Paraficz}}]{2007arXiv0710.5636E}
{El{\'\i}asd{\'o}ttir}, {\'A}., {Limousin}, M., {Richard}, J., {et~al.} 2007,
  arXiv e-prints, arXiv:0710.5636

\bibitem[{{Girardi} {et~al.}(2015){Girardi}, {Mercurio}, {Balestra}, {Nonino},
  {Biviano}, {Grillo}, {Rosati}, {Annunziatella}, {Demarco}, {Fritz}, {Gobat},
  {Lemze}, {Presotto}, {Scodeggio}, {Tozzi}, {Bartosch Caminha}, {Brescia},
  {Coe}, {Kelson}, {Koekemoer}, {Lombardi}, {Medezinski}, {Postman},
  {Sartoris}, {Umetsu}, {Zitrin}, {Boschin}, {Czoske}, {De Lucia}, {Kuchner},
  {Maier}, {Meneghetti}, {Monaco}, {Monna}, {Munari}, {Seitz}, {Verdugo}, \&
  {Ziegler}}]{2015A&A...579A...4G}
{Girardi}, M., {Mercurio}, A., {Balestra}, I., {et~al.} 2015, \aap, 579, A4

\bibitem[{{Gladders} \& {Yee}(2000)}]{2000AJ....120.2148G}
{Gladders}, M.~D. \& {Yee}, H.~K.~C. 2000, \aj, 120, 2148

\bibitem[{{Grillo}(2012)}]{2012ApJ...747L..15G}
{Grillo}, C. 2012, \apjl, 747, L15

\bibitem[{{Grillo} {et~al.}(2016){Grillo}, {Karman}, {Suyu}, {Rosati},
  {Balestra}, {Mercurio}, {Lombardi}, {Treu}, {Caminha}, {Halkola}, {Rodney},
  {Gavazzi}, \& {Caputi}}]{2016ApJ...822...78G}
{Grillo}, C., {Karman}, W., {Suyu}, S.~H., {et~al.} 2016, \apj, 822, 78

\bibitem[{{Grillo} {et~al.}(2015){Grillo}, {Suyu}, {Rosati}, {Mercurio},
  {Balestra}, {Munari}, {Nonino}, {Caminha}, {Lombardi}, {De Lucia}, {Borgani},
  {Gobat}, {Biviano}, {Girardi}, {Umetsu}, {Coe}, {Koekemoer}, {Postman},
  {Zitrin}, {Halkola}, {Broadhurst}, {Sartoris}, {Presotto}, {Annunziatella},
  {Maier}, {Fritz}, {Vanzella}, \& {Frye}}]{2015ApJ...800...38G}
{Grillo}, C., {Suyu}, S.~H., {Rosati}, P., {et~al.} 2015, \apj, 800, 38

\bibitem[{Harris {et~al.}(2020)Harris, Millman, van~der Walt, Gommers,
  Virtanen, Cournapeau, Wieser, Taylor, Berg, Smith, Kern, Picus, Hoyer, van
  Kerkwijk, Brett, Haldane, del R{\'{i}}o, Wiebe, Peterson,
  G{\'{e}}rard-Marchant, Sheppard, Reddy, Weckesser, Abbasi, Gohlke, \&
  Oliphant}]{harris2020array}
Harris, C.~R., Millman, K.~J., van~der Walt, S.~J., {et~al.} 2020, Nature, 585,
  357

\bibitem[{{HI4PI Collaboration} {et~al.}(2016){HI4PI Collaboration}, {Ben
  Bekhti}, {Fl{\"o}er}, {Keller}, {Kerp}, {Lenz}, {Winkel}, {Bailin},
  {Calabretta}, {Dedes}, {Ford}, {Gibson}, {Haud}, {Janowiecki}, {Kalberla},
  {Lockman}, {McClure-Griffiths}, {Murphy}, {Nakanishi}, {Pisano}, \&
  {Staveley-Smith}}]{2016HI4PI}
{HI4PI Collaboration}, {Ben Bekhti}, N., {Fl{\"o}er}, L., {et~al.} 2016, \aap,
  594, A116

\bibitem[{{Host}(2012)}]{2012MNRAS.420L..18H}
{Host}, O. 2012, \mnras, 420, L18

\bibitem[{{Hudson} {et~al.}(2010){Hudson}, {Mittal}, {Reiprich}, {Nulsen},
  {Andernach}, \& {Sarazin}}]{2010A&A...513A..37H}
{Hudson}, D.~S., {Mittal}, R., {Reiprich}, T.~H., {et~al.} 2010, \aap, 513, A37

\bibitem[{Hunter(2007)}]{Hunter:2007}
Hunter, J.~D. 2007, Computing in Science \& Engineering, 9, 90

\bibitem[{{Inami} {et~al.}(2017){Inami}, {Bacon}, {Brinchmann}, {Richard},
  {Contini}, {Conseil}, {Hamer}, {Akhlaghi}, {Bouch{\'e}}, {Cl{\'e}ment},
  {Desprez}, {Drake}, {Hashimoto}, {Leclercq}, {Maseda}, {Michel-Dansac},
  {Paalvast}, {Tresse}, {Ventou}, {Kollatschny}, {Boogaard}, {Finley},
  {Marino}, {Schaye}, \& {Wisotzki}}]{2017A&A...608A...2I}
{Inami}, H., {Bacon}, R., {Brinchmann}, J., {et~al.} 2017, \aap, 608, A2

\bibitem[{{Jee} {et~al.}(2014){Jee}, {Hughes}, {Menanteau}, {Sif{\'o}n},
  {Mandelbaum}, {Barrientos}, {Infante}, \& {Ng}}]{2014ApJ...785...20J}
{Jee}, M.~J., {Hughes}, J.~P., {Menanteau}, F., {et~al.} 2014, \apj, 785, 20

\bibitem[{{Johnson} \& {Sharon}(2016)}]{2016ApJ...832...82J}
{Johnson}, T.~L. \& {Sharon}, K. 2016, \apj, 832, 82

\bibitem[{{Jullo} \& {Kneib}(2009)}]{2009MNRAS.395.1319J}
{Jullo}, E. \& {Kneib}, J.~P. 2009, \mnras, 395, 1319

\bibitem[{{Jullo} {et~al.}(2007){Jullo}, {Kneib}, {Limousin},
  {El{\'\i}asd{\'o}ttir}, {Marshall}, \& {Verdugo}}]{2007NJPh....9..447J}
{Jullo}, E., {Kneib}, J.~P., {Limousin}, M., {et~al.} 2007, New Journal of
  Physics, 9, 447

\bibitem[{{Karman} {et~al.}(2017){Karman}, {Caputi}, {Caminha}, {Gronke},
  {Grillo}, {Balestra}, {Rosati}, {Vanzella}, {Coe}, {Dijkstra}, {Koekemoer},
  {McLeod}, {Mercurio}, \& {Nonino}}]{2017A&A...599A..28K}
{Karman}, W., {Caputi}, K.~I., {Caminha}, G.~B., {et~al.} 2017, \aap, 599, A28

\bibitem[{{Kassiola} \& {Kovner}(1993)}]{1993ApJ...417..450K}
{Kassiola}, A. \& {Kovner}, I. 1993, \apj, 417, 450

\bibitem[{{Kim} {et~al.}(2021){Kim}, {Jee}, {Hughes}, {Yoon}, {HyeongHan},
  {Menanteau}, {Sif{\'o}n}, {Hovey}, \& {Arunachalam}}]{2021arXiv210600031K}
{Kim}, J., {Jee}, M.~J., {Hughes}, J.~P., {et~al.} 2021, \apj, 923, 101

\bibitem[{{Kneib} {et~al.}(1996){Kneib}, {Ellis}, {Smail}, {Couch}, \&
  {Sharples}}]{1996ApJ...471..643K}
{Kneib}, J.~P., {Ellis}, R.~S., {Smail}, I., {Couch}, W.~J., \& {Sharples},
  R.~M. 1996, \apj, 471, 643

\bibitem[{{Kneib} \& {Natarajan}(2011)}]{2011A&ARv..19...47K}
{Kneib}, J.-P. \& {Natarajan}, P. 2011, \aapr, 19, 47

\bibitem[{{Kurk} {et~al.}(2013){Kurk}, {Cimatti}, {Daddi}, {Mignoli},
  {Pozzetti}, {Dickinson}, {Bolzonella}, {Zamorani}, {Cassata}, {Rodighiero},
  {Franceschini}, {Renzini}, {Rosati}, {Halliday}, \&
  {Berta}}]{2013A&A...549A..63K}
{Kurk}, J., {Cimatti}, A., {Daddi}, E., {et~al.} 2013, \aap, 549, A63

\bibitem[{{Lagattuta} {et~al.}(2019){Lagattuta}, {Richard}, {Bauer},
  {Cl{\'e}ment}, {Mahler}, {Soucail}, {Carton}, {Kneib}, {Laporte}, {Martinez},
  {Patr{\'\i}cio}, {Payne}, {Pell{\'o}}, {Schmidt}, \& {de la
  Vieuville}}]{2019MNRAS.485.3738L}
{Lagattuta}, D.~J., {Richard}, J., {Bauer}, F.~E., {et~al.} 2019, \mnras, 485,
  3738

\bibitem[{{Lagattuta} {et~al.}(2017){Lagattuta}, {Richard}, {Cl{\'e}ment},
  {Mahler}, {Patr{\'\i}cio}, {Pell{\'o}}, {Soucail}, {Schmidt}, {Wisotzki},
  {Martinez}, \& {Bina}}]{2017MNRAS.469.3946L}
{Lagattuta}, D.~J., {Richard}, J., {Cl{\'e}ment}, B., {et~al.} 2017, \mnras,
  469, 3946

\bibitem[{{Le F{\`e}vre} {et~al.}(2013){Le F{\`e}vre}, {Cassata}, {Cucciati},
  {Garilli}, {Ilbert}, {Le Brun}, {Maccagni}, {Moreau}, {Scodeggio}, {Tresse},
  {Zamorani}, {Adami}, {Arnouts}, {Bardelli}, {Bolzonella}, {Bondi},
  {Bongiorno}, {Bottini}, {Cappi}, {Charlot}, {Ciliegi}, {Contini}, {de la
  Torre}, {Foucaud}, {Franzetti}, {Gavignaud}, {Guzzo}, {Iovino}, {Lemaux},
  {L{\'o}pez-Sanjuan}, {McCracken}, {Marano}, {Marinoni}, {Mazure}, {Mellier},
  {Merighi}, {Merluzzi}, {Paltani}, {Pell{\`o}}, {Pollo}, {Pozzetti},
  {Scaramella}, {Tasca}, {Vergani}, {Vettolani}, {Zanichelli}, \&
  {Zucca}}]{2013A&A...559A..14L}
{Le F{\`e}vre}, O., {Cassata}, P., {Cucciati}, O., {et~al.} 2013, \aap, 559,
  A14

\bibitem[{{Lilly} {et~al.}(2007){Lilly}, {Le F{\`e}vre}, {Renzini}, {Zamorani},
  {Scodeggio}, {Contini}, {Carollo}, {Hasinger}, {Kneib}, {Iovino}, {Le Brun},
  {Maier}, {Mainieri}, {Mignoli}, {Silverman}, {Tasca}, {Bolzonella},
  {Bongiorno}, {Bottini}, {Capak}, {Caputi}, {Cimatti}, {Cucciati}, {Daddi},
  {Feldmann}, {Franzetti}, {Garilli}, {Guzzo}, {Ilbert}, {Kampczyk}, {Kovac},
  {Lamareille}, {Leauthaud}, {Le Borgne}, {McCracken}, {Marinoni}, {Pello},
  {Ricciardelli}, {Scarlata}, {Vergani}, {Sanders}, {Schinnerer}, {Scoville},
  {Taniguchi}, {Arnouts}, {Aussel}, {Bardelli}, {Brusa}, {Cappi}, {Ciliegi},
  {Finoguenov}, {Foucaud}, {Franceschini}, {Halliday}, {Impey}, {Knobel},
  {Koekemoer}, {Kurk}, {Maccagni}, {Maddox}, {Marano}, {Marconi}, {Meneux},
  {Mobasher}, {Moreau}, {Peacock}, {Porciani}, {Pozzetti}, {Scaramella},
  {Schiminovich}, {Shopbell}, {Smail}, {Thompson}, {Tresse}, {Vettolani},
  {Zanichelli}, \& {Zucca}}]{2007ApJS..172...70L}
{Lilly}, S.~J., {Le F{\`e}vre}, O., {Renzini}, A., {et~al.} 2007, \apjs, 172,
  70

\bibitem[{{Limousin} {et~al.}(2022){Limousin}, {Beauchesne}, \&
  {Jullo}}]{2022arXiv220202992L}
{Limousin}, M., {Beauchesne}, B., \& {Jullo}, E. 2022, \aap, 664, A90

\bibitem[{{Lindner} {et~al.}(2014){Lindner}, {Baker}, {Hughes}, {Battaglia},
  {Gupta}, {Knowles}, {Marriage}, {Menanteau}, {Moodley}, {Reese}, \&
  {Srianand}}]{2014ApJ...786...49L}
{Lindner}, R.~R., {Baker}, A.~J., {Hughes}, J.~P., {et~al.} 2014, \apj, 786, 49

\bibitem[{{Mahler} {et~al.}(2020){Mahler}, {Sharon}, {Gladders}, {Bleem},
  {Bayliss}, {Calzadilla}, {Floyd}, {Khullar}, {McDonald}, {Remolina
  Gonz{\'a}lez}, {Schrabback}, {Stark}, \& {van den
  Busch}}]{2020ApJ...894..150M}
{Mahler}, G., {Sharon}, K., {Gladders}, M.~D., {et~al.} 2020, \apj, 894, 150

\bibitem[{{Menanteau} {et~al.}(2010){Menanteau}, {Gonz{\'a}lez}, {Juin},
  {Marriage}, {Reese}, {Acquaviva}, {Aguirre}, {Appel}, {Baker}, {Barrientos},
  {Battistelli}, {Bond}, {Das}, {Deshpande}, {Devlin}, {Dicker}, {Dunkley},
  {D{\"u}nner}, {Essinger-Hileman}, {Fowler}, {Hajian}, {Halpern},
  {Hasselfield}, {Hern{\'a}ndez-Monteagudo}, {Hilton}, {Hincks}, {Hlozek},
  {Huffenberger}, {Hughes}, {Infante}, {Irwin}, {Klein}, {Kosowsky}, {Lin},
  {Marsden}, {Moodley}, {Niemack}, {Nolta}, {Page}, {Parker}, {Partridge},
  {Sehgal}, {Sievers}, {Spergel}, {Staggs}, {Swetz}, {Switzer}, {Thornton},
  {Trac}, {Warne}, \& {Wollack}}]{2010ApJ...723.1523M}
{Menanteau}, F., {Gonz{\'a}lez}, J., {Juin}, J.-B., {et~al.} 2010, \apj, 723,
  1523

\bibitem[{{Menanteau} {et~al.}(2012){Menanteau}, {Hughes}, {Sif{\'o}n},
  {Hilton}, {Gonz{\'a}lez}, {Infante}, {Barrientos}, {Baker}, {Bond}, {Das},
  {Devlin}, {Dunkley}, {Hajian}, {Hincks}, {Kosowsky}, {Marsden}, {Marriage},
  {Moodley}, {Niemack}, {Nolta}, {Page}, {Reese}, {Sehgal}, {Sievers},
  {Spergel}, {Staggs}, \& {Wollack}}]{2012ApJ...748....7M}
{Menanteau}, F., {Hughes}, J.~P., {Sif{\'o}n}, C., {et~al.} 2012, \apj, 748, 7

\bibitem[{{Meneghetti}(2021)}]{2021LNP...956.....M}
{Meneghetti}, M. 2021, {Introduction to Gravitational Lensing; With Python
  Examples}, Vol. 956

\bibitem[{{Meneghetti} {et~al.}(2020){Meneghetti}, {Davoli}, {Bergamini},
  {Rosati}, {Natarajan}, {Giocoli}, {Caminha}, {Metcalf}, {Rasia}, {Borgani},
  {Calura}, {Grillo}, {Mercurio}, \& {Vanzella}}]{Meneghetti_2020}
{Meneghetti}, M., {Davoli}, G., {Bergamini}, P., {et~al.} 2020, Science, 369,
  1347

\bibitem[{{Meneghetti} {et~al.}(2022){Meneghetti}, {Ragagnin}, {Borgani},
  {Calura}, {Despali}, {Giocoli}, {Granato}, {Grillo}, {Moscardini}, {Rasia},
  {Rosati}, {Angora}, {Bassini}, {Bergamini}, {Caminha}, {Granata}, {Mercurio},
  {Metcalf}, {Natarajan}, {Nonino}, {Venusta Pignataro}, {Ragone-Figueroa},
  {Vanzella}, {Acebron}, {Dolag}, {Murante}, {Taffoni}, {Tornatore},
  {Tortorelli}, \& {Valentini}}]{2022arXiv220409065M}
{Meneghetti}, M., {Ragagnin}, A., {Borgani}, S., {et~al.} 2022, arXiv e-prints,
  arXiv:2204.09065

\bibitem[{{Mercurio} {et~al.}(2021){Mercurio}, {Rosati}, {Biviano},
  {Annunziatella}, {Girardi}, {Sartoris}, {Nonino}, {Brescia}, {Riccio},
  {Grillo}, {Balestra}, {Caminha}, {De Lucia}, {Gobat}, {Seitz}, {Tozzi},
  {Scodeggio}, {Vanzella}, {Angora}, {Bergamini}, {Borgani}, {Demarco},
  {Meneghetti}, {Strazzullo}, {Tortorelli}, {Umetsu}, {Fritz}, {Gruen},
  {Kelson}, {Lombardi}, {Maier}, {Postman}, {Rodighiero}, \&
  {Ziegler}}]{2021A&A...656A.147M}
{Mercurio}, A., {Rosati}, P., {Biviano}, A., {et~al.} 2021, \aap, 656, A147

\bibitem[{{Navarro} {et~al.}(1996){Navarro}, {Frenk}, \&
  {White}}]{1996ApJ...462..563N}
{Navarro}, J.~F., {Frenk}, C.~S., \& {White}, S. D.~M. 1996, \apj, 462, 563

\bibitem[{{Oguri}(2010)}]{2010PASJ...62.1017O}
{Oguri}, M. 2010, \pasj, 62, 1017

\bibitem[{{Oguri} {et~al.}(2014){Oguri}, {Rusu}, \&
  {Falco}}]{2014MNRAS.439.2494O}
{Oguri}, M., {Rusu}, C.~E., \& {Falco}, E.~E. 2014, \mnras, 439, 2494

\bibitem[{{Okabe} {et~al.}(2020){Okabe}, {Oguri}, {Peirani}, {Suto}, {Dubois},
  {Pichon}, {Kitayama}, {Sasaki}, \& {Nishimichi}}]{2020MNRAS.496.2591O}
{Okabe}, T., {Oguri}, M., {Peirani}, S., {et~al.} 2020, \mnras, 496, 2591

\bibitem[{{Paterno-Mahler} {et~al.}(2018){Paterno-Mahler}, {Sharon}, {Coe},
  {Mahler}, {Cerny}, {Johnson}, {Schrabback}, {Andrade-Santos}, {Avila},
  {Brada{\v{c}}}, {Bradley}, {Carrasco}, {Czakon}, {Dawson}, {Frye}, {Hoag},
  {Huang}, {Jones}, {Lam}, {Livermore}, {Lovisari}, {Mainali}, {Oesch}, {Ogaz},
  {Past}, {Peterson}, {Ryan}, {Salmon}, {Sendra-Server}, {Stark}, {Umetsu},
  {Vulcani}, \& {Zitrin}}]{2018ApJ...863..154P}
{Paterno-Mahler}, R., {Sharon}, K., {Coe}, D., {et~al.} 2018, \apj, 863, 154

\bibitem[{{Ragagnin} {et~al.}(2022){Ragagnin}, {Meneghetti}, {Bassini},
  {Ragone-Figueroa}, {Granato}, {Despali}, {Giocoli}, {Granata}, {Moscardini},
  {Bergamini}, {Rasia}, {Valentini}, {Borgani}, {Calura}, {Dolag}, {Grillo},
  {Mercurio}, {Murante}, {Natarajan}, {Rosati}, {Taffoni}, {Tornatore}, \&
  {Tortorelli}}]{2022arXiv220409067R}
{Ragagnin}, A., {Meneghetti}, M., {Bassini}, L., {et~al.} 2022, arXiv e-prints,
  arXiv:2204.09067

\bibitem[{{Remolina Gonz{\'a}lez} {et~al.}(2021{\natexlab{a}}){Remolina
  Gonz{\'a}lez}, {Sharon}, {Li}, {Mahler}, {Bleem}, {Gladders}, \&
  {Niemiec}}]{2021ApJ...910..146R}
{Remolina Gonz{\'a}lez}, J.~D., {Sharon}, K., {Li}, N., {et~al.}
  2021{\natexlab{a}}, \apj, 910, 146

\bibitem[{{Remolina Gonz{\'a}lez} {et~al.}(2021{\natexlab{b}}){Remolina
  Gonz{\'a}lez}, {Sharon}, {Mahler}, {Fox}, {Garcia Diaz}, {Napier}, {Bleem},
  {Gladders}, {Li}, \& {Niemiec}}]{2021ApJ...920...98R}
{Remolina Gonz{\'a}lez}, J.~D., {Sharon}, K., {Mahler}, G., {et~al.}
  2021{\natexlab{b}}, \apj, 920, 98

\bibitem[{{Rosati} {et~al.}(2002){Rosati}, {Borgani}, \& {Norman}}]{Rosati_02}
{Rosati}, P., {Borgani}, S., \& {Norman}, C. 2002, \araa, 40, 539

\bibitem[{{Rossetti} {et~al.}(2016){Rossetti}, {Gastaldello}, {Ferioli},
  {Bersanelli}, {De Grandi}, {Eckert}, {Ghizzardi}, {Maino}, \&
  {Molendi}}]{2016MNRAS.457.4515R}
{Rossetti}, M., {Gastaldello}, F., {Ferioli}, G., {et~al.} 2016, \mnras, 457,
  4515

\bibitem[{Schwarz(1978)}]{schwarz1978}
Schwarz, G. 1978, Ann. Statist., 6, 461

\bibitem[{{Smith} {et~al.}(2001){Smith}, {Brickhouse}, {Liedahl}, \&
  {Raymond}}]{2001Smith}
{Smith}, R.~K., {Brickhouse}, N.~S., {Liedahl}, D.~A., \& {Raymond}, J.~C.
  2001, \apjl, 556, L91

\bibitem[{{Soto} {et~al.}(2016){Soto}, {Lilly}, {Bacon}, {Richard}, \&
  {Conseil}}]{2016MNRAS.458.3210S}
{Soto}, K.~T., {Lilly}, S.~J., {Bacon}, R., {Richard}, J., \& {Conseil}, S.
  2016, \mnras, 458, 3210

\bibitem[{{Suyu} \& {Halkola}(2010)}]{2010A&A...524A..94S}
{Suyu}, S.~H. \& {Halkola}, A. 2010, \aap, 524, A94

\bibitem[{{Umetsu} {et~al.}(2018){Umetsu}, {Sereno}, {Tam}, {Chiu}, {Fan},
  {Ettori}, {Gruen}, {Okumura}, {Medezinski}, {Donahue}, {Meneghetti}, {Frye},
  {Koekemoer}, {Broadhurst}, {Zitrin}, {Balestra}, {Ben{\'\i}tez}, {Higuchi},
  {Melchior}, {Mercurio}, {Merten}, {Molino}, {Nonino}, {Postman}, {Rosati},
  {Sayers}, \& {Seitz}}]{2018ApJ...860..104U}
{Umetsu}, K., {Sereno}, M., {Tam}, S.-I., {et~al.} 2018, \apj, 860, 104

\bibitem[{{Umetsu} {et~al.}(2016){Umetsu}, {Zitrin}, {Gruen}, {Merten},
  {Donahue}, \& {Postman}}]{2016ApJ...821..116U}
{Umetsu}, K., {Zitrin}, A., {Gruen}, D., {et~al.} 2016, \apj, 821, 116

\bibitem[{van~der Walt {et~al.}(2014)van~der Walt, {S}ch\"onberger,
  {Nunez-Iglesias}, {B}oulogne, {W}arner, {Y}ager, {G}ouillart, {Y}u, \& the
  scikit-image contributors}]{scikit-image}
van~der Walt, S., {S}ch\"onberger, J.~L., {Nunez-Iglesias}, J., {et~al.} 2014,
  PeerJ, 2, e453

\bibitem[{{Wang} \& {Steinhardt}(1998)}]{1998ApJ...508..483W}
{Wang}, L. \& {Steinhardt}, P.~J. 1998, \apj, 508, 483

\bibitem[{{Weilbacher} {et~al.}(2020){Weilbacher}, {Palsa}, {Streicher},
  {Bacon}, {Urrutia}, {Wisotzki}, {Conseil}, {Husemann}, {Jarno}, {Kelz},
  {P{\'e}contal-Rousset}, {Richard}, {Roth}, {Selman}, \&
  {Vernet}}]{2020A&A...641A..28W}
{Weilbacher}, P.~M., {Palsa}, R., {Streicher}, O., {et~al.} 2020, \aap, 641,
  A28

\bibitem[{{White} \& {Frenk}(1991)}]{1991ApJ...379...52W}
{White}, S. D.~M. \& {Frenk}, C.~S. 1991, \apj, 379, 52

\bibitem[{{Wright} {et~al.}(2010){Wright}, {Eisenhardt}, {Mainzer}, {Ressler},
  {Cutri}, {Jarrett}, {Kirkpatrick}, {Padgett}, {McMillan}, {Skrutskie},
  {Stanford}, {Cohen}, {Walker}, {Mather}, {Leisawitz}, {Gautier}, {McLean},
  {Benford}, {Lonsdale}, {Blain}, {Mendez}, {Irace}, {Duval}, {Liu}, {Royer},
  {Heinrichsen}, {Howard}, {Shannon}, {Kendall}, {Walsh}, {Larsen}, {Cardon},
  {Schick}, {Schwalm}, {Abid}, {Fabinsky}, {Naes}, \&
  {Tsai}}]{2010AJ....140.1868W}
{Wright}, E.~L., {Eisenhardt}, P. R.~M., {Mainzer}, A.~K., {et~al.} 2010, \aj,
  140, 1868

\bibitem[{{Zarattini} {et~al.}(2021){Zarattini}, {Biviano}, {Aguerri},
  {Girardi}, \& {D'Onghia}}]{2021A&A...655A.103Z}
{Zarattini}, S., {Biviano}, A., {Aguerri}, J.~A.~L., {Girardi}, M., \&
  {D'Onghia}, E. 2021, \aap, 655, A103

\bibitem[{{Zhang} {et~al.}(2015){Zhang}, {Yu}, \& {Lu}}]{2015ApJ...813..129Z}
{Zhang}, C., {Yu}, Q., \& {Lu}, Y. 2015, \apj, 813, 129

\bibitem[{{Zhang} {et~al.}(2018){Zhang}, {Yu}, \& {Lu}}]{2018ApJ...855...36Z}
{Zhang}, C., {Yu}, Q., \& {Lu}, Y. 2018, \apj, 855, 36

\bibitem[{{Zitrin} {et~al.}(2013){Zitrin}, {Menanteau}, {Hughes}, {Coe},
  {Barrientos}, {Infante}, \& {Mandelbaum}}]{2013ApJ...770L..15Z}
{Zitrin}, A., {Menanteau}, F., {Hughes}, J.~P., {et~al.} 2013, \apjl, 770, L15

\end{thebibliography}

\begin{appendix}
\section{Multiple image spectra}
\label{ap:multiple_image_spectra}

Figure \ref{fig:specs} shows the MUSE spectra of all the confirmed multiple images.
The coordinates and redshift values are listed in Table \ref{tab:multiple_images} and are also included in the redshift catalogue available in the electronic version of this manuscript.

\begin{figure*}
   \label{fig:specs}
Family 1

   \includegraphics[width = 0.666\columnwidth]{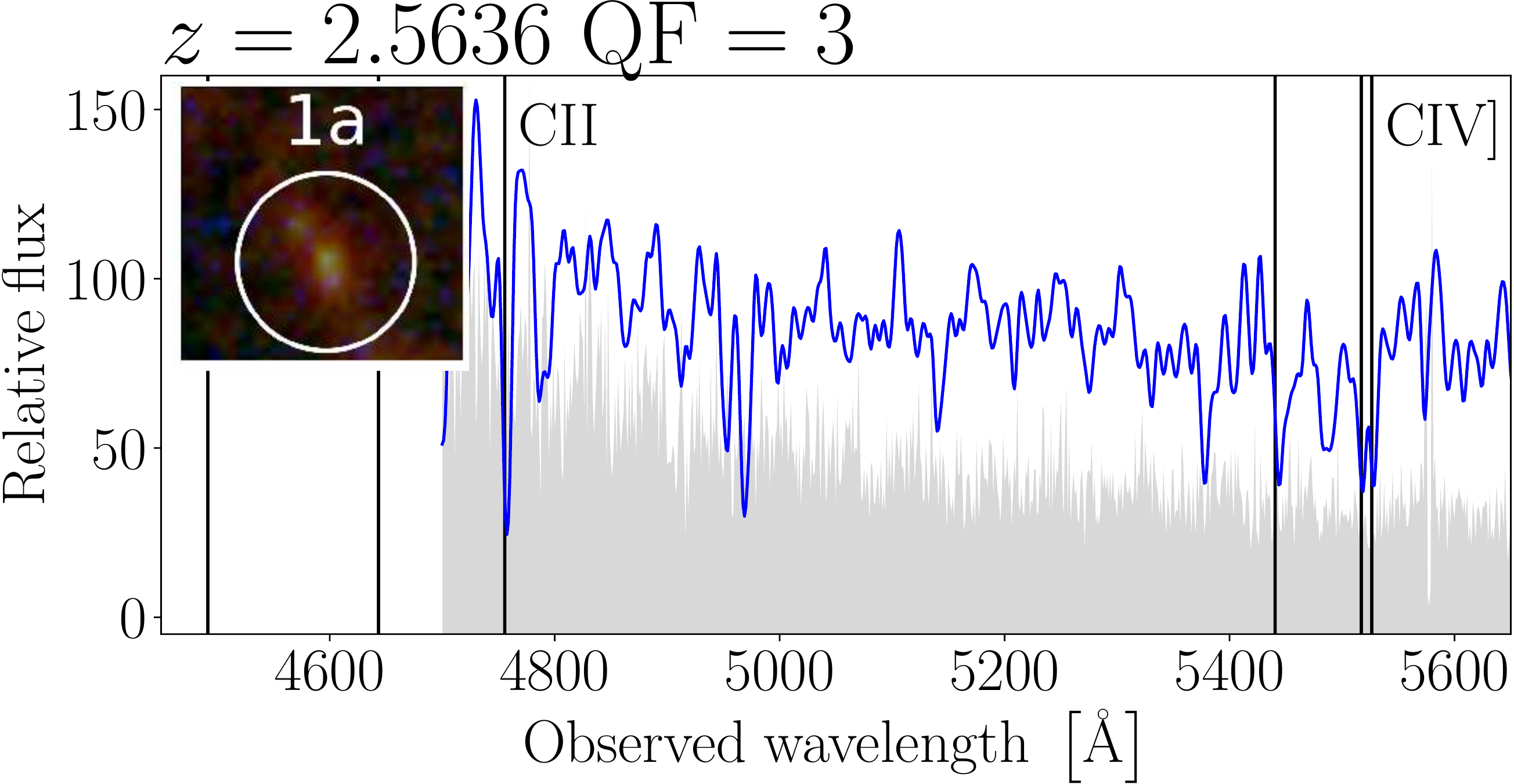}
   \includegraphics[width = 0.666\columnwidth]{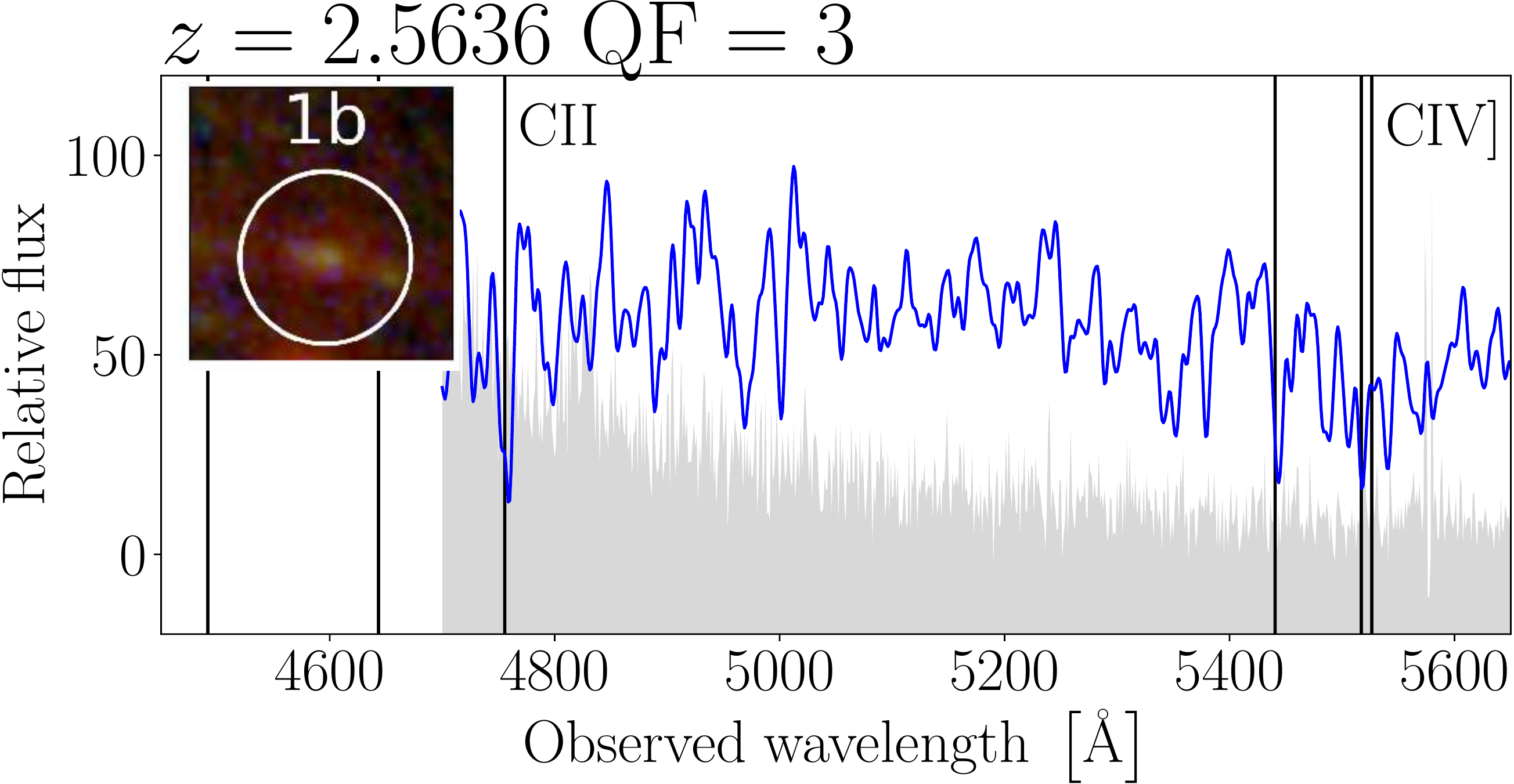}
   \includegraphics[width = 0.666\columnwidth]{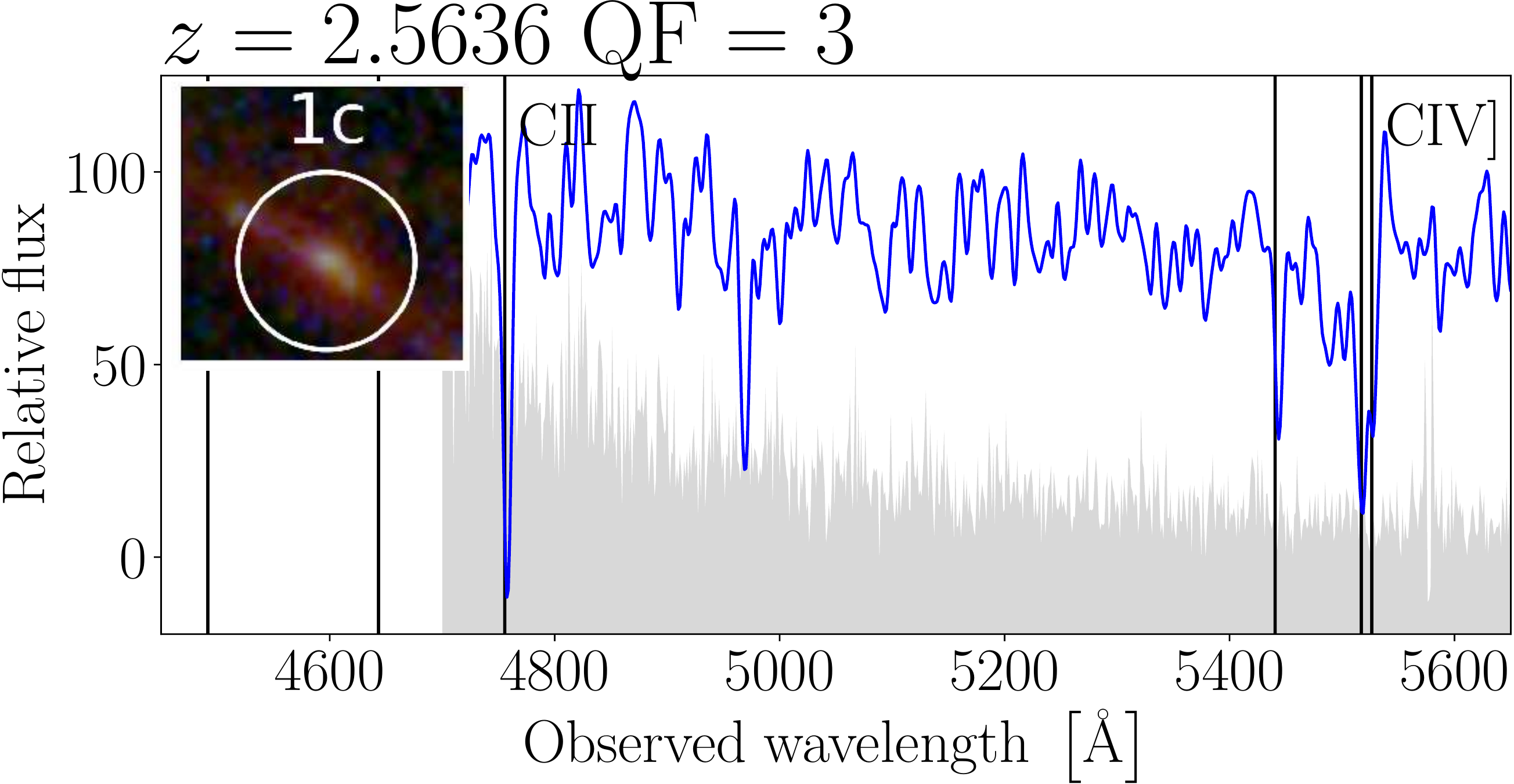}

Family 2
        
   \includegraphics[width = 0.666\columnwidth]{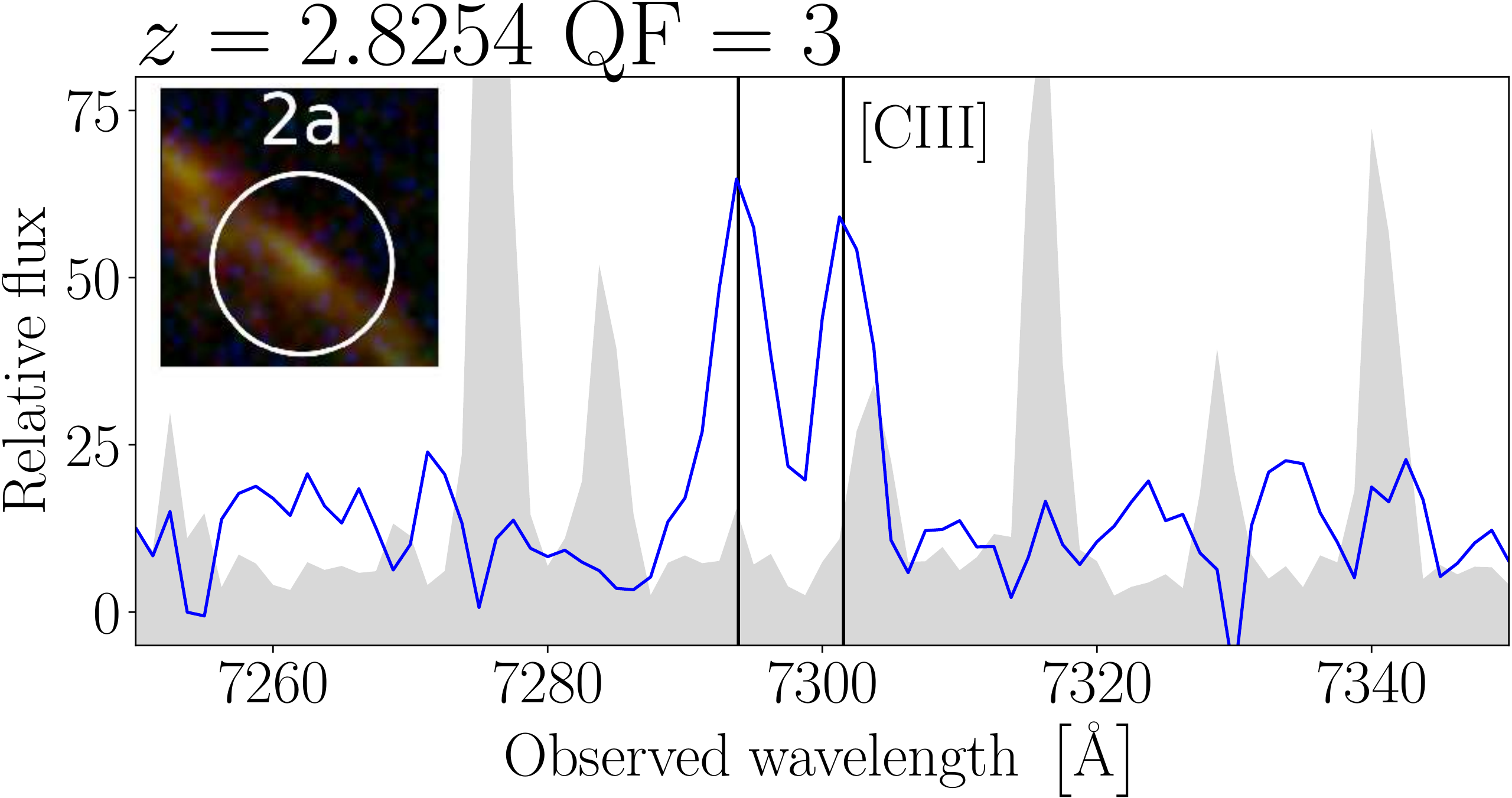}
   \includegraphics[width = 0.666\columnwidth]{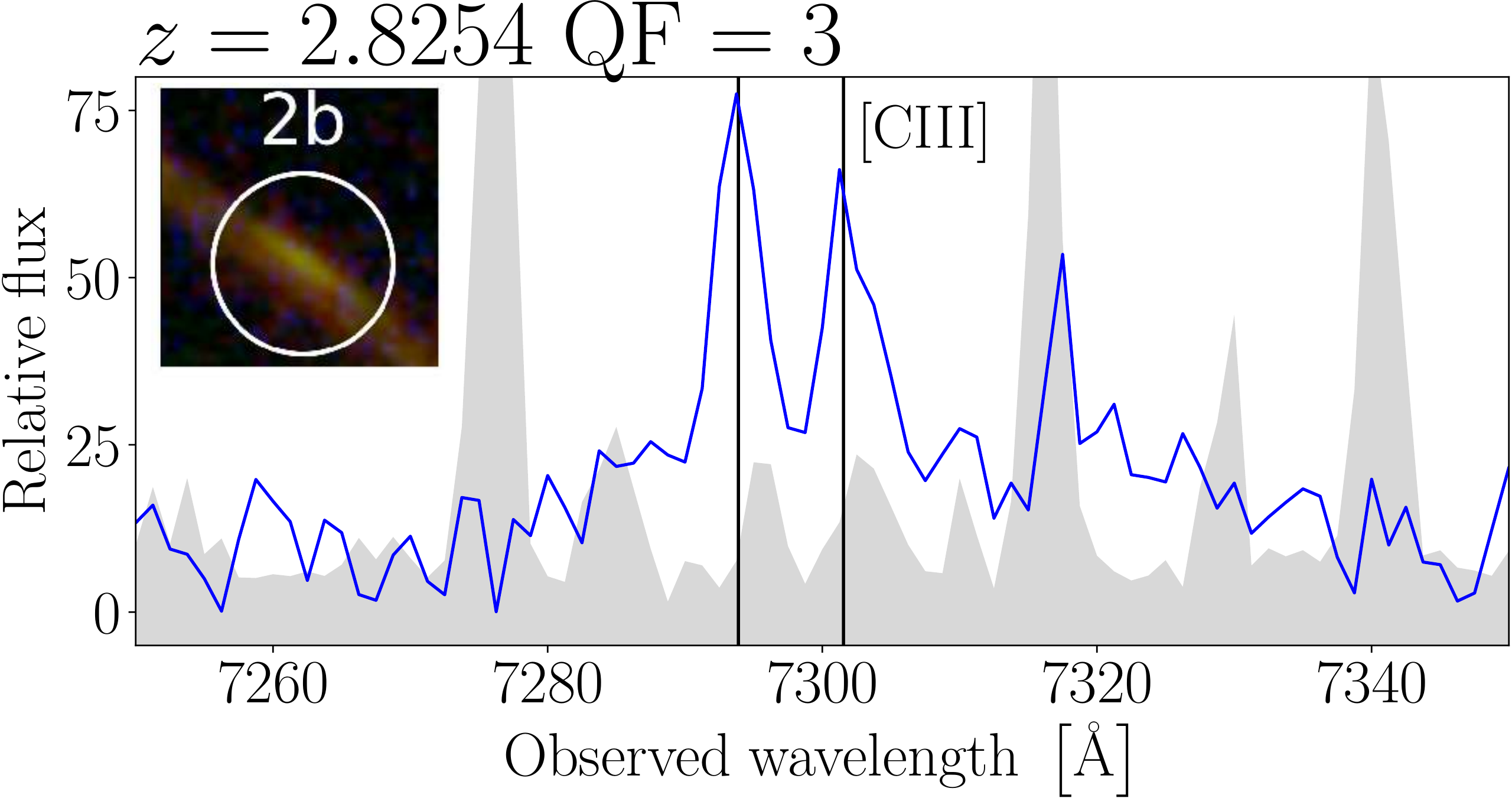}
   \includegraphics[width = 0.666\columnwidth]{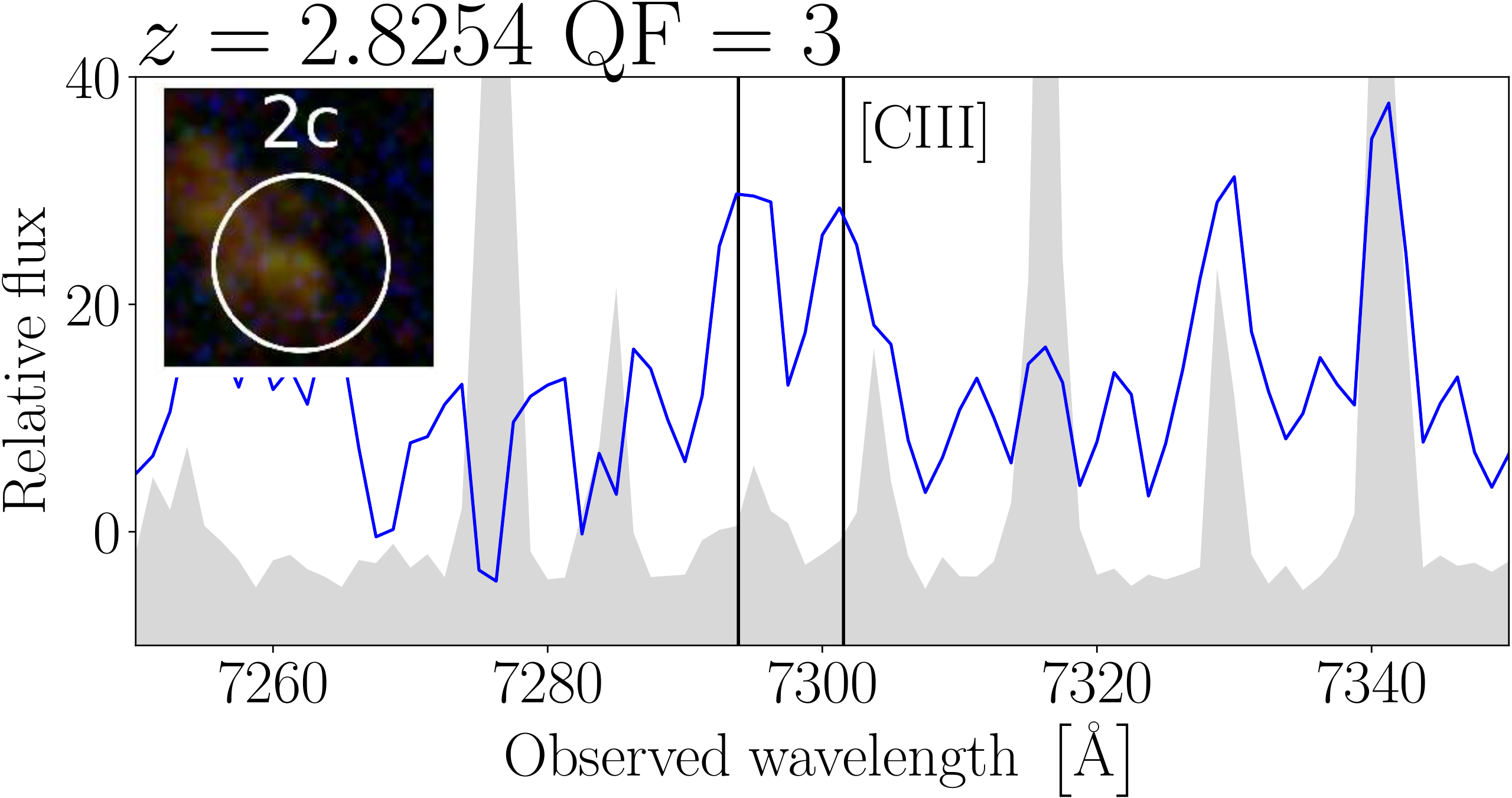}

Family 3

   \includegraphics[width = 0.666\columnwidth]{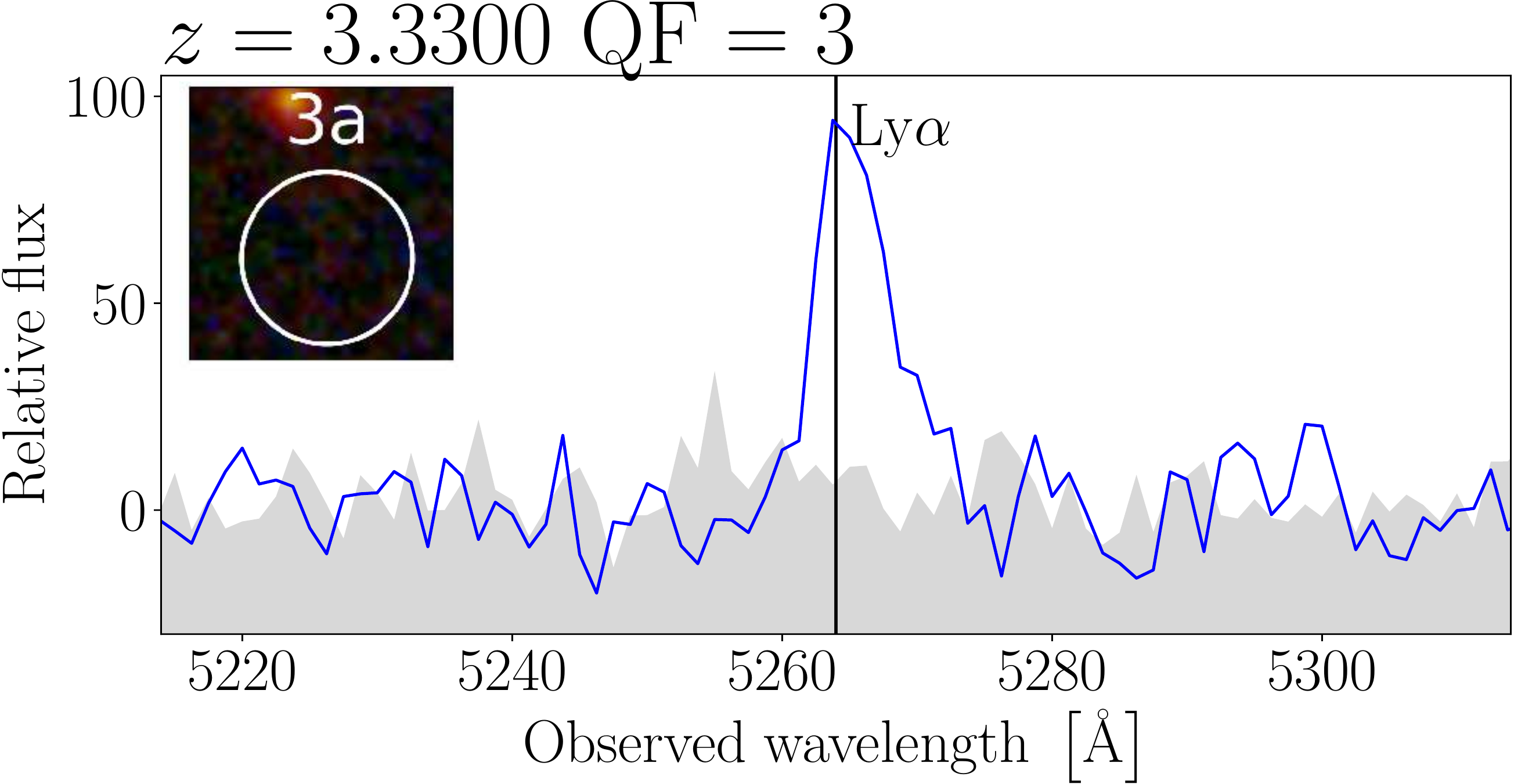}
   \includegraphics[width = 0.666\columnwidth]{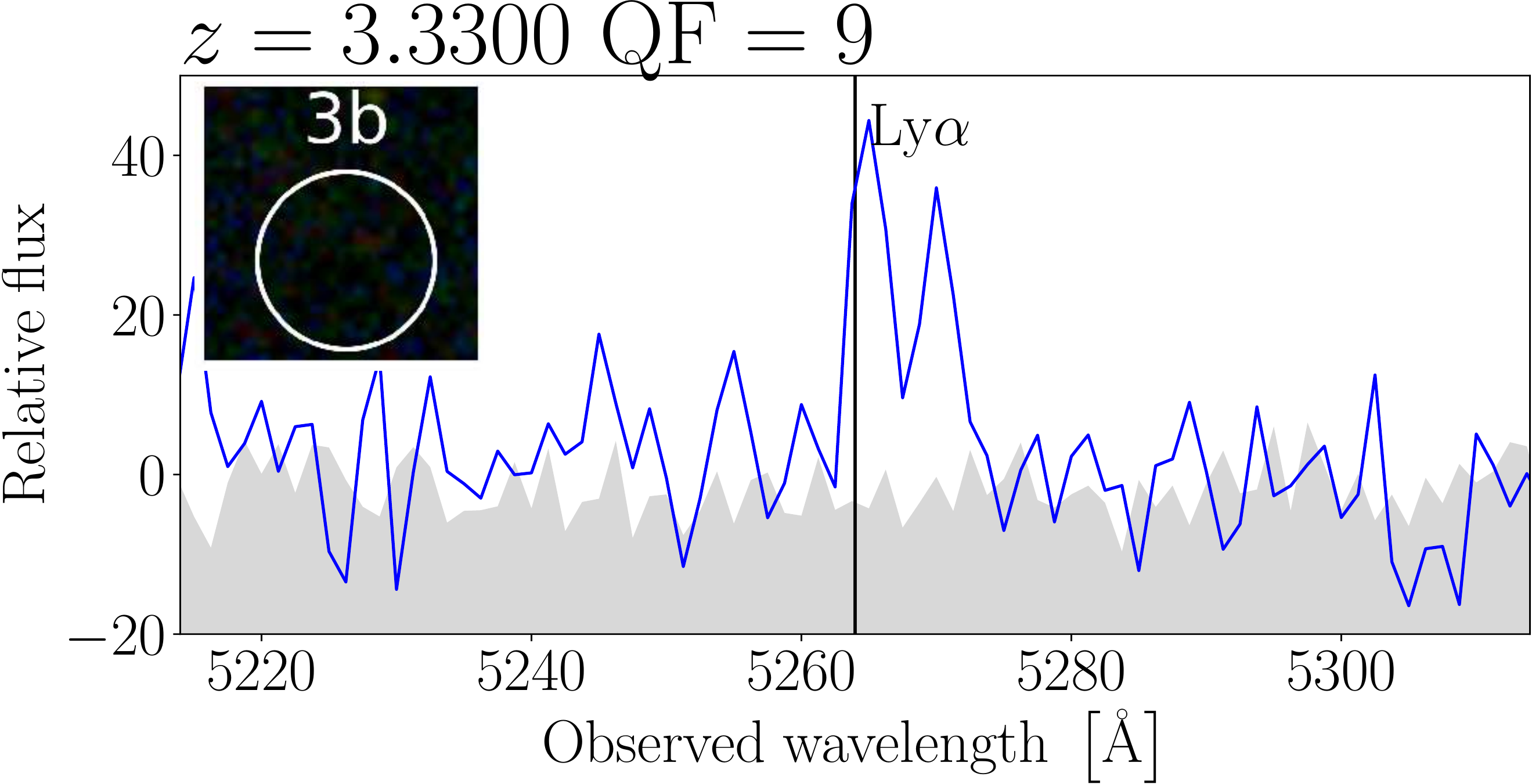}

Family 4

   \includegraphics[width = 0.666\columnwidth]{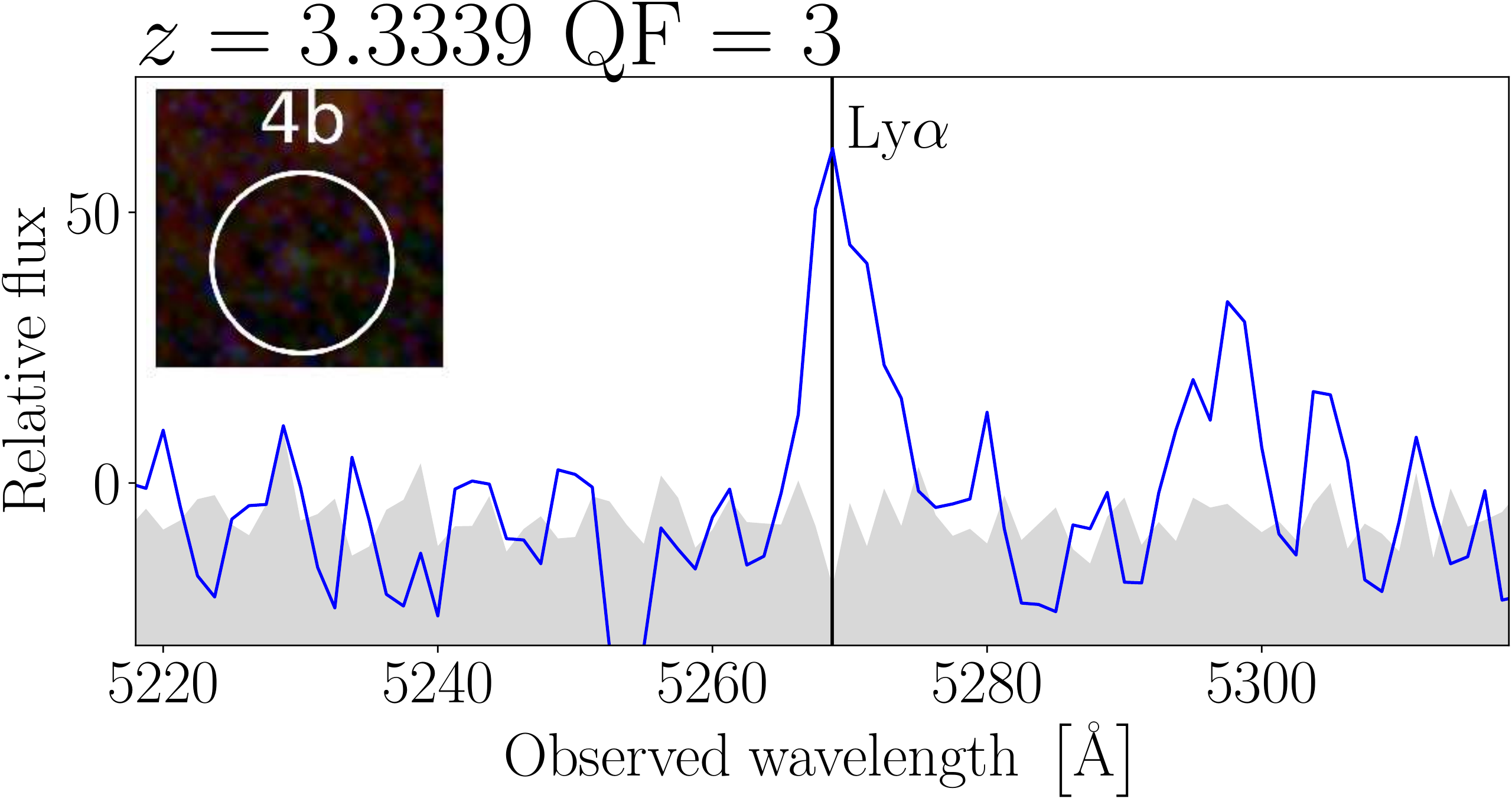}
   \includegraphics[width = 0.666\columnwidth]{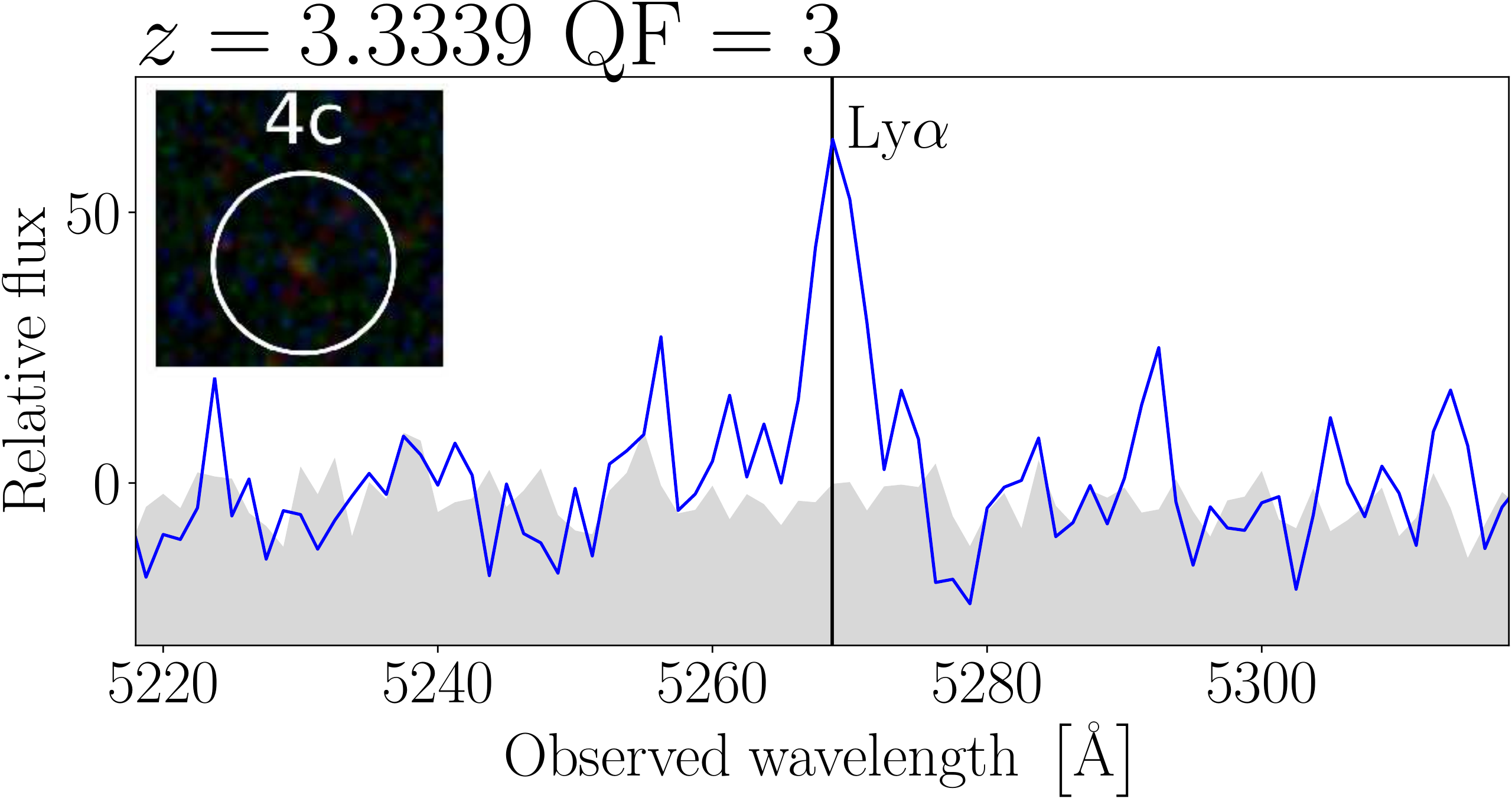}

Family 5
        
   \includegraphics[width = 0.666\columnwidth]{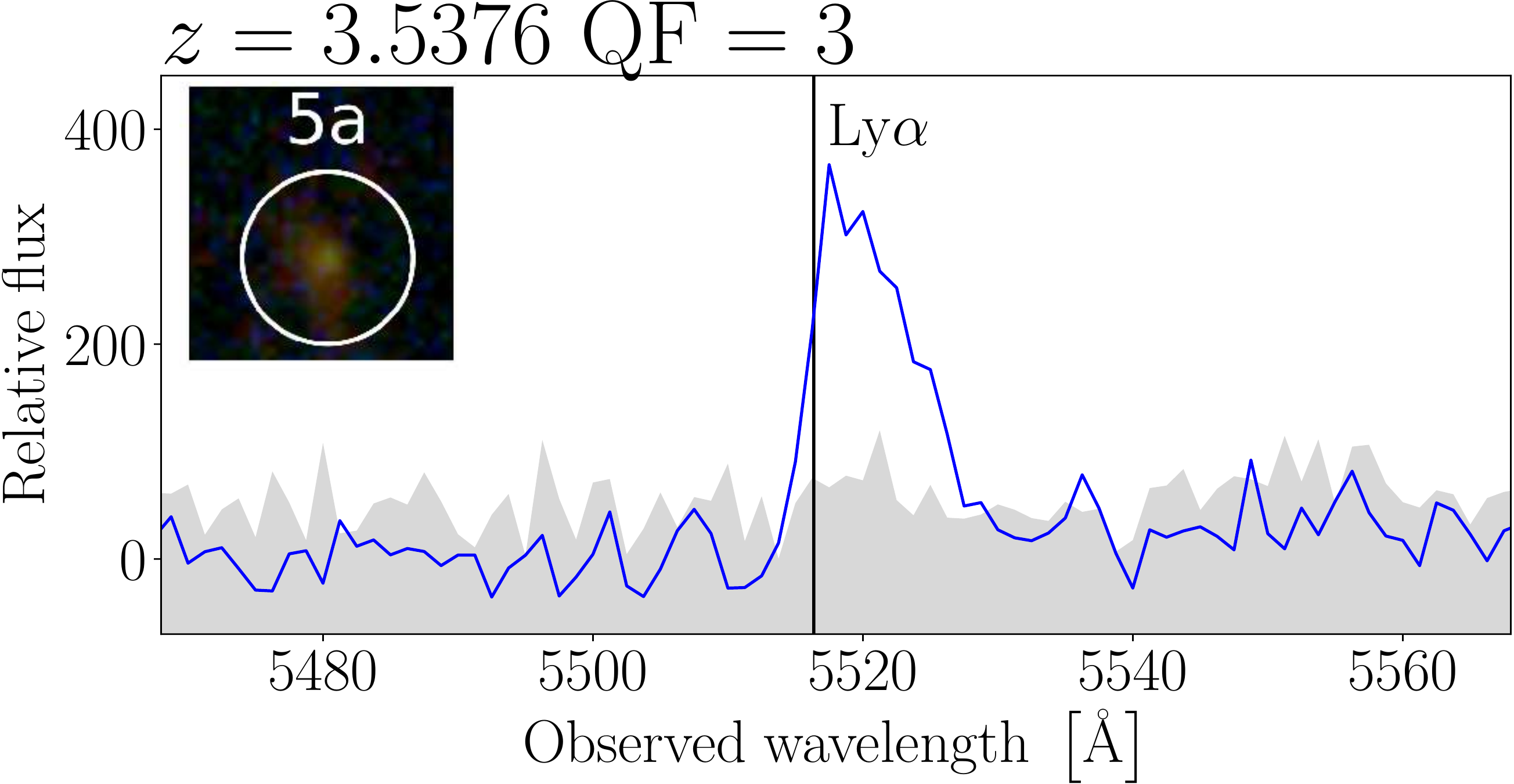}
   \includegraphics[width = 0.666\columnwidth]{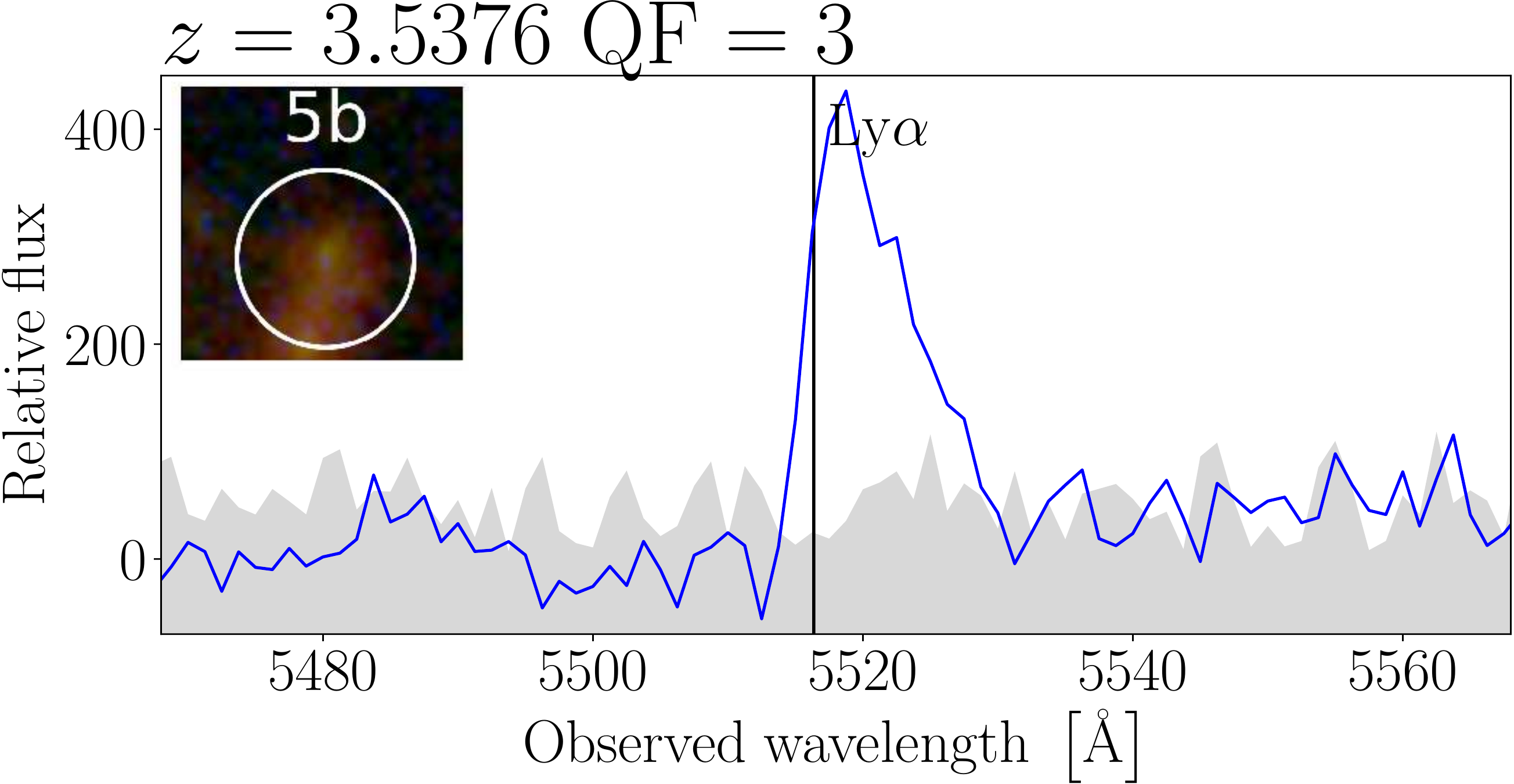}
   \includegraphics[width = 0.666\columnwidth]{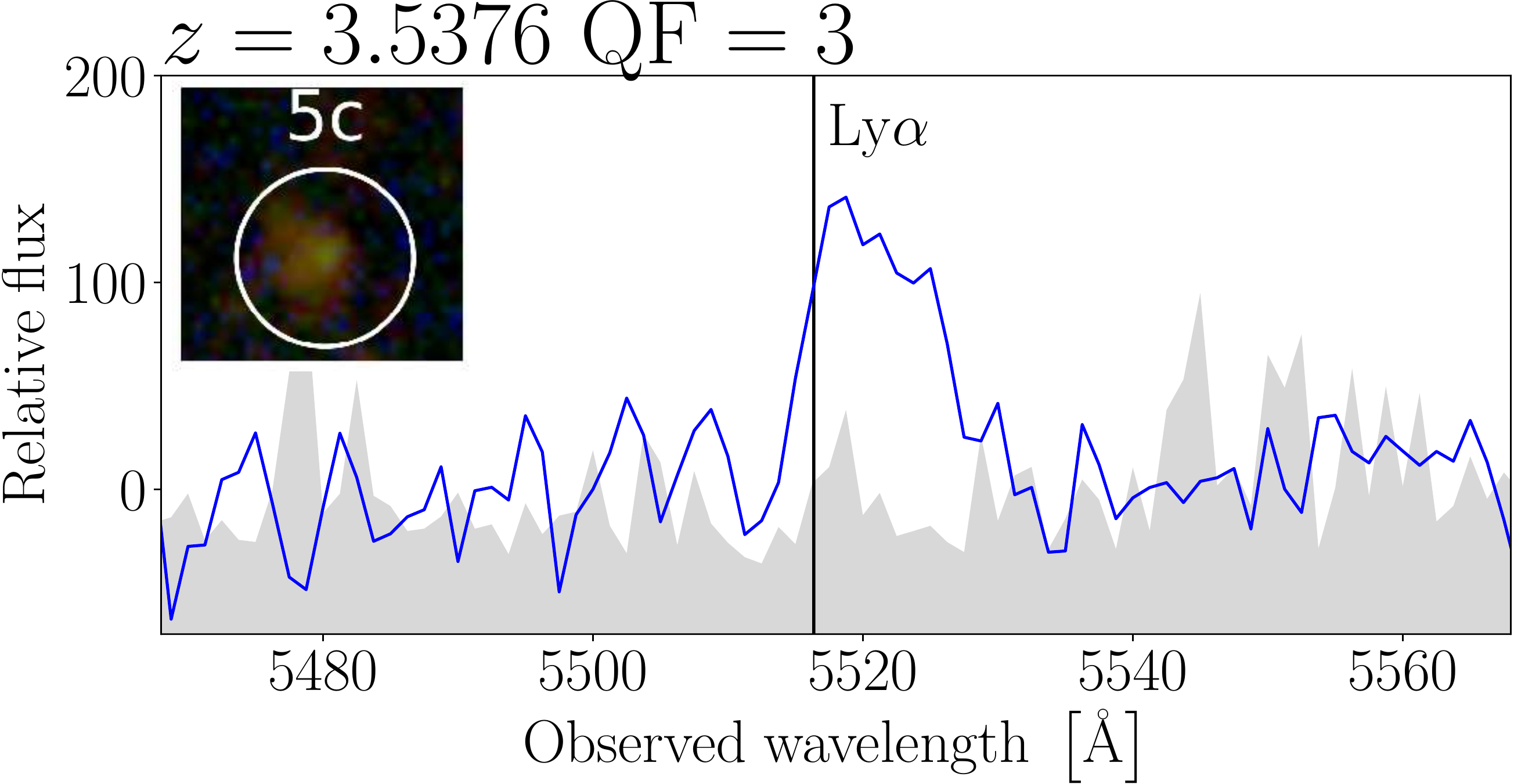}

Family 6
        
   \includegraphics[width = 0.666\columnwidth]{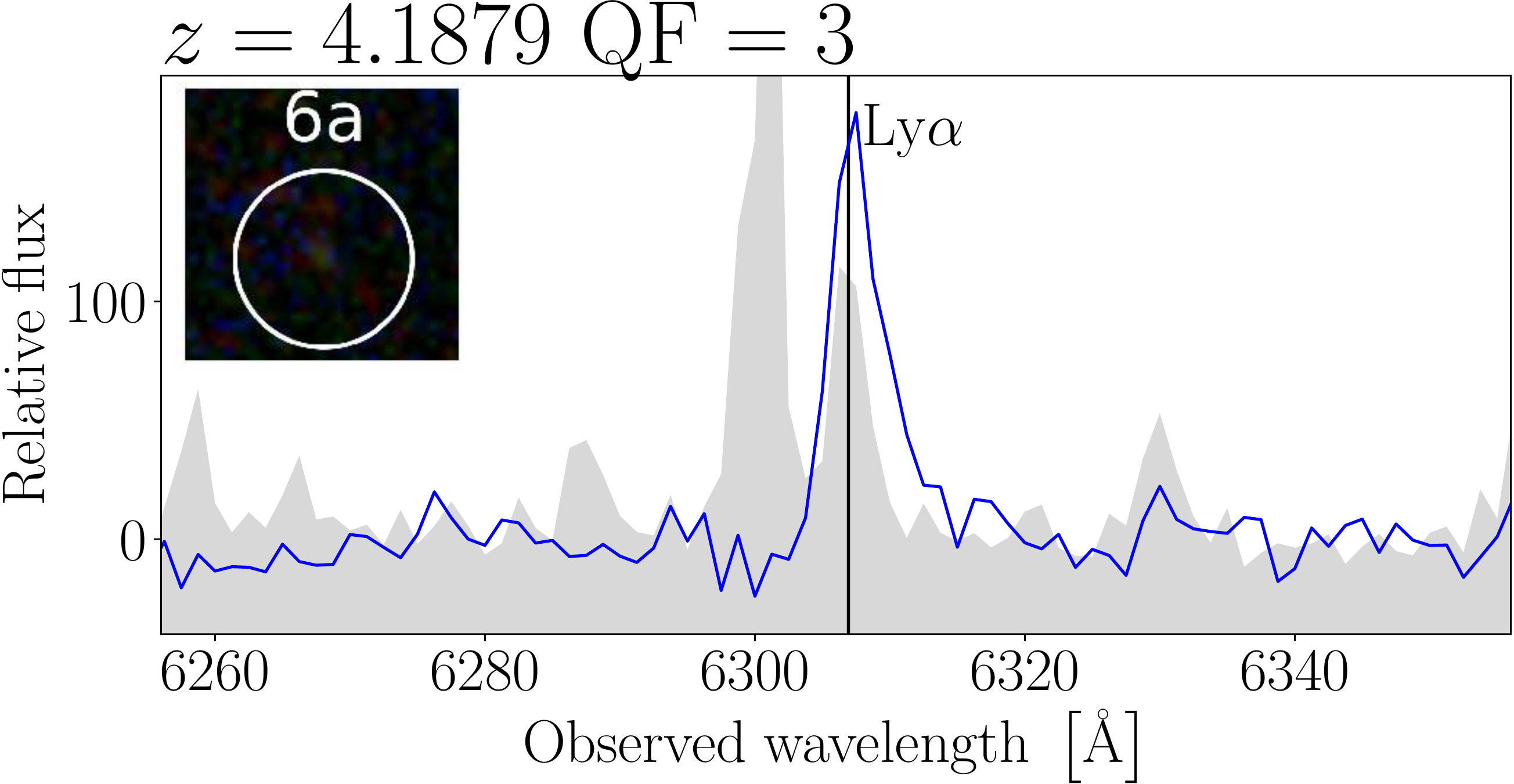}
   \includegraphics[width = 0.666\columnwidth]{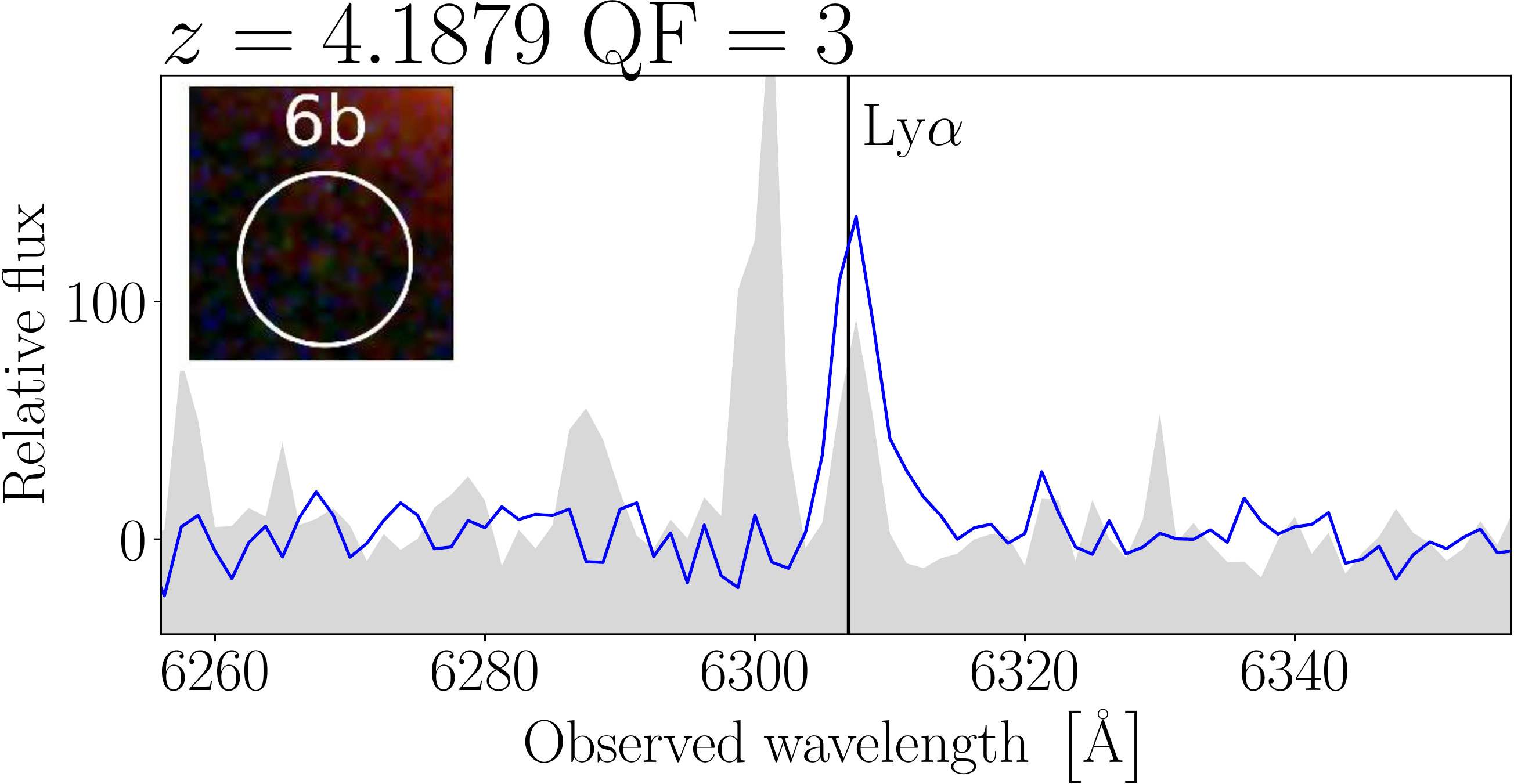}
   \caption{Multiple image spectra of all the confirmed multiple images. Vertical lines indicate spectral features at the source redshift, and the grey curves the scaled data variance. The cut-out images are composed with the same \hst filters as in Fig. \ref{fig:members_all}, and the circles have $1\arcsec$ diameters.}
\end{figure*}

\begin{figure*}
  \setcounter{figure}{\value{figure}-1}

Family 7
        
   \includegraphics[width = 0.666\columnwidth]{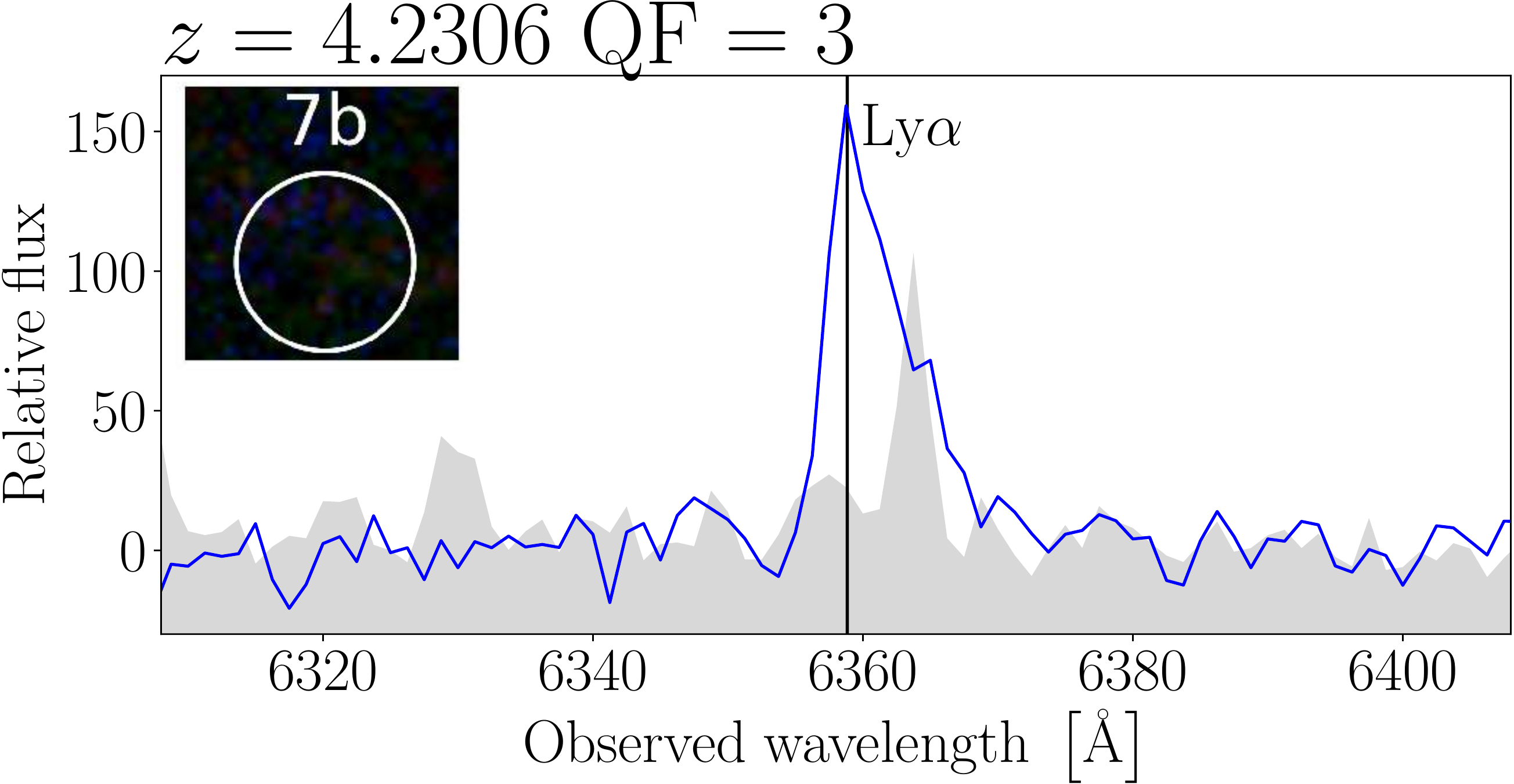}
   \includegraphics[width = 0.666\columnwidth]{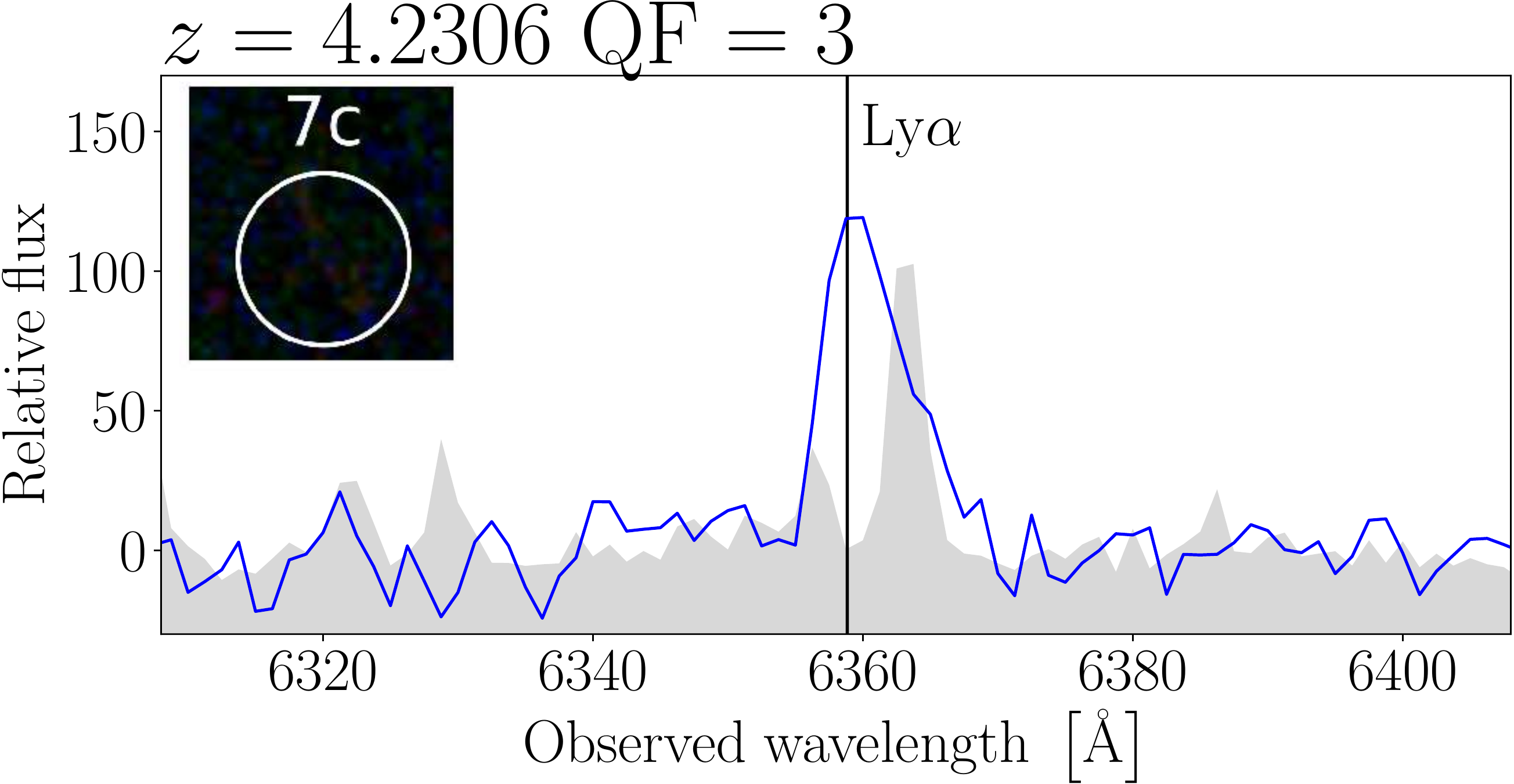}
   
Family 8

   \includegraphics[width = 0.666\columnwidth]{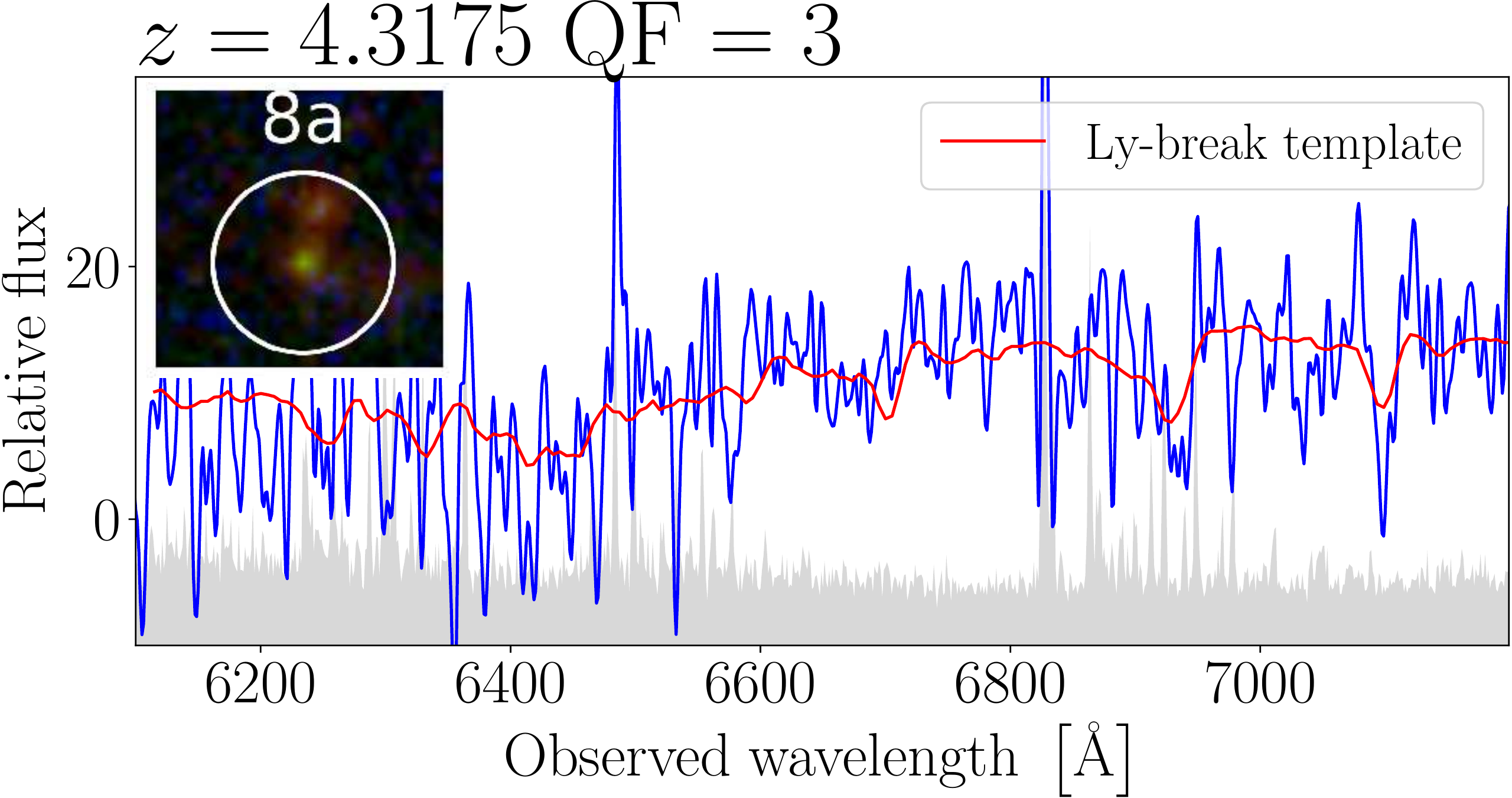}

Family 9

   \includegraphics[width = 0.666\columnwidth]{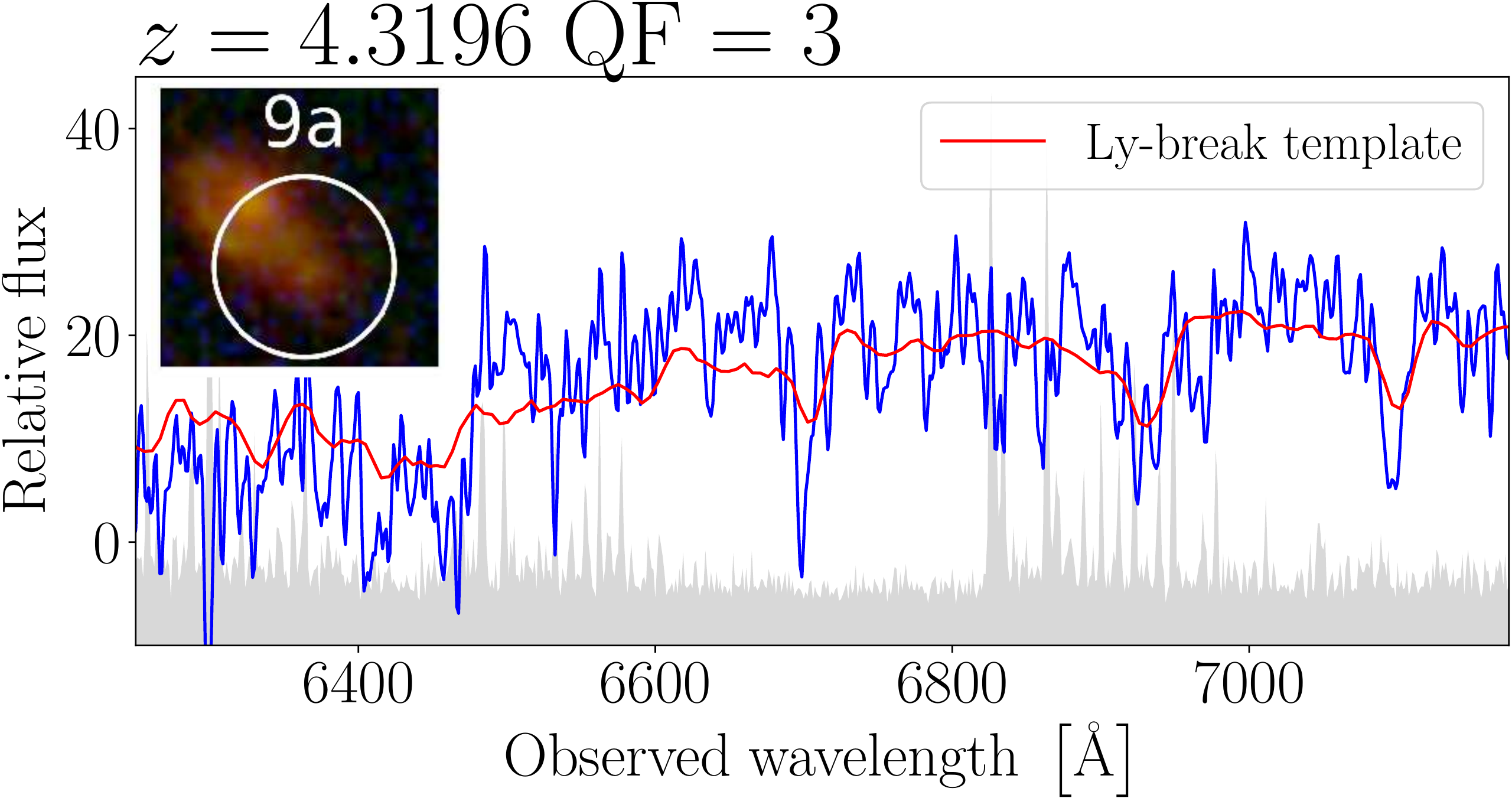}
   
Family 10

   \includegraphics[width = 0.666\columnwidth]{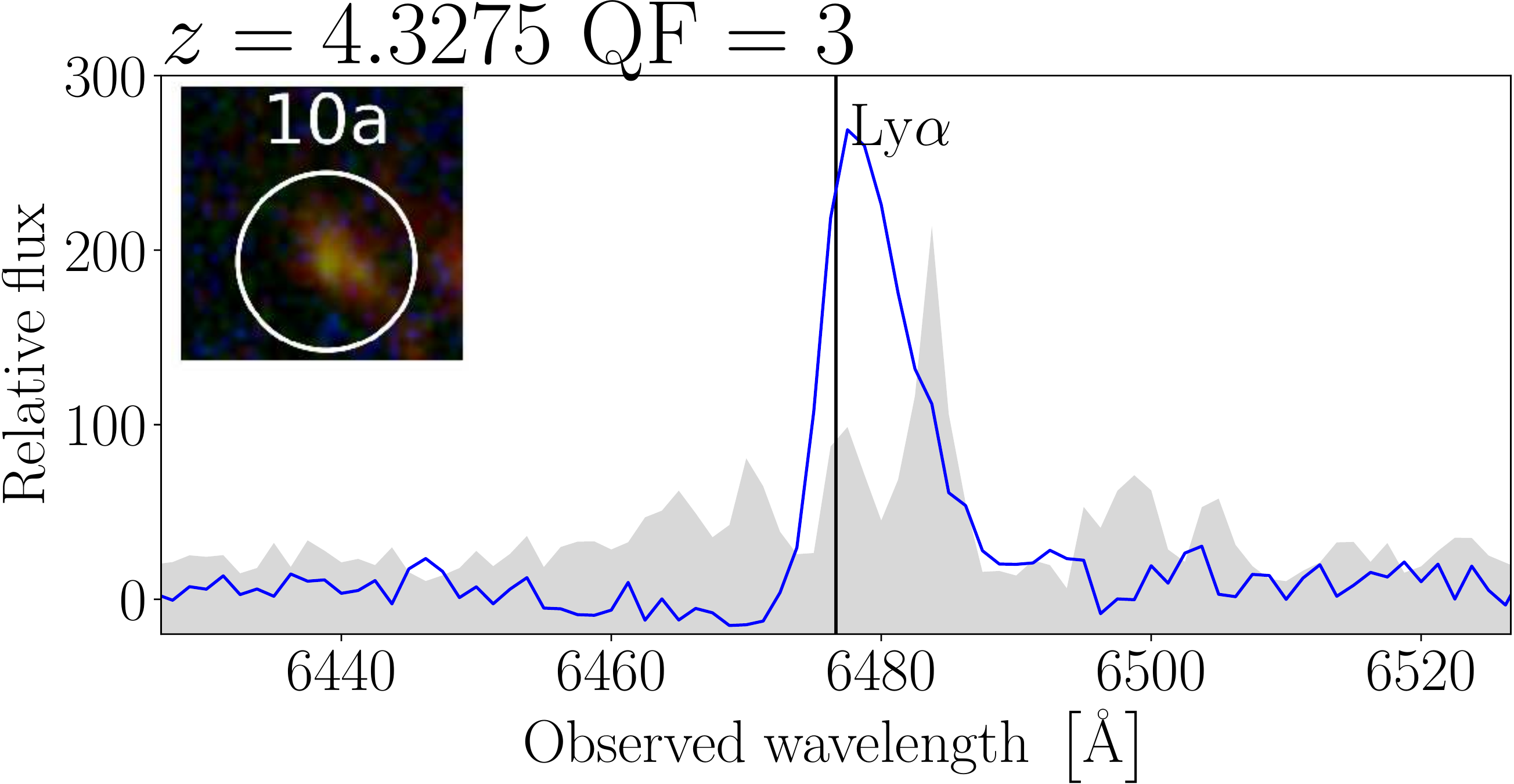}
   \includegraphics[width = 0.666\columnwidth]{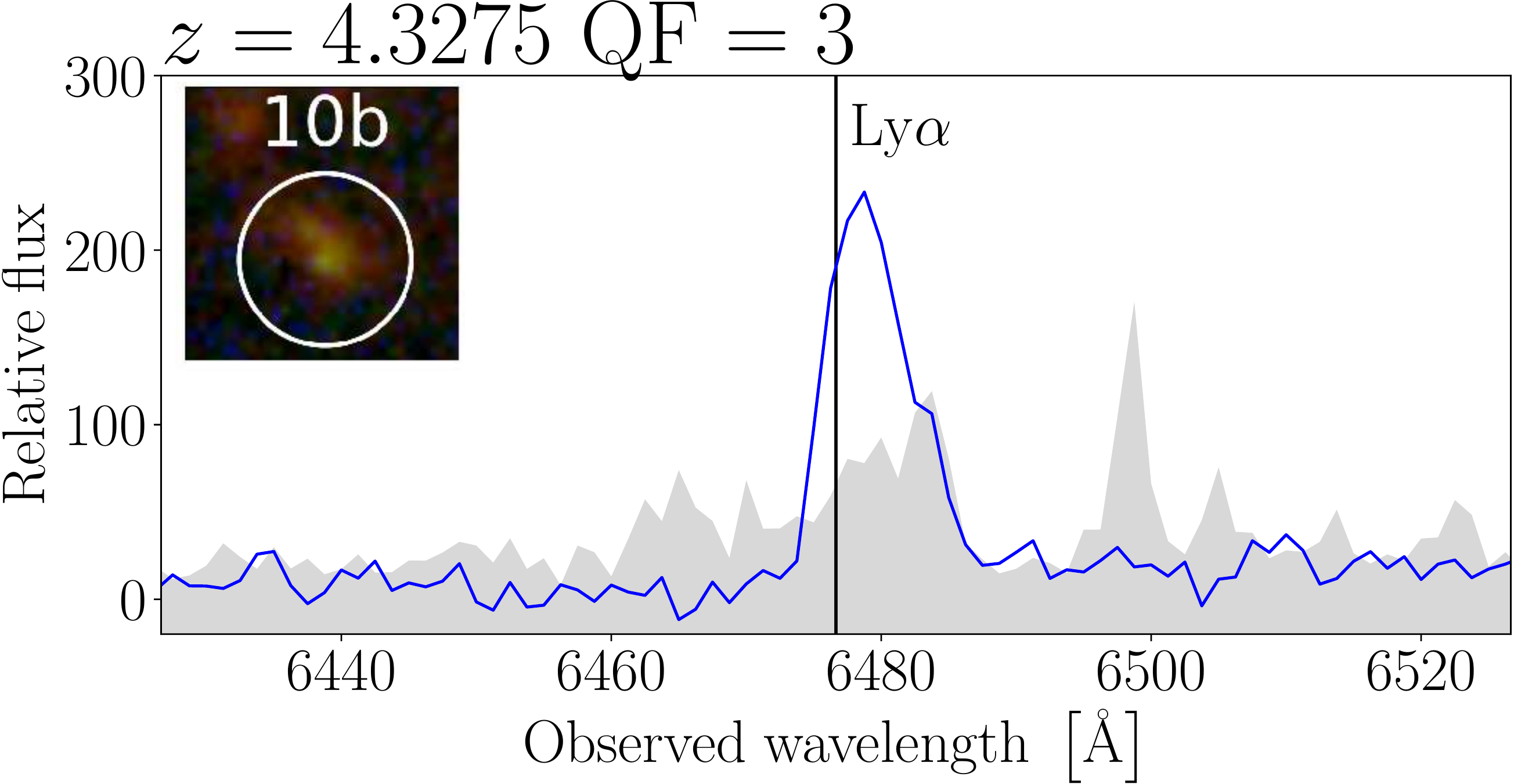}   
   \includegraphics[width = 0.666\columnwidth]{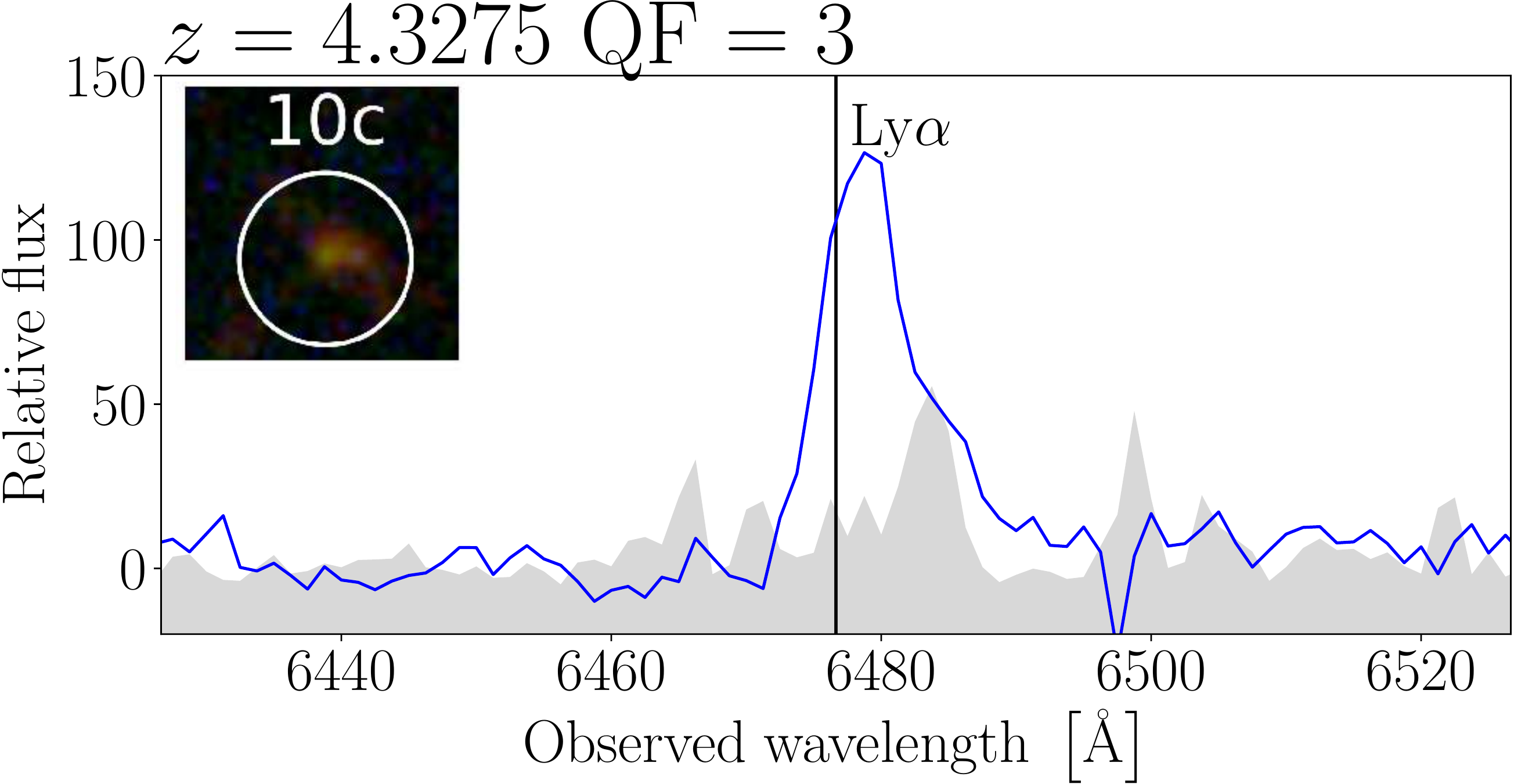}   
   
Family 11

   \includegraphics[width = 0.666\columnwidth]{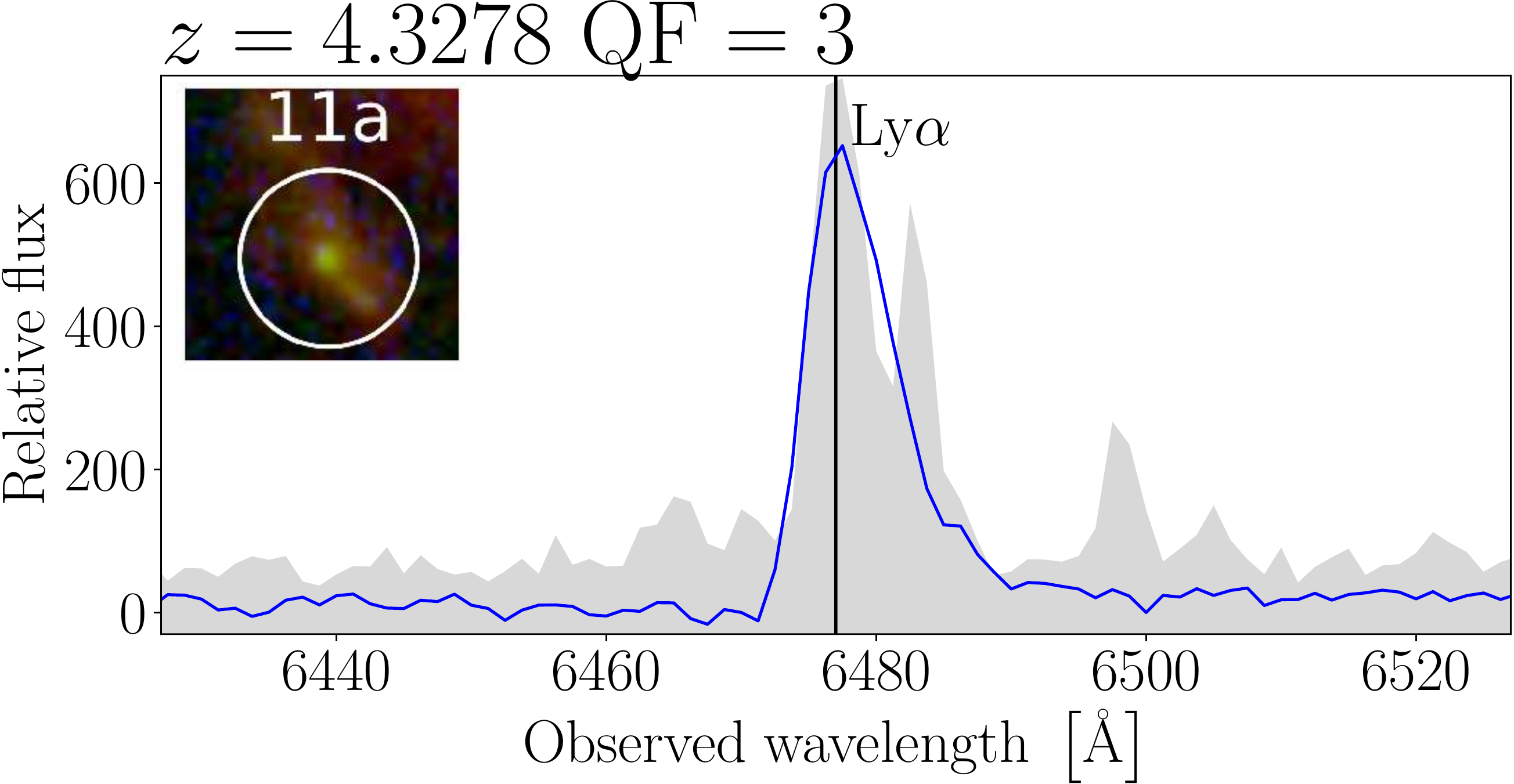}
   \includegraphics[width = 0.666\columnwidth]{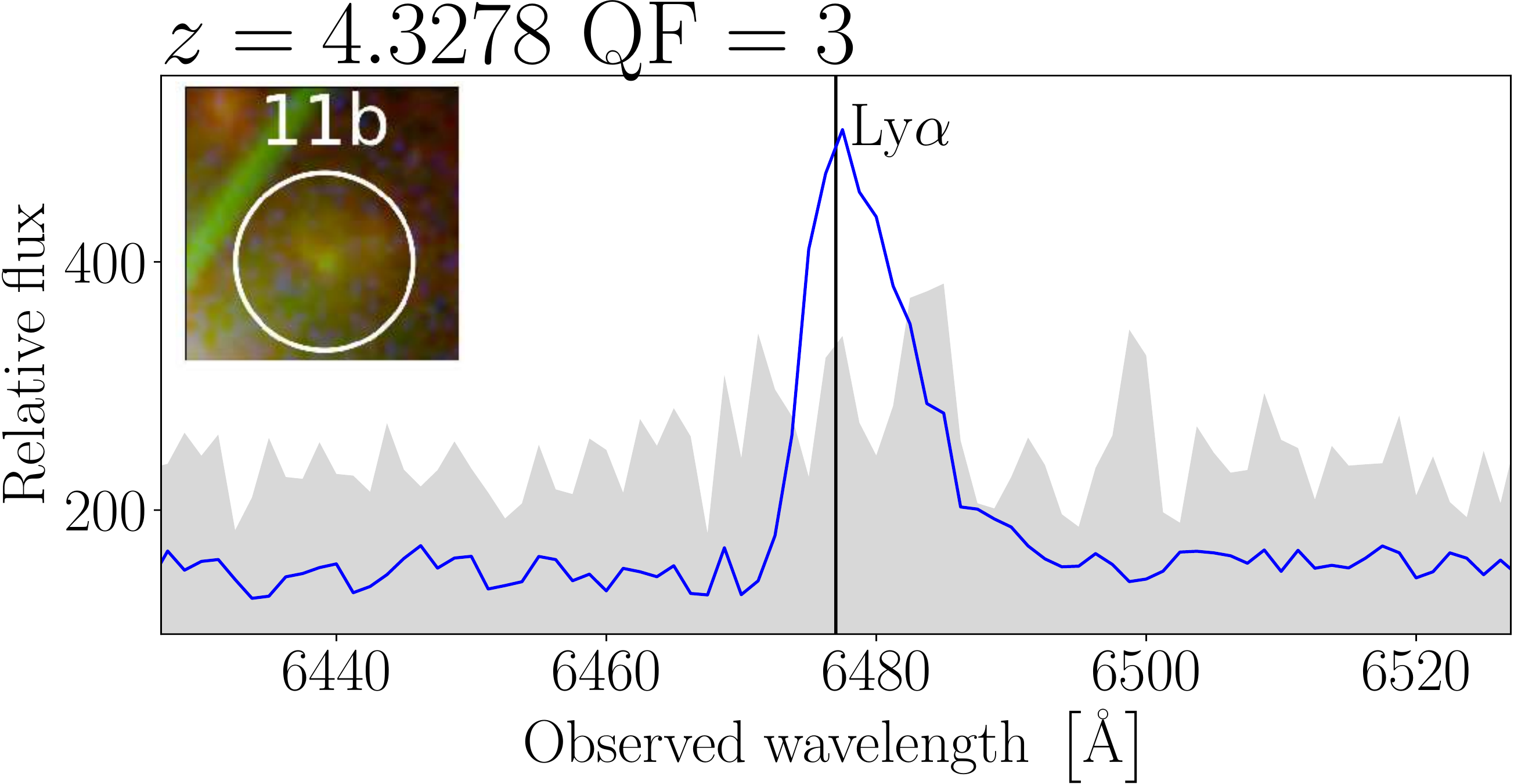}   
   \includegraphics[width = 0.666\columnwidth]{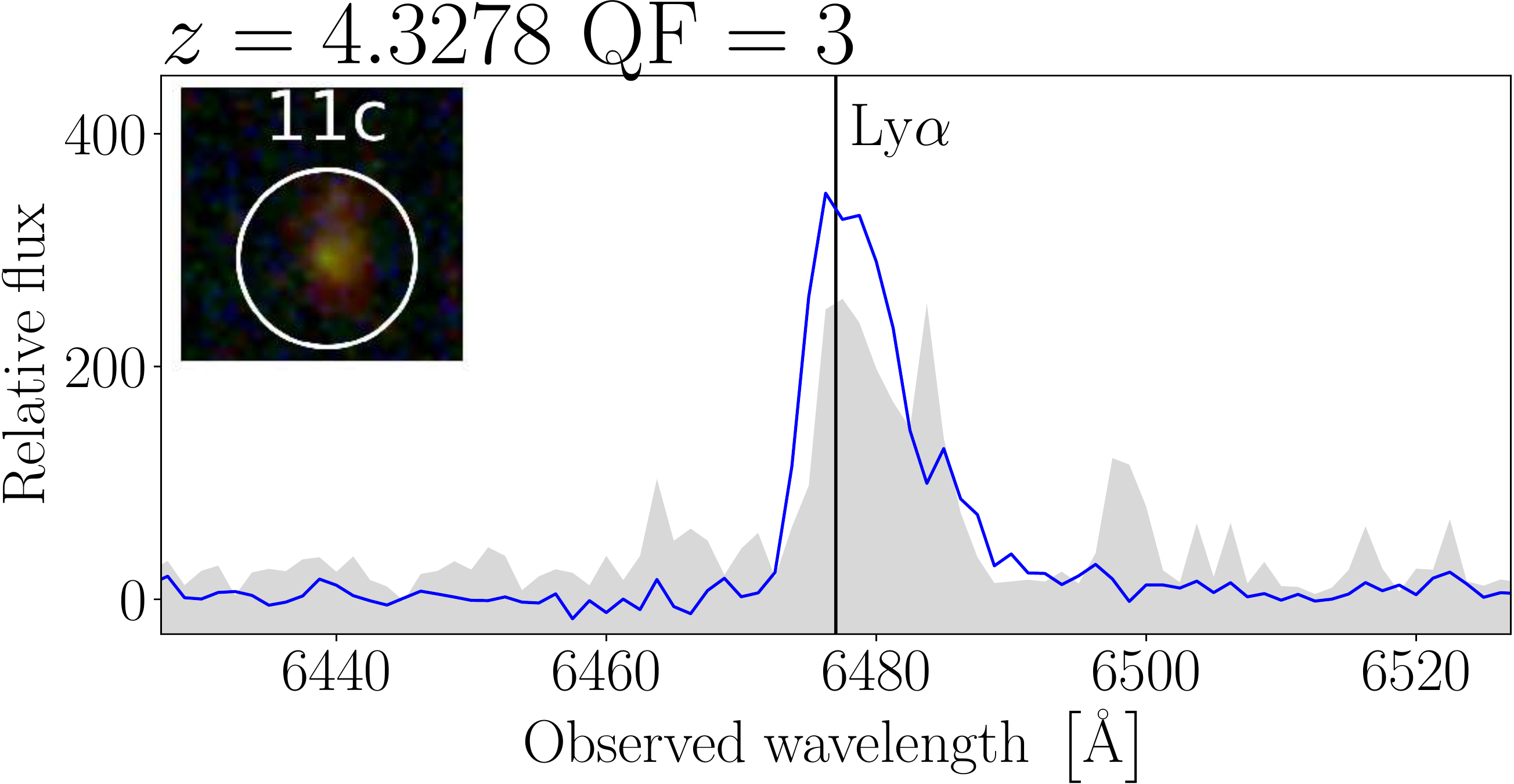}      
   
Family 12

   \includegraphics[width = 0.666\columnwidth]{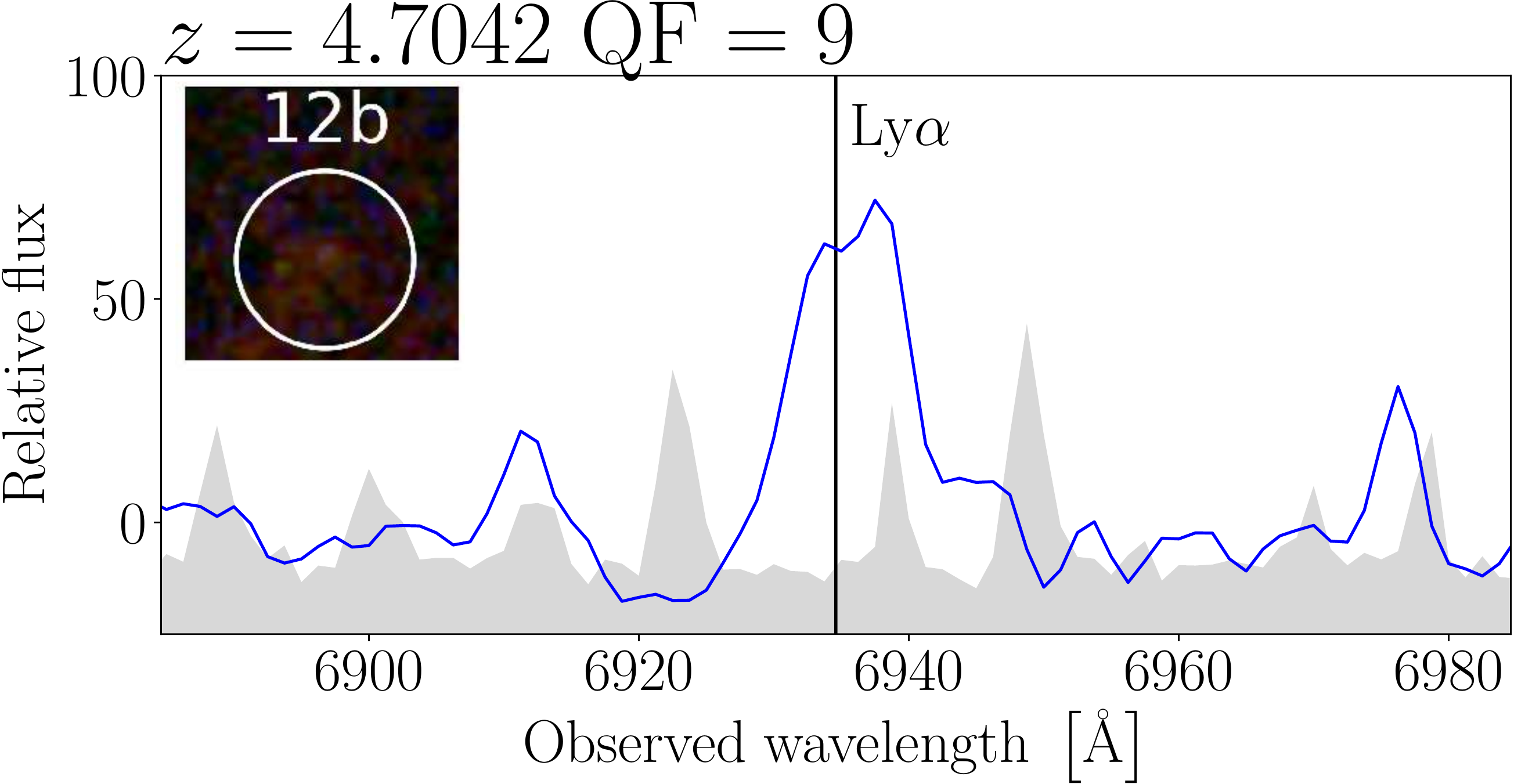}
   \includegraphics[width = 0.666\columnwidth]{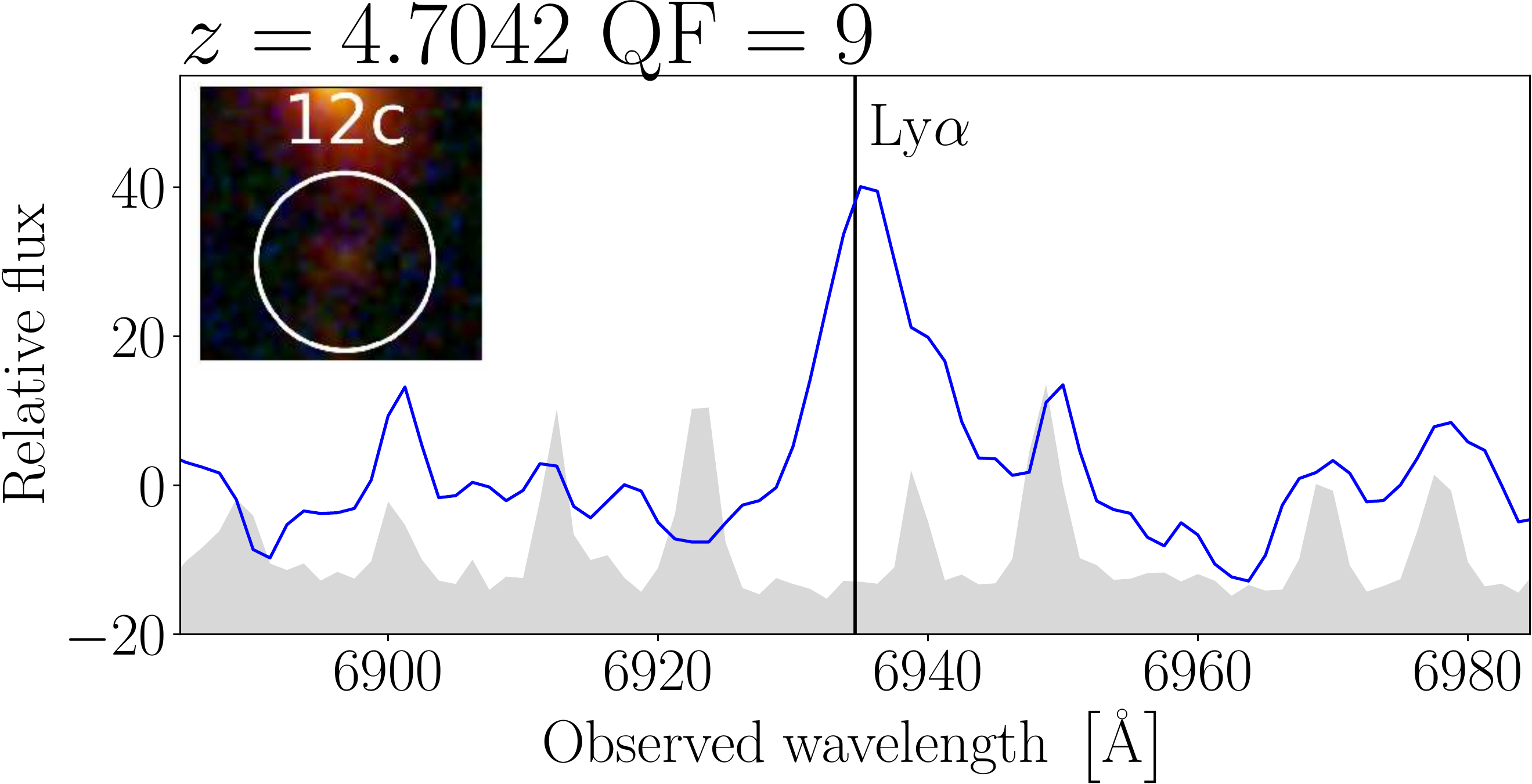}      

  \caption{(Continued)}

\end{figure*}

\begin{figure*}
   \setcounter{figure}{\value{figure}-1}
Family 13

   \includegraphics[width = 0.666\columnwidth]{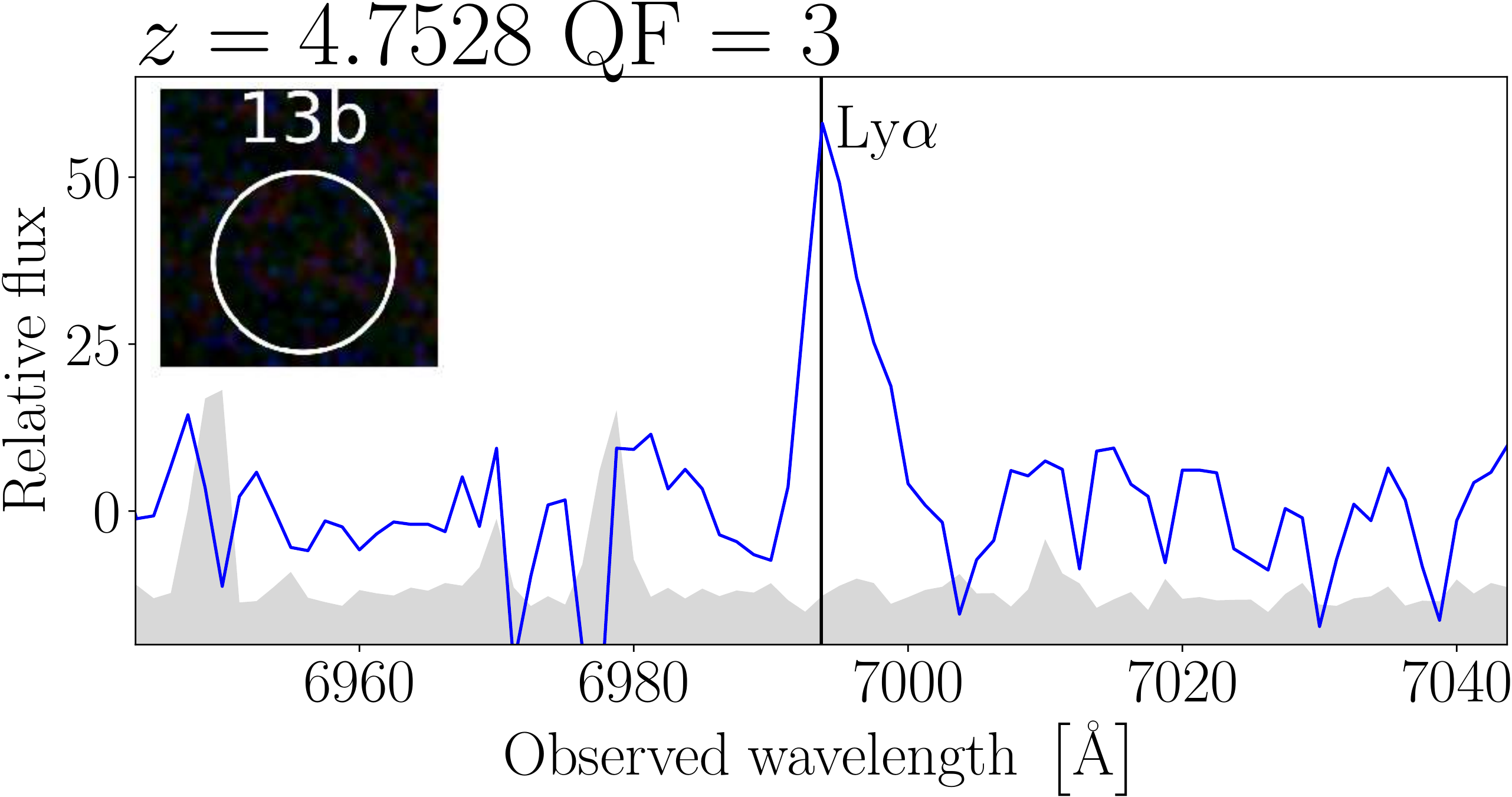}
   \includegraphics[width = 0.666\columnwidth]{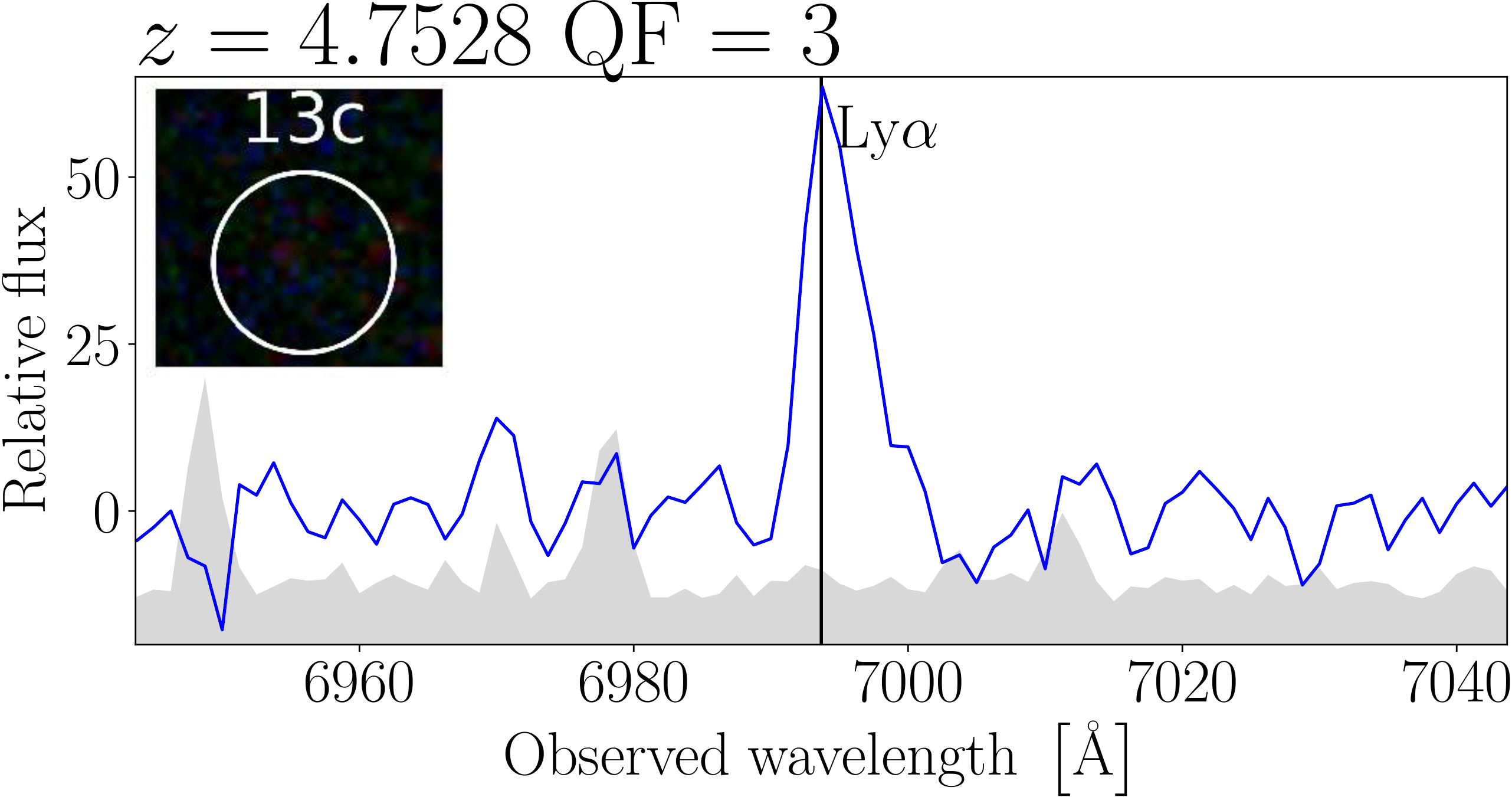} 
   
Family 14

   \includegraphics[width = 0.666\columnwidth]{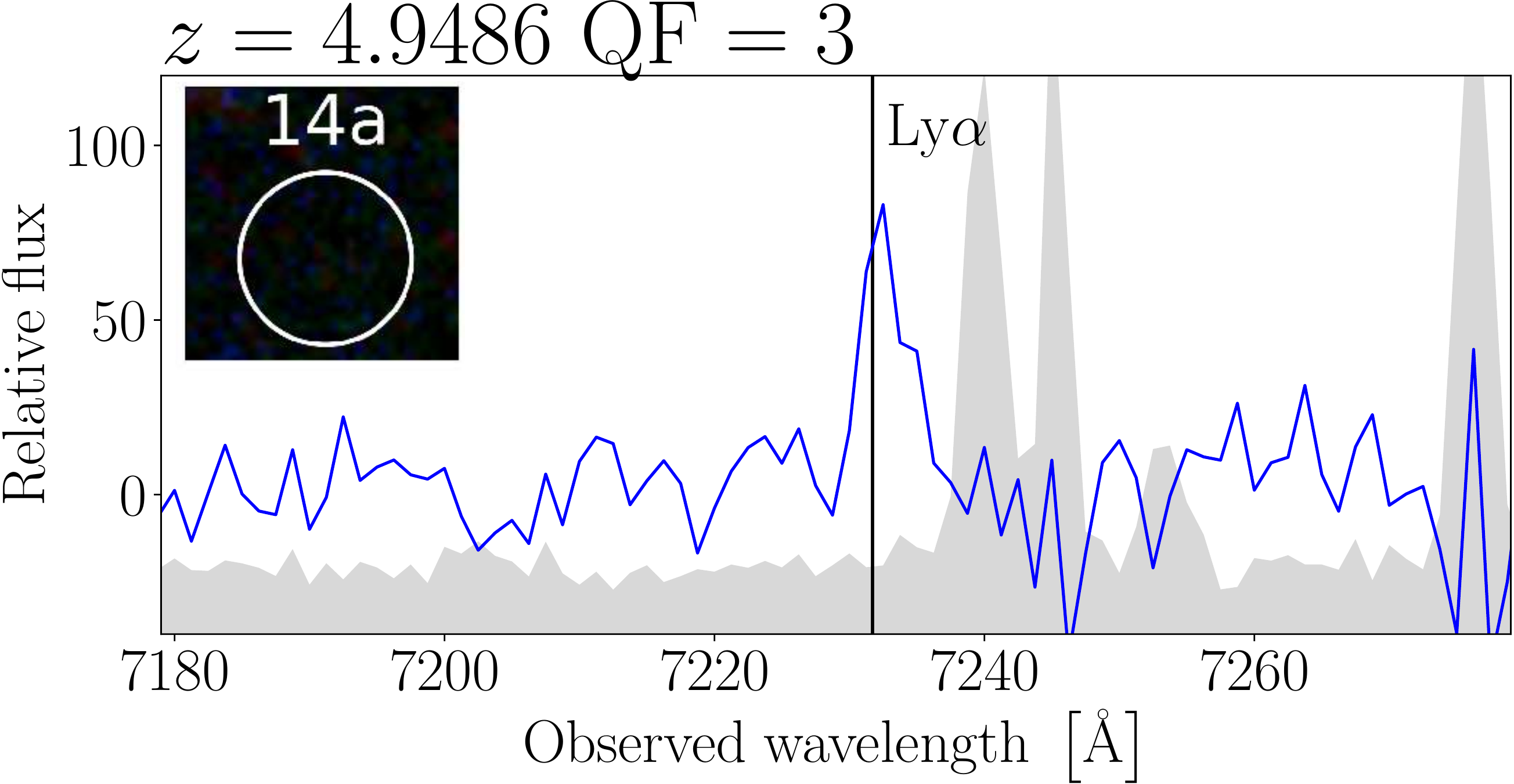}
   \includegraphics[width = 0.666\columnwidth]{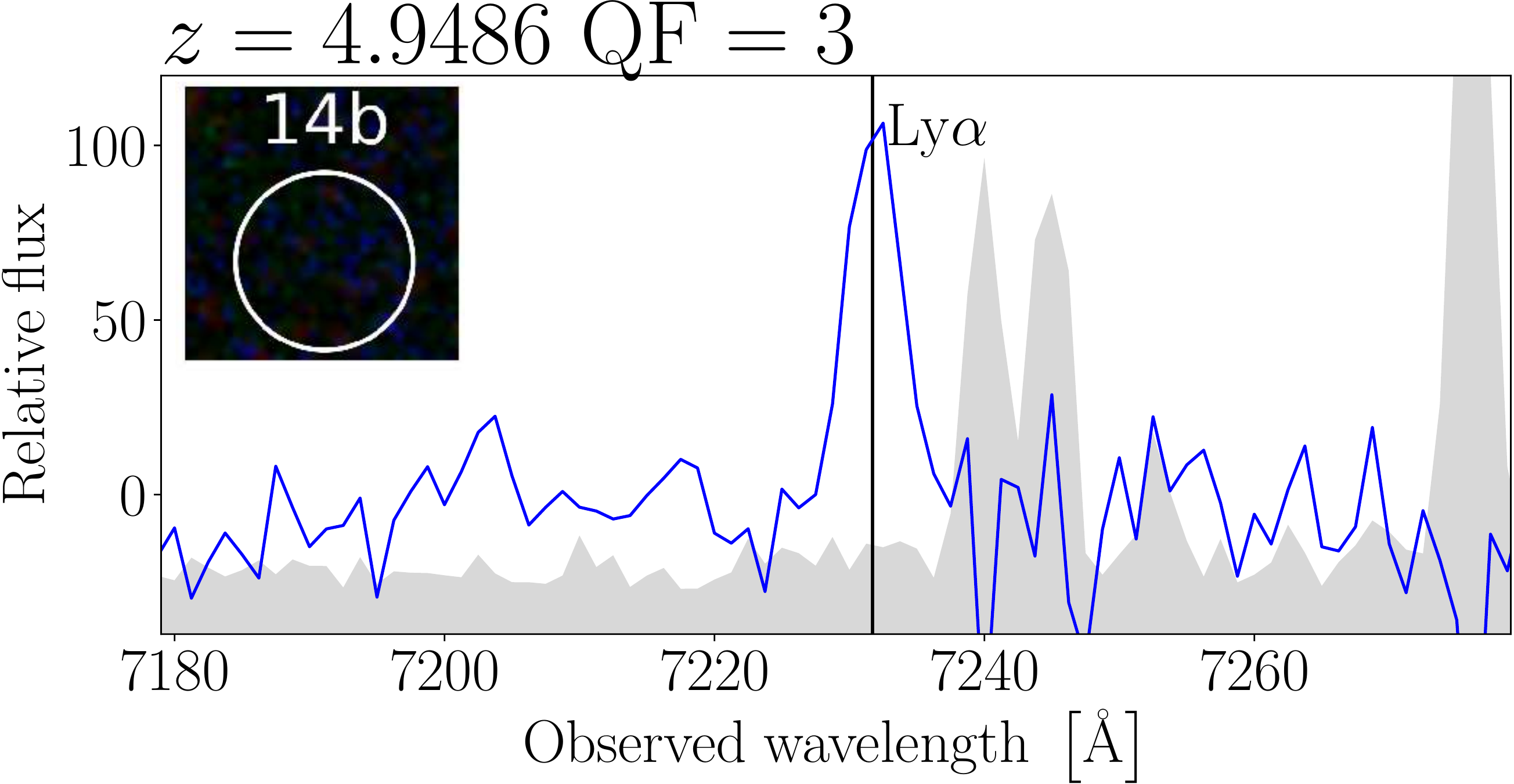} 

Family 15

   \includegraphics[width = 0.666\columnwidth]{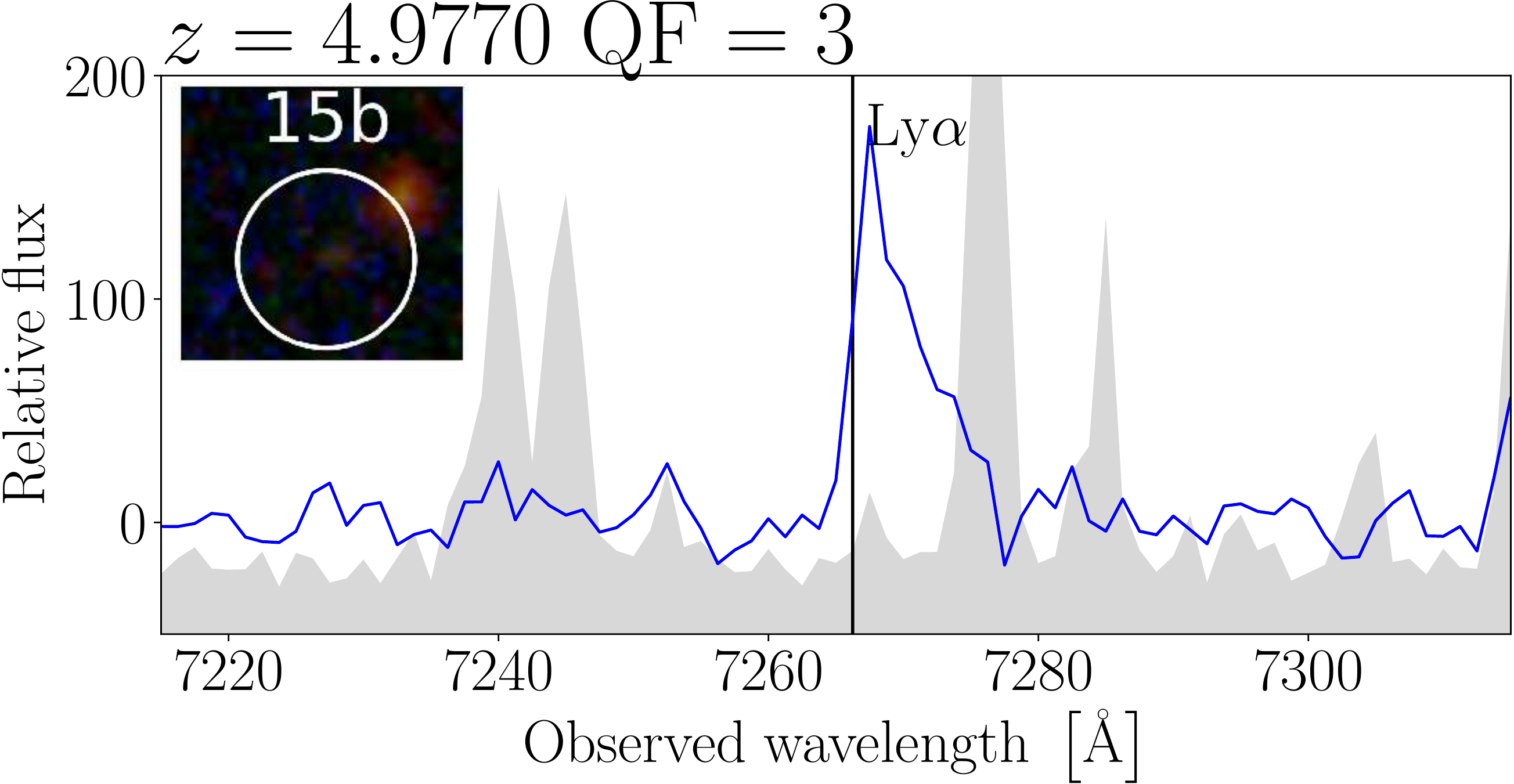}
   \includegraphics[width = 0.666\columnwidth]{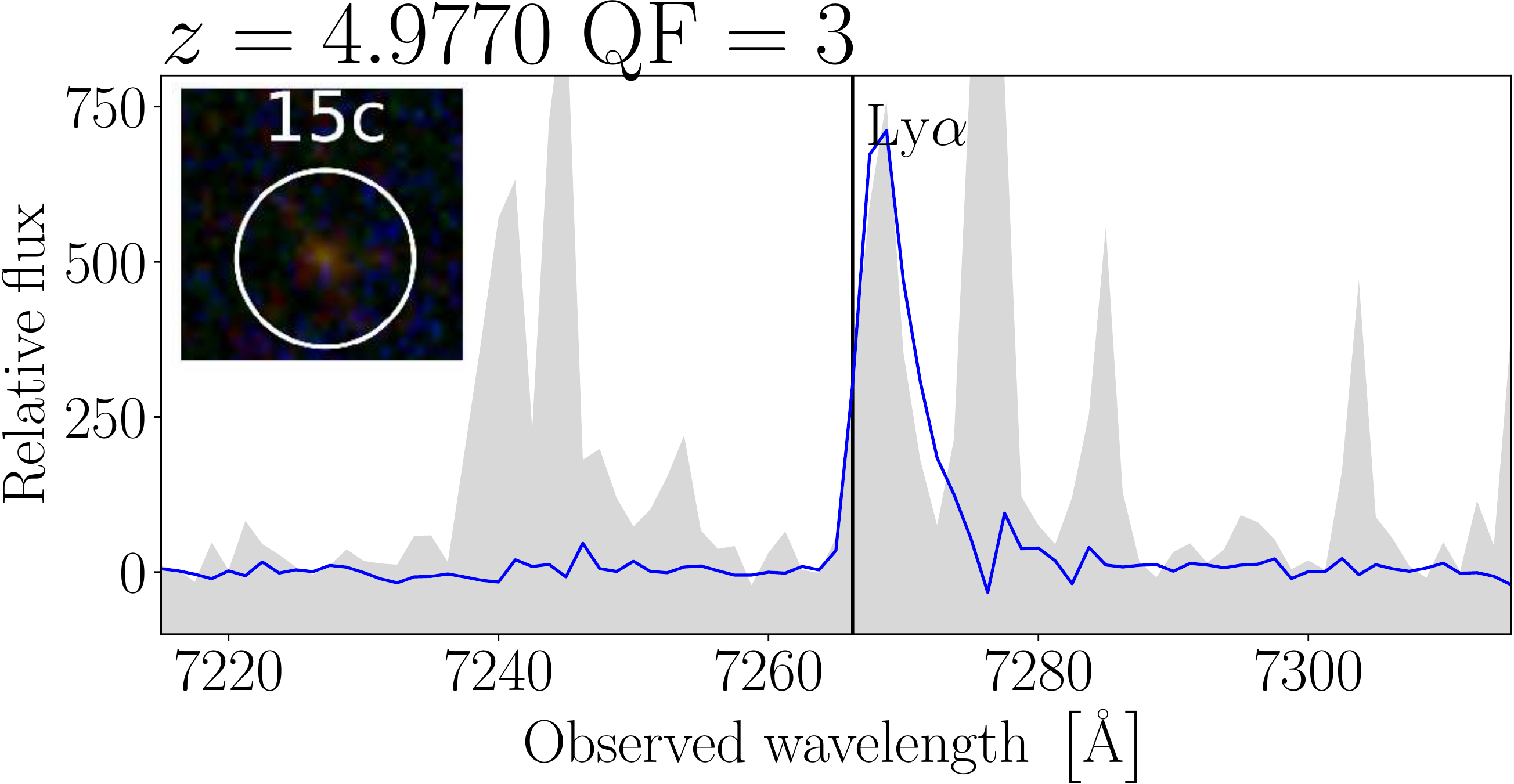} 

Family 16

   \includegraphics[width = 0.666\columnwidth]{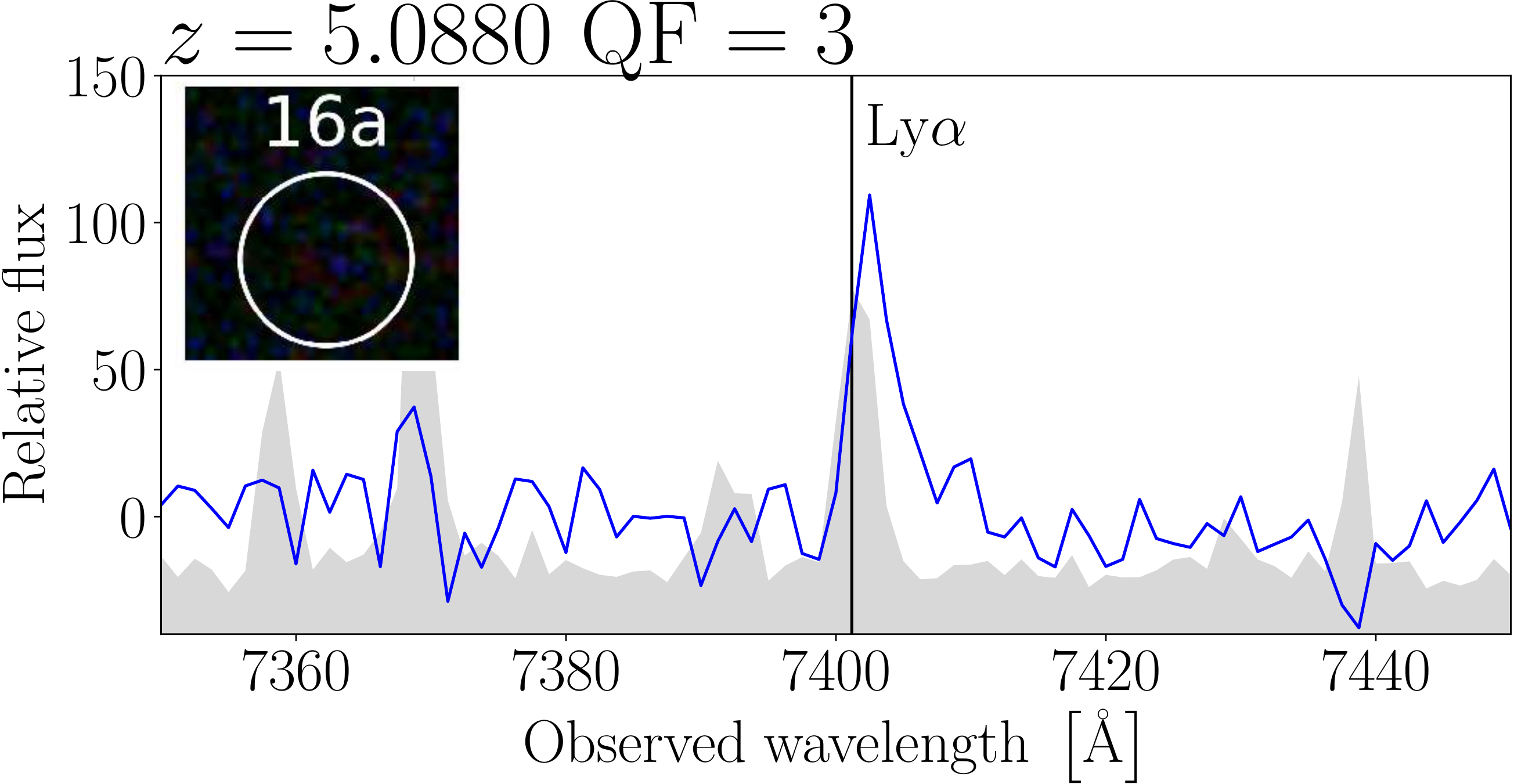}
   \includegraphics[width = 0.666\columnwidth]{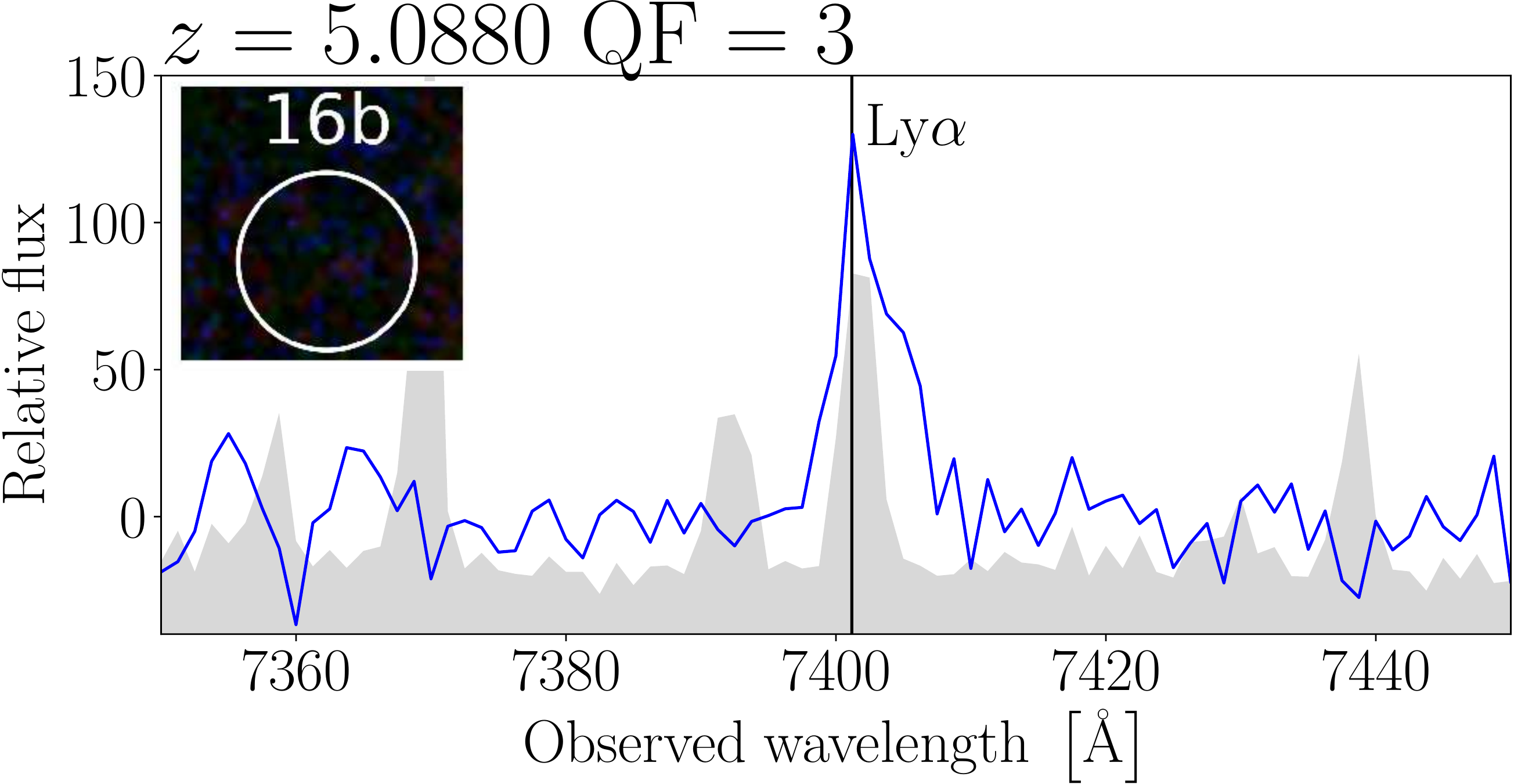} 

Family 17

   \includegraphics[width = 0.666\columnwidth]{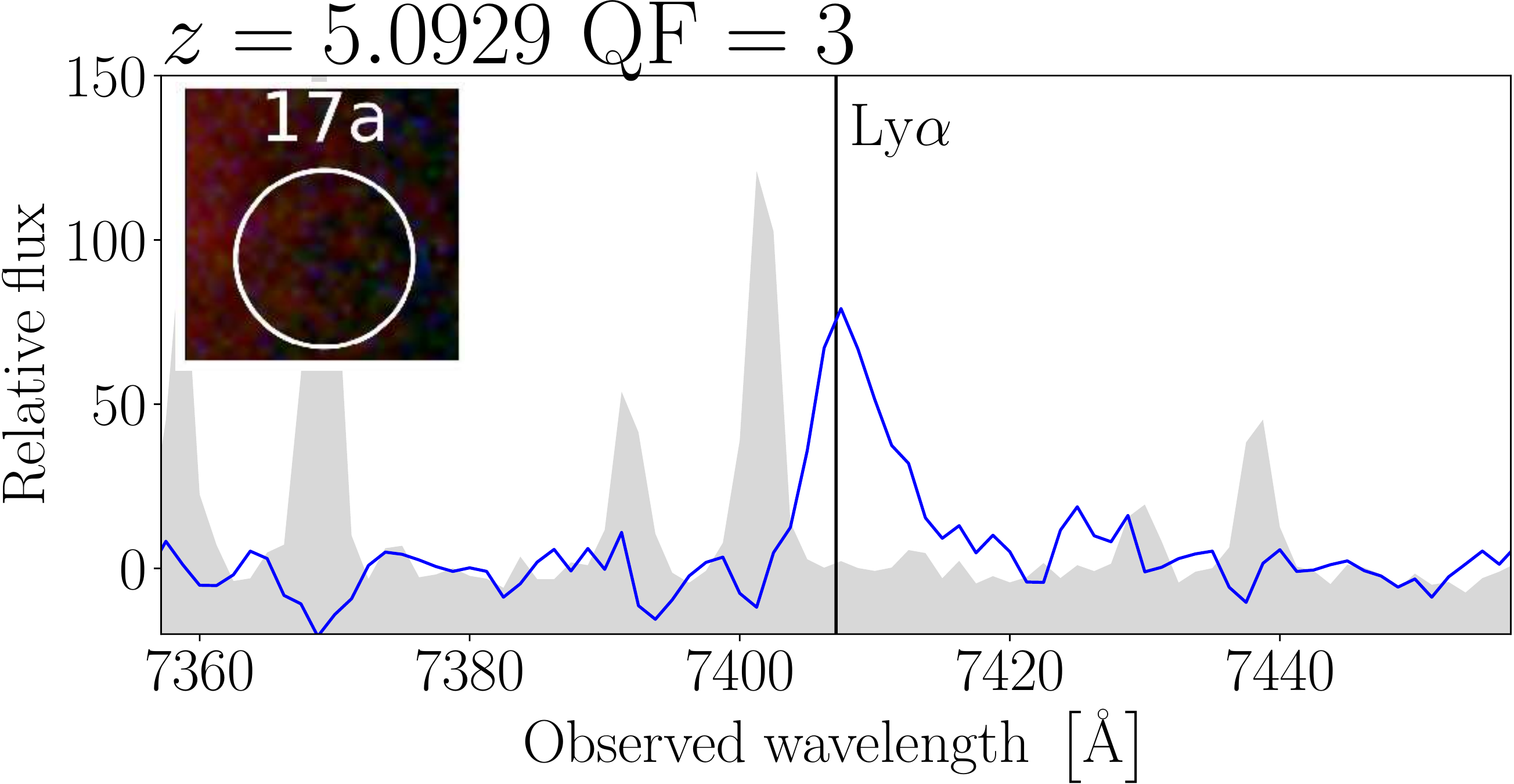}
   \includegraphics[width = 0.666\columnwidth]{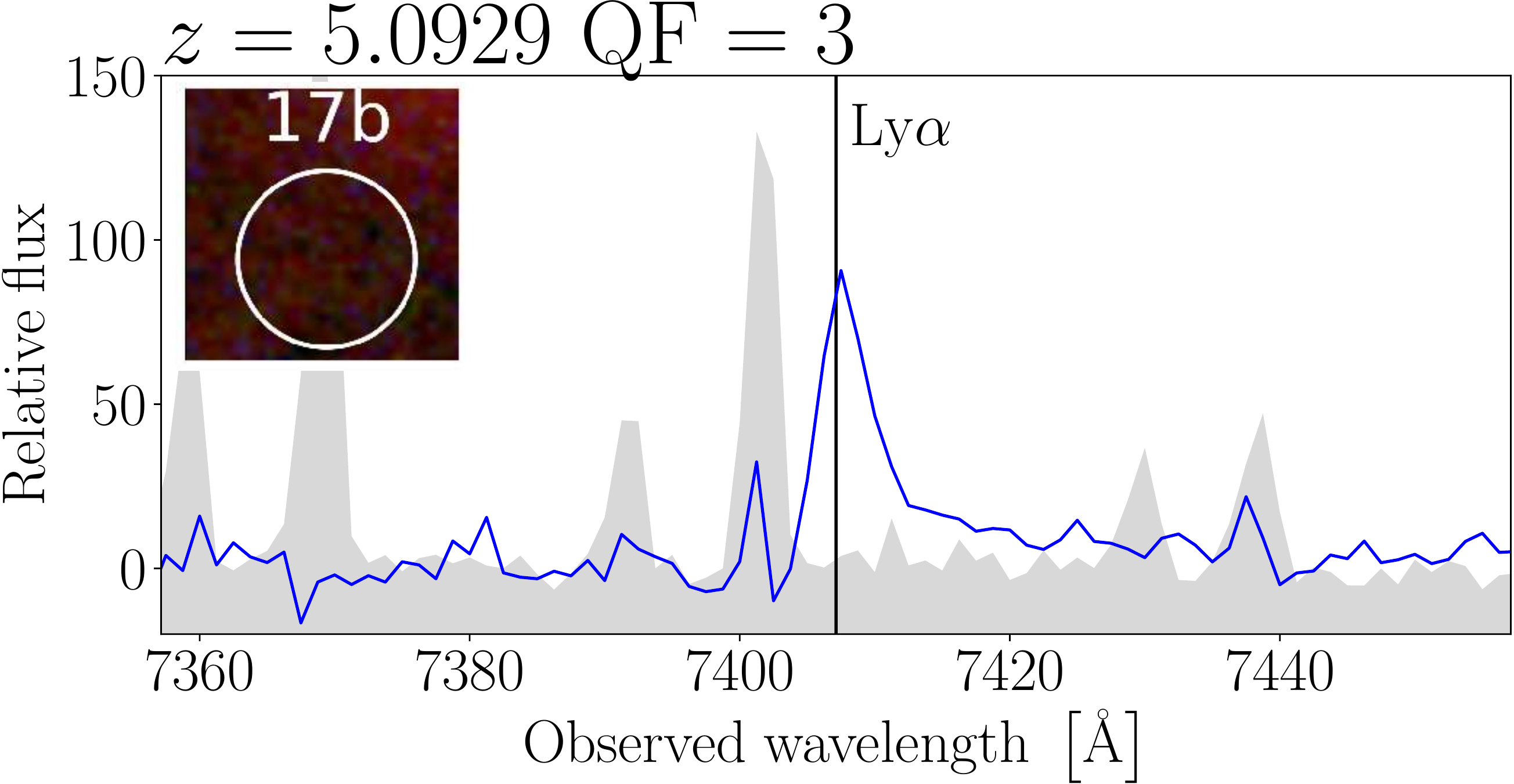} 
   \includegraphics[width = 0.666\columnwidth]{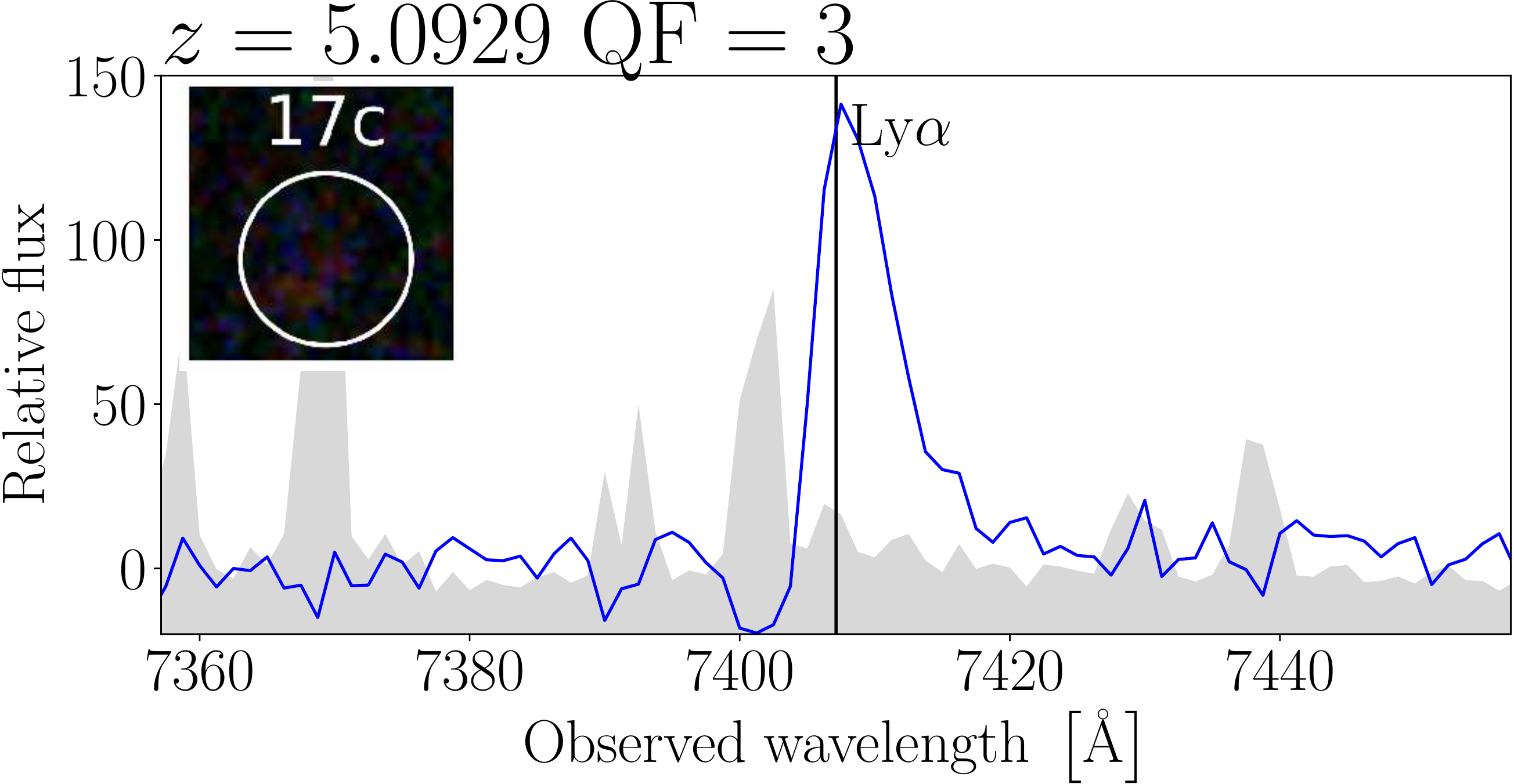} 

Family 18

   \includegraphics[width = 0.666\columnwidth]{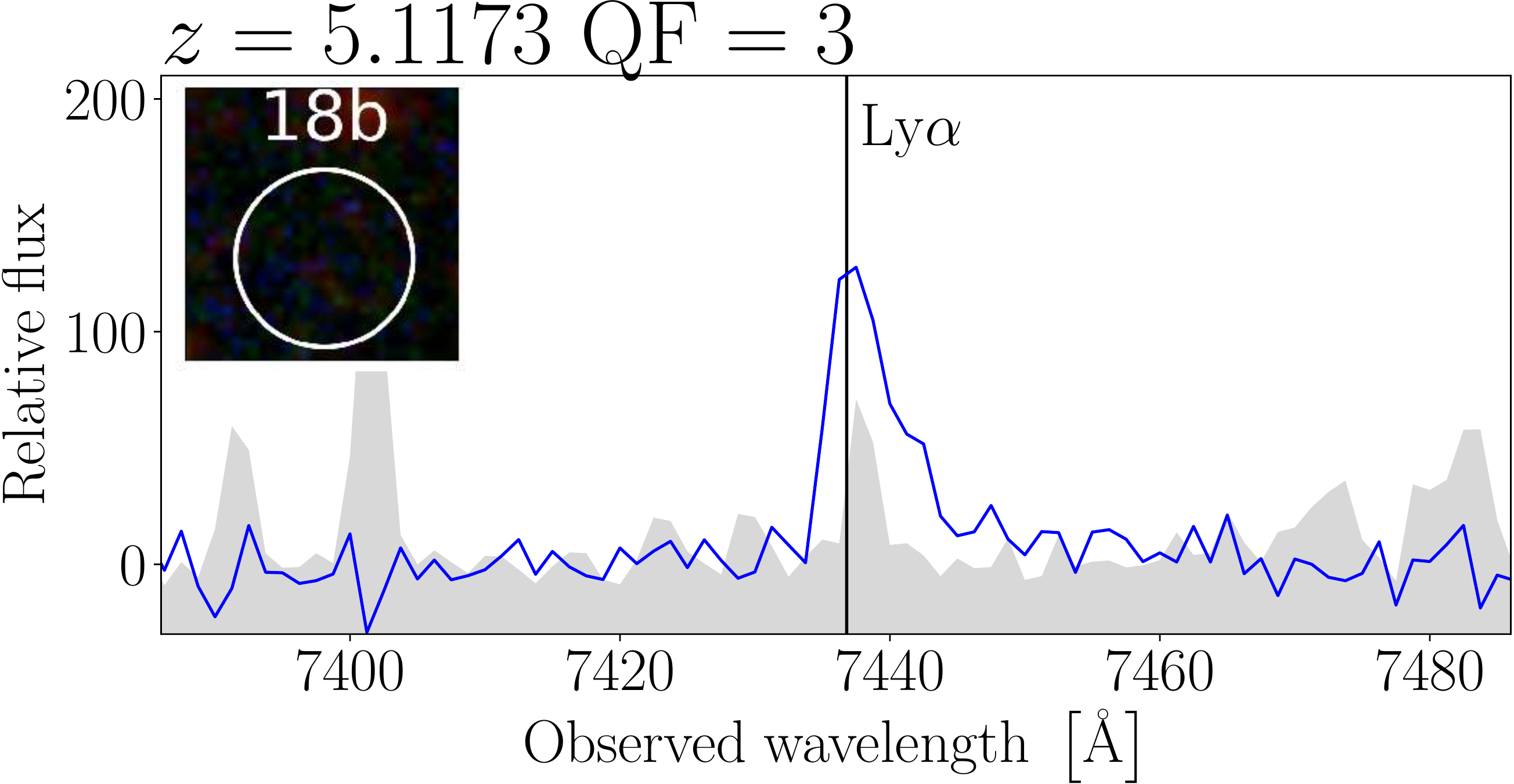}
   \includegraphics[width = 0.666\columnwidth]{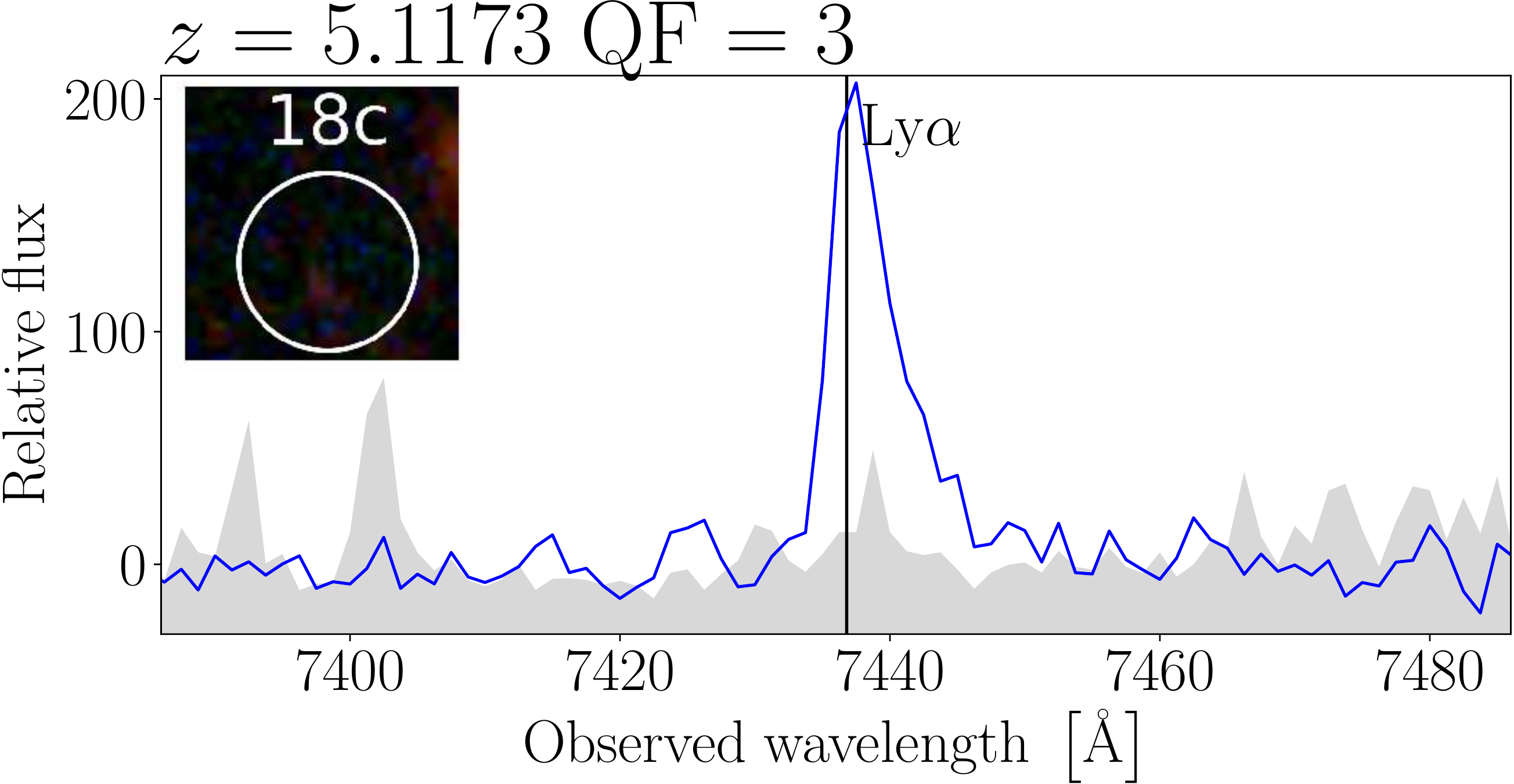} 

  \caption{(Continued)}
\end{figure*}

\begin{figure*}
   \setcounter{figure}{\value{figure}-1}

Family 19

   \includegraphics[width = 0.666\columnwidth]{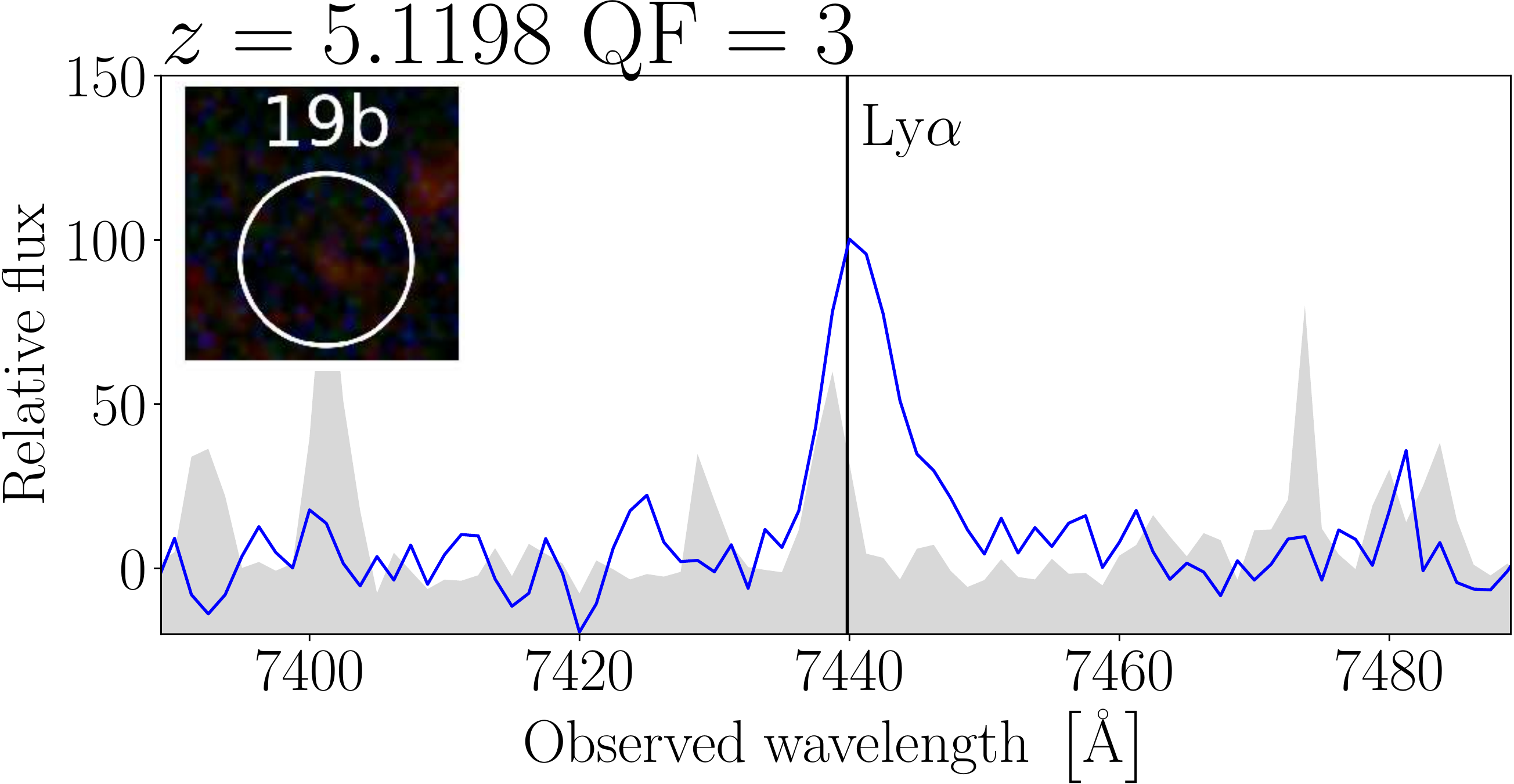}
   \includegraphics[width = 0.666\columnwidth]{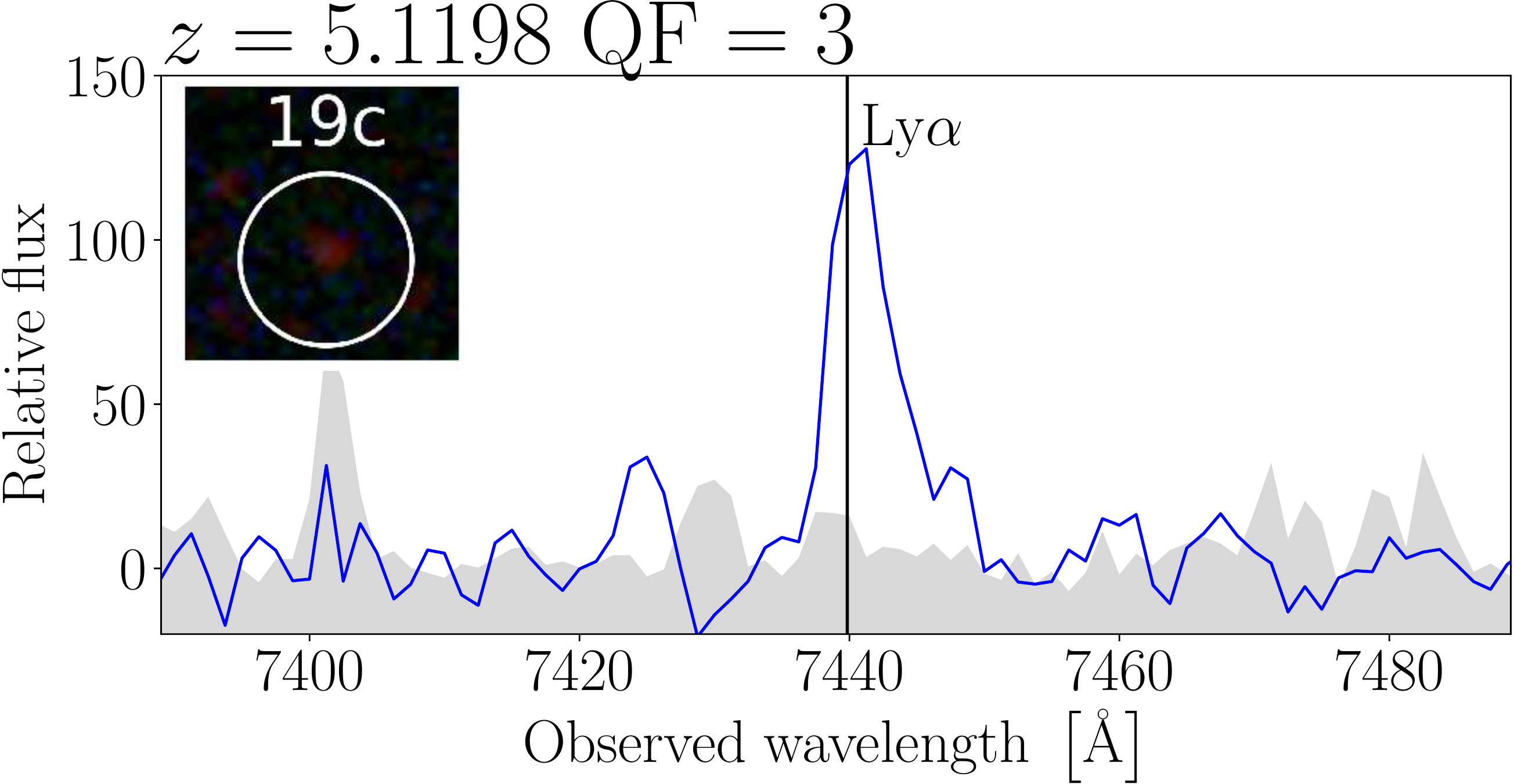} 

Family 20

   \includegraphics[width = 0.666\columnwidth]{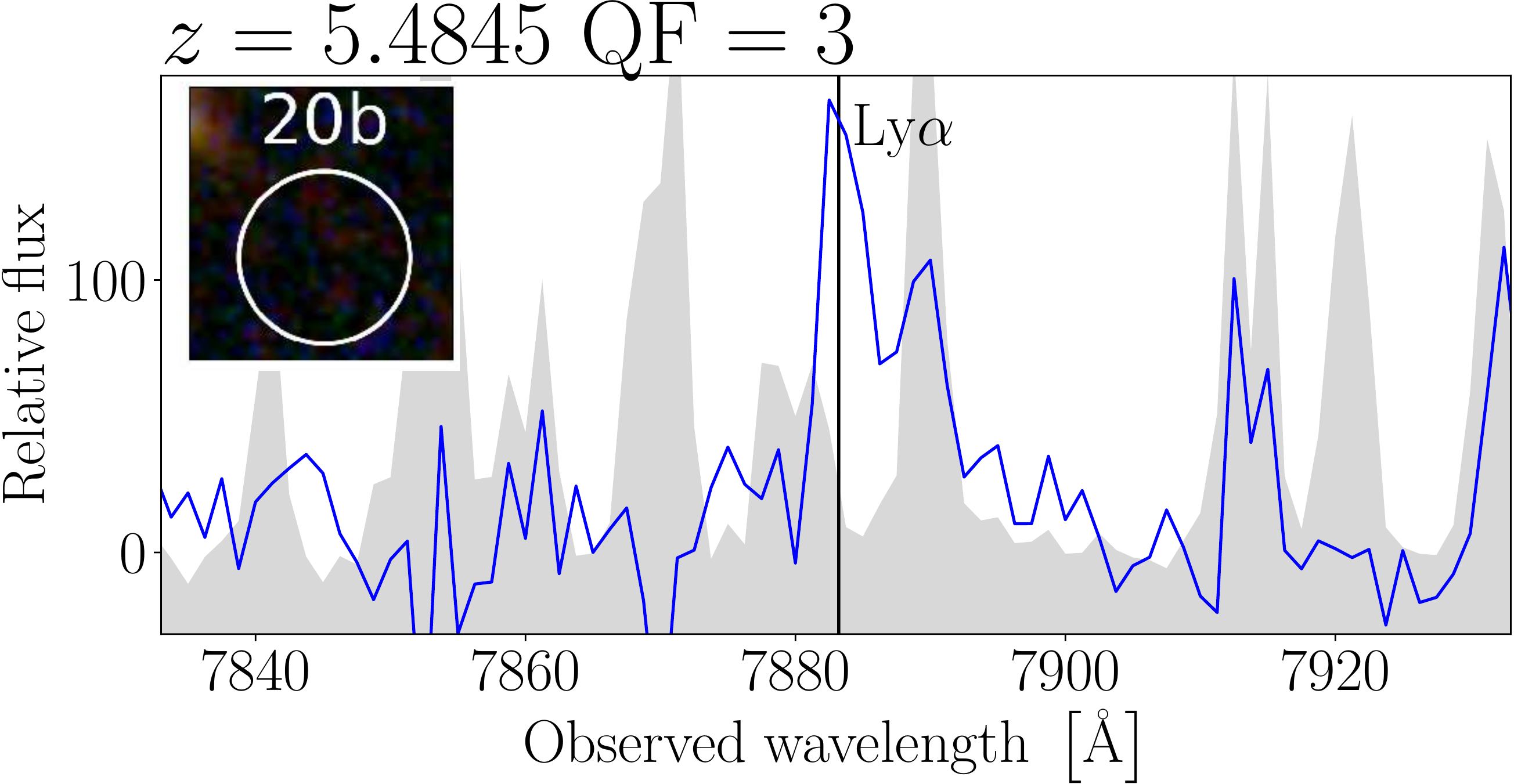}
   \includegraphics[width = 0.666\columnwidth]{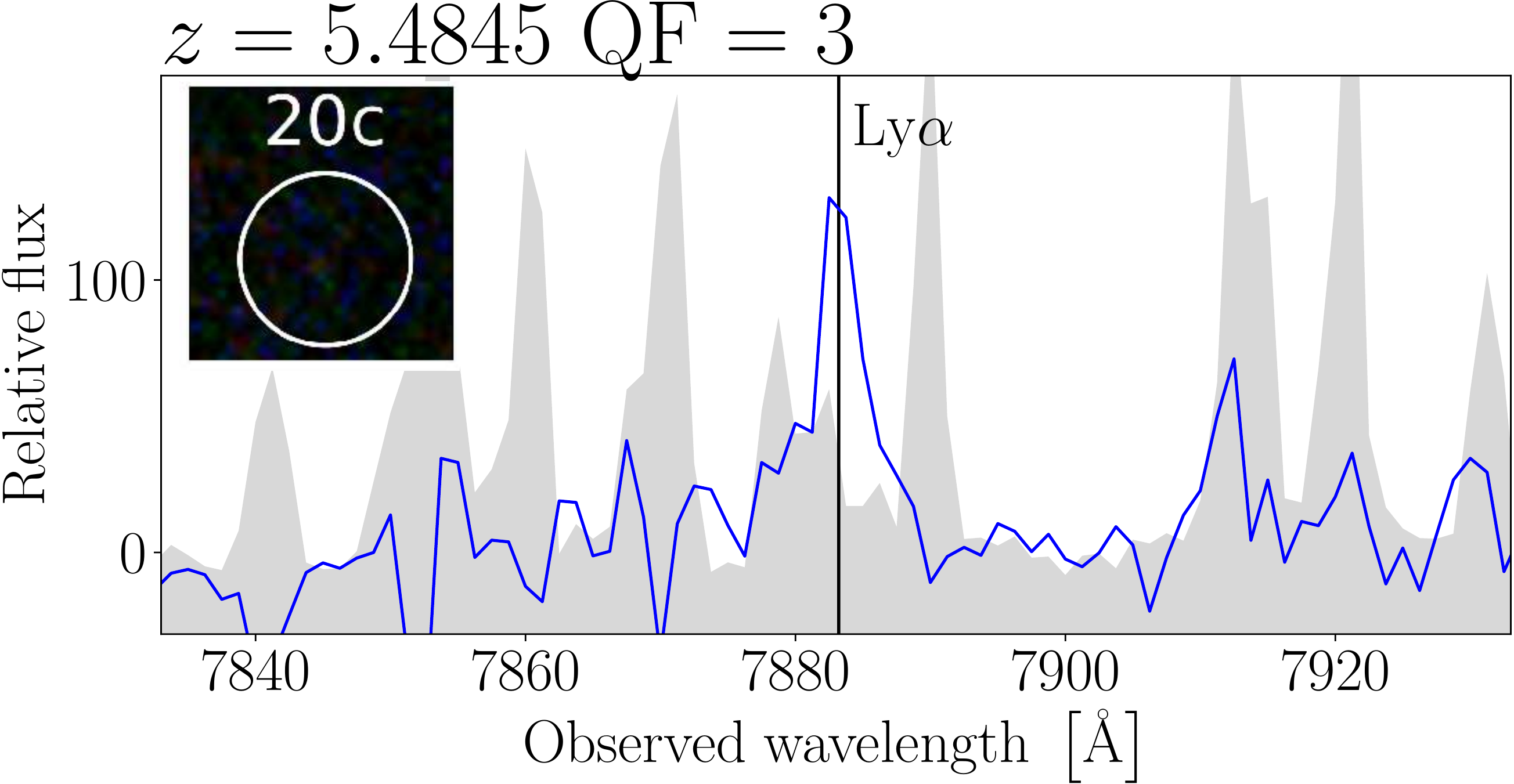}

Family 21

   \includegraphics[width = 0.666\columnwidth]{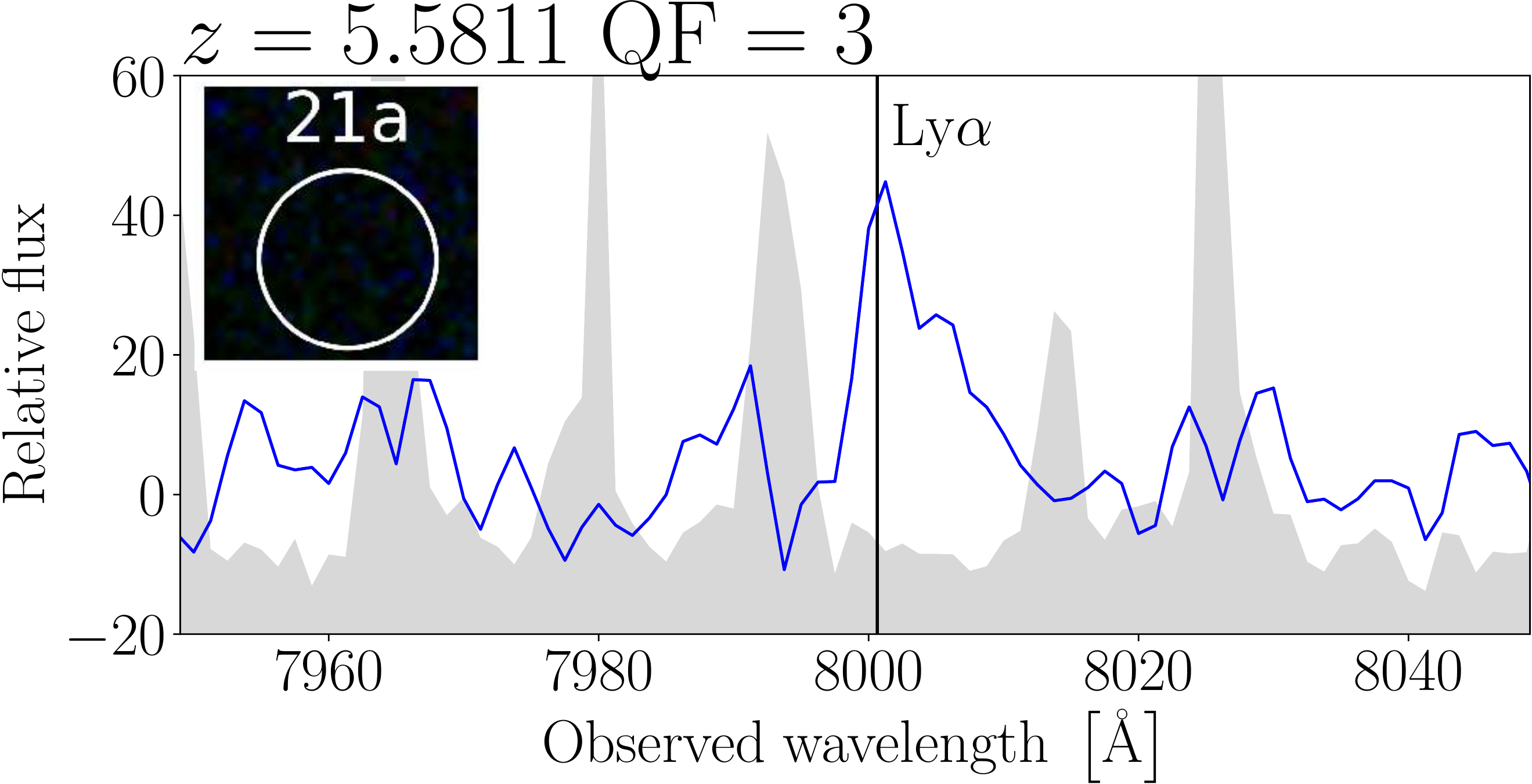}
   \includegraphics[width = 0.666\columnwidth]{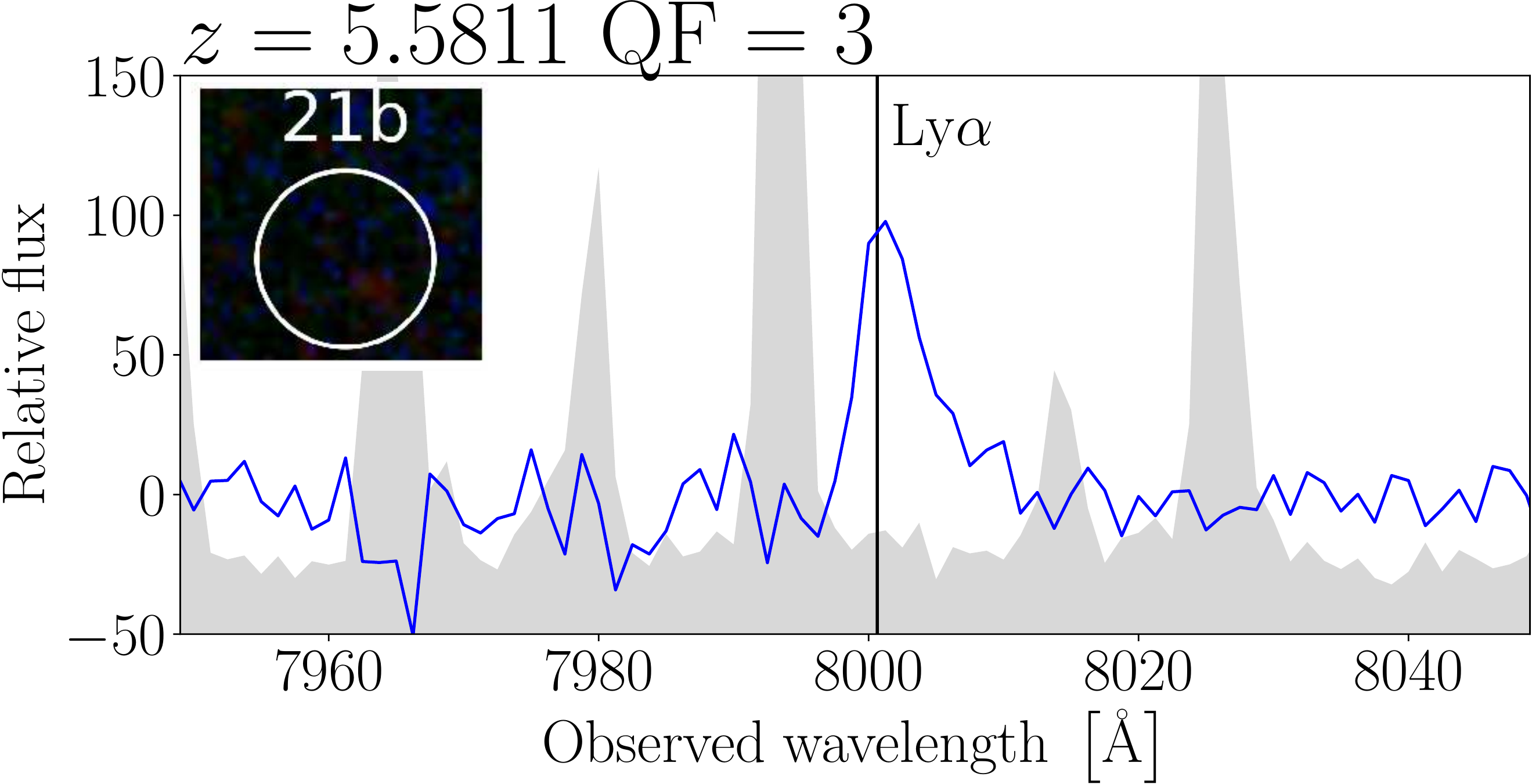}

Family 22

   \includegraphics[width = 0.666\columnwidth]{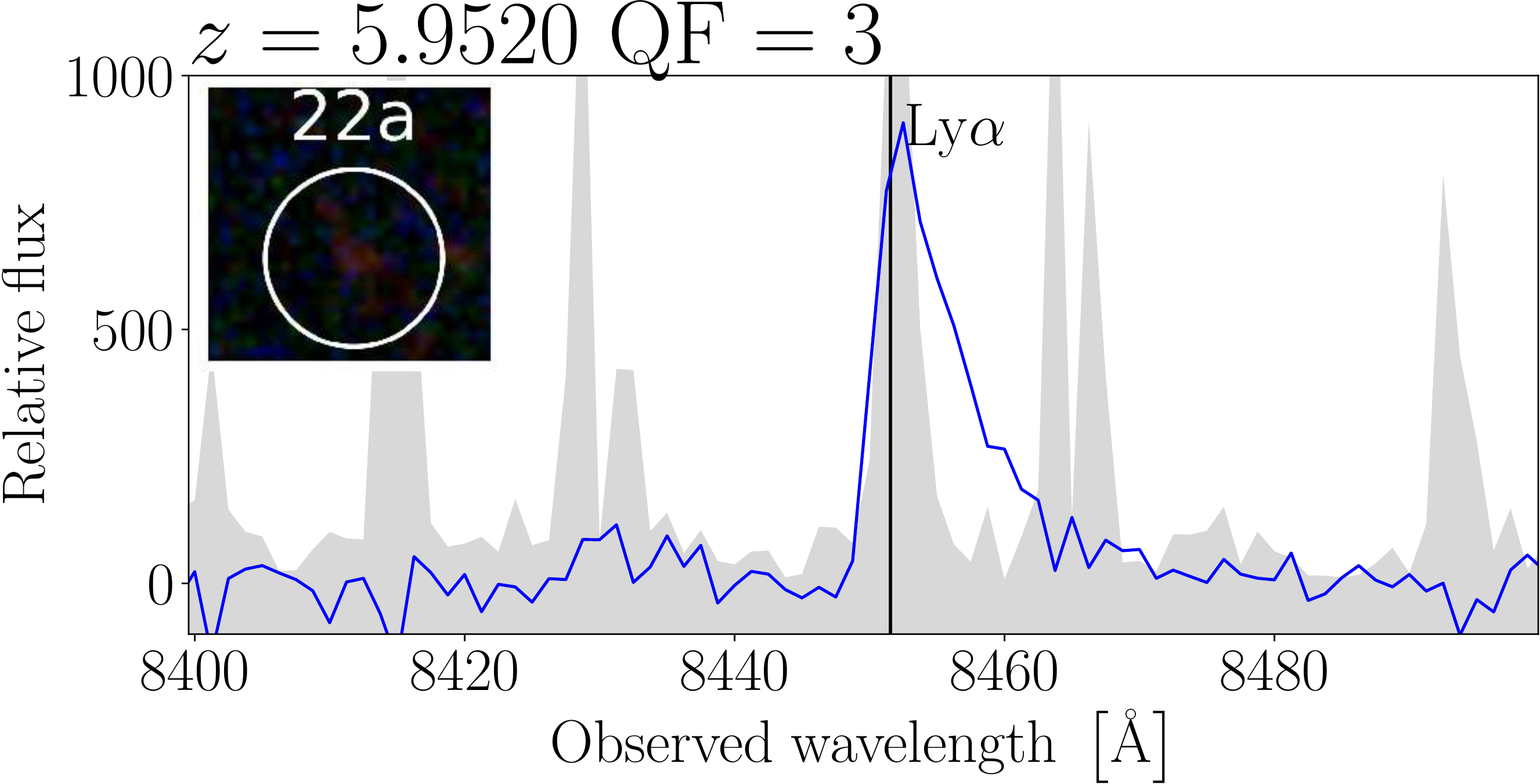}
   \includegraphics[width = 0.666\columnwidth]{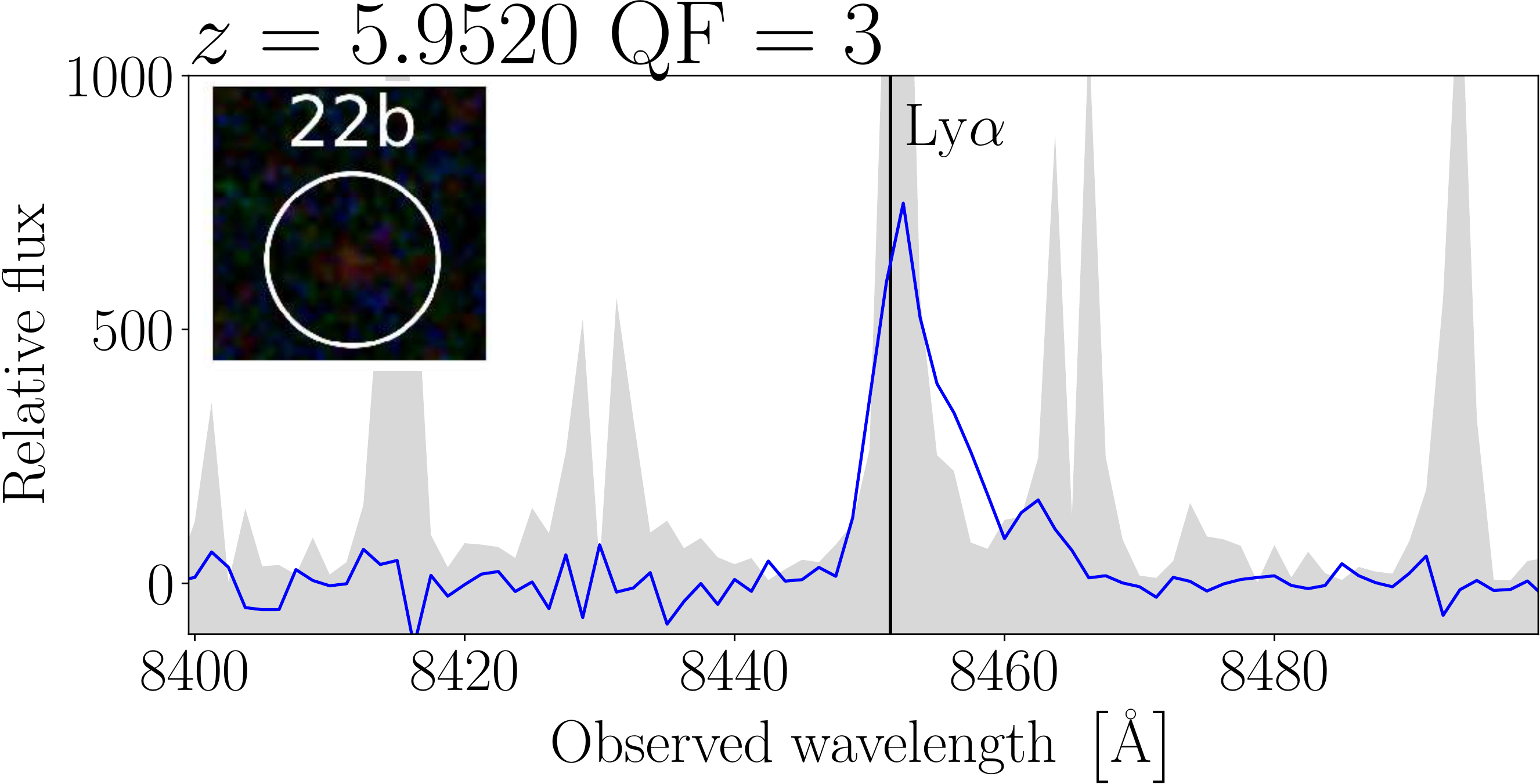} 
   
Family 23

   \includegraphics[width = 0.666\columnwidth]{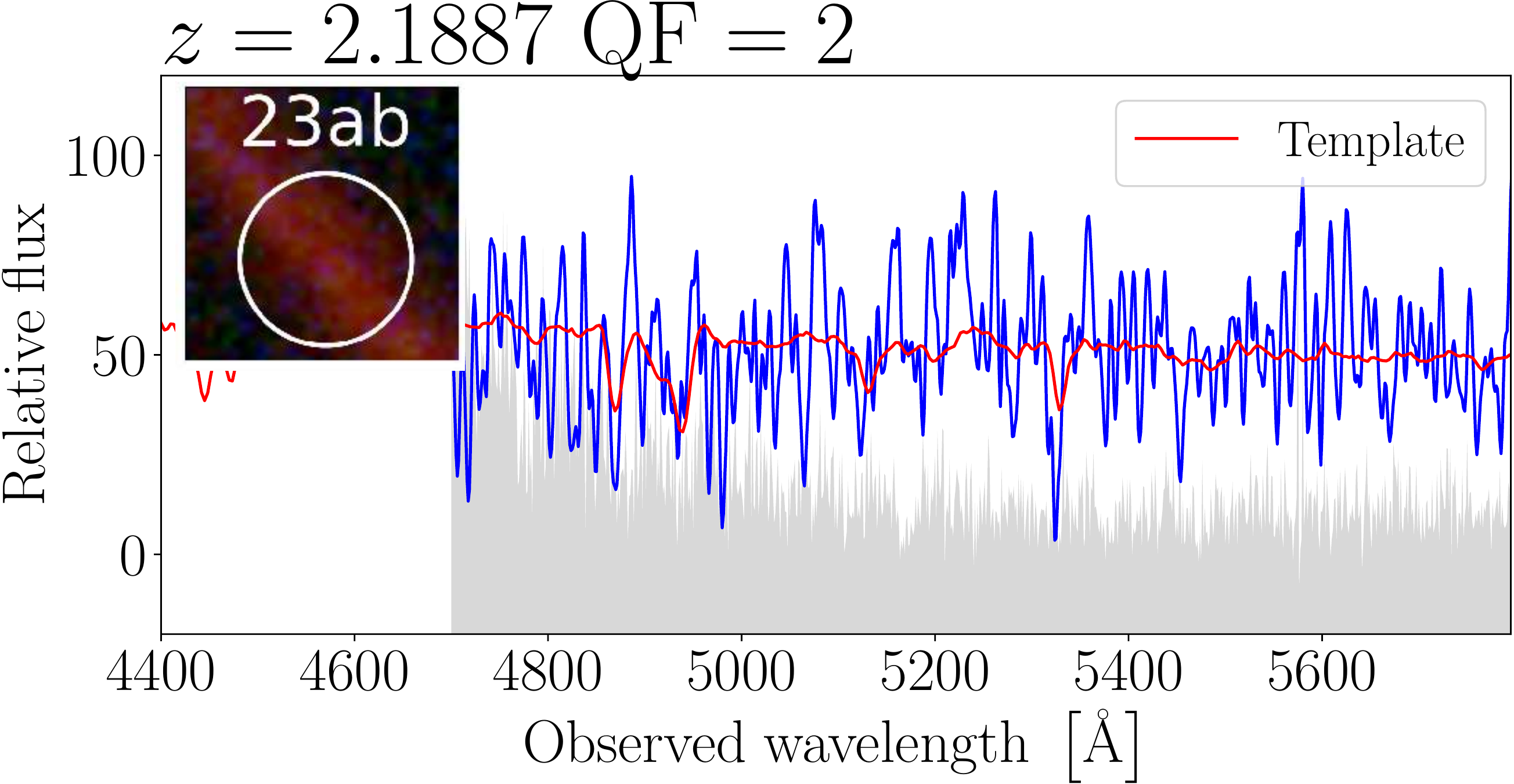}
   \includegraphics[width = 0.666\columnwidth]{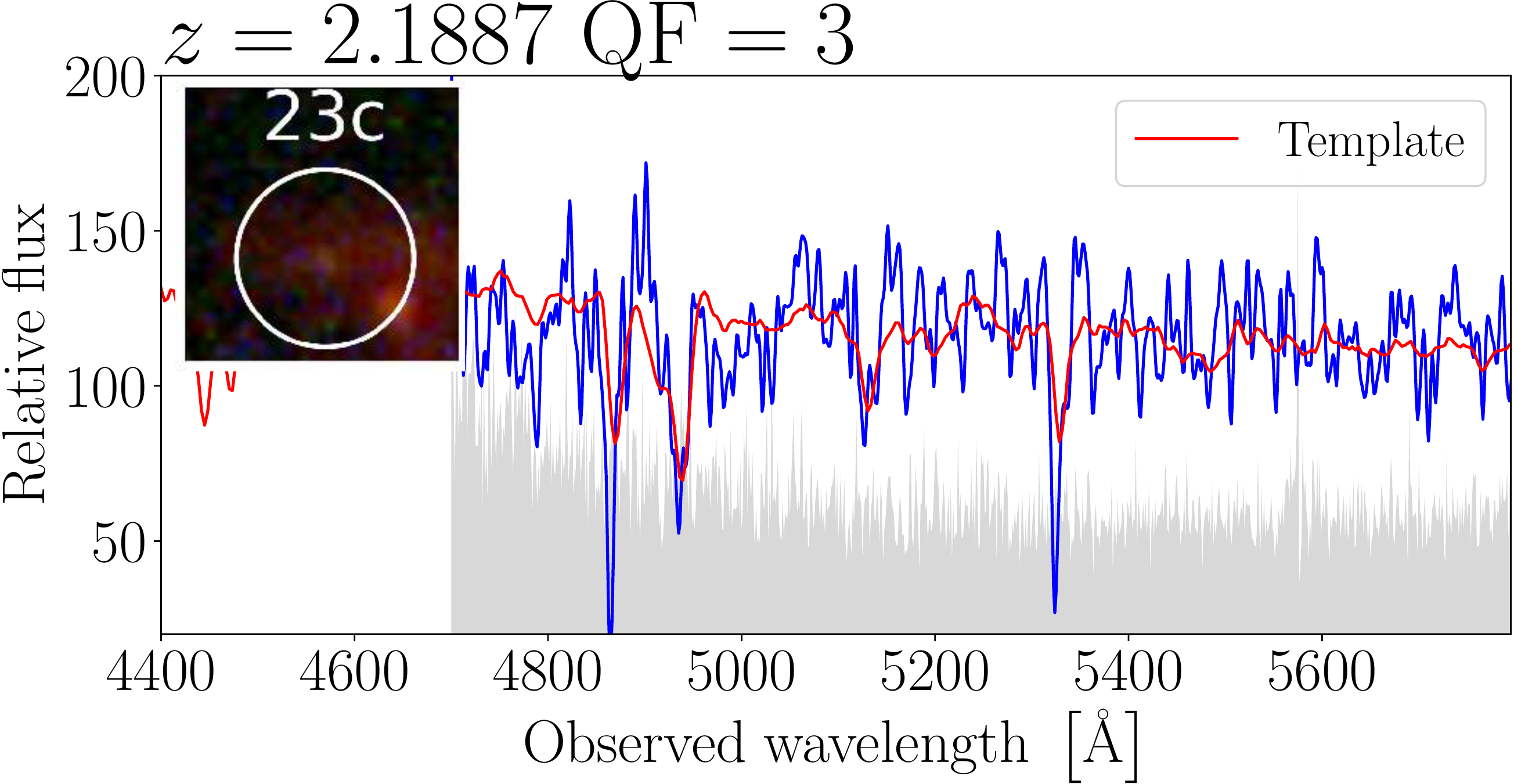} 

\caption{(Continued)}
\end{figure*}

\end{appendix}

\end{document}